\documentclass[]{pasj01}
\begin{document} 
\Received{}%{yyyy/mm/dd}
\Accepted{}%{yyyy/mm/dd}
%\Published{yyyy/mm/dd}

\title{A 16 deg$^2$ survey of emission-line galaxies at z$<$1.5 in HSC-SSP PDR1}

%%% begin:list of authors
% Do NOT capitalize all letters in "textsc".
\author{Masao \textsc{Hayashi}\altaffilmark{1}%
%\thanks{Example: Present Address is xxxxxxxxxx}
}
\altaffiltext{1}{National Astronomical Observatory of Japan, 2-21-1 Osawa, Mitaka, Tokyo 181-8588, Japan}
\email{masao.hayashi@nao.ac.jp}

\author{Masayuki \textsc{Tanaka}\altaffilmark{1}}
\author{Rhythm \textsc{Shimakawa}\altaffilmark{1,2}}
\author{Hisanori \textsc{Furusawa}\altaffilmark{1}}
\author{Rieko \textsc{Momose}\altaffilmark{3}}
\author{Yusei \textsc{Koyama}\altaffilmark{4,5}}
\author{John D. \textsc{Silverman}\altaffilmark{6}}
\author{Tadayuki \textsc{Kodama}\altaffilmark{1,5,7}}
\author{Yutaka \textsc{Komiyama}\altaffilmark{1,5}}
\author{Alexie \textsc{Leauthaud}\altaffilmark{8}}
\author{Yen-Ting \textsc{Lin}\altaffilmark{9}}
\author{Satoshi \textsc{Miyazaki}\altaffilmark{1,5}}
\author{Tohru \textsc{Nagao}\altaffilmark{10}}
\author{Atsushi J. \textsc{Nishizawa}\altaffilmark{11}}
\author{Masami \textsc{Ouchi}\altaffilmark{6,12}}
\author{Takatoshi \textsc{Shibuya}\altaffilmark{12}}
\author{Ken-ichi \textsc{Tadaki}\altaffilmark{1,13}}
\author{Kiyoto \textsc{Yabe}\altaffilmark{6}}

\altaffiltext{2}{UCO/Lick Observatory, University of California, 1156 High Street, Santa Cruz, CA 95064, USA}
\altaffiltext{3}{Institute of Astronomy, National Tsing Hua University, 101, Section 2 Kuang-Fu Road, Hsinchu, Taiwan, 30013, R.O.C.}
\altaffiltext{4}{Subaru Telescope, National Astronomical Observatory of Japan, 650 N Aohoku Pl, Hilo, HI 96720}
\altaffiltext{5}{Department of Astronomy, School of Science, Graduate University for Advanced Studies (SOKENDAI), 2-21-1, Osawa, Mitaka,Tokyo 181-8588, Japan}
\altaffiltext{6}{Kavli Institute for the Physics and Mathematics of the Universe (Kavli IPMU, WPI), The University of Tokyo, 5-1-5 Kashiwanoha, Kashiwa, Chiba, 277-8583, Japan}
\altaffiltext{7}{Astronomical Institute, Tohoku University, Aramaki, Aoba-ku, Sendai 980-8578, Japan}
\altaffiltext{8}{Department of Astronomy and Astrophysics, University of California, Santa Cruz, 1156 High Street, Santa Cruz, CA 95064 USA}
\altaffiltext{9}{Academia Sinica Institute of Astronomy and Astrophysics, P.O. Box 23-141, Taipei 10617, Taiwan}
\altaffiltext{10}{Research Center for Space and Cosmic Evolution, Ehime University, 2-5 Bunkyo-cho, Matsuyama, Ehime 790-8577, Japan}
\altaffiltext{11}{Institute for Advanced Research, Nagoya University, Chikusaku, Nagoya 464-8602, Japan} 
\altaffiltext{12}{Institute for Cosmic Ray Research, The University of Tokyo, 5-1-5 Kashiwanoha, Kashiwa, Chiba 277-8582, Japan} 
\altaffiltext{13}{Max-Planck-Institut f$\rm\ddot{u}$r Extraterrestrische Physik, Giessenbachstrasse, D-85748 Garching, Germany}

%% `\KeyWords{}' always has to be placed before `\maketitle'.
 %Do NOT move this preamble from here! 

\maketitle

%%%%%%%%%%%%%%%%%%%%%%%%%%%%%%%%%%%%%%%%%%%%%%%%%%%%%%%%%%%%%%%%
%%%%%%%%%%%%%%%%%%%%%%%%%%%%%%%%%%%%%%%%%%%%%%%%%%%%%%%%%%%%%%%%

\begin{abstract}
  We present initial results from the Subaru Strategic Program (SSP)
  with Hyper Suprime-Cam (HSC) on a comprehensive survey of
  emission-line galaxies at $z<1.5$ based on narrowband imaging.
  The first Public Data Release (PDR1) provides us with data from two
  narrowband filters, specifically NB816 and NB921 over 5.7 deg$^2$
  and 16.2 deg$^2$ respectively. The $5 \sigma$ limiting magnitudes
  are 25.2 (UltraDeep layer, 1.4 deg$^2$) and 24.8 (Deep layer, 4.3
  deg$^2$) mag in NB816, and 25.1 (UltraDeep, 2.9 deg$^2$) and
  24.6--24.8 (Deep, 13.3 deg$^2$) mag in NB921. The wide-field imaging
  allows us to construct unprecedentedly large samples of 8,054
  H$\alpha$ emitters at $z \approx$ 0.25 and 0.40, 8,656 [OIII]
  emitters at $z \approx$ 0.63 and 0.84, and 16,877 [OII] emitters at
  $z \approx$ 1.19 and 1.47.
  We map the cosmic web on scales out to about 50 comoving Mpc that
  includes galaxy clusters, identified by red sequence galaxies,
  located at the intersection of filamentary structures of
  star-forming galaxies.  
  The luminosity functions of emission-line galaxies are measured with
  precision and consistent with published studies. The wide field
  coverage of the data enables us to measure the luminosity functions
  up to brighter luminosities than previous studies. The comparison of
  the luminosity functions between the different HSC-SSP fields
  suggests that a survey volume of $>5\times10^5$ Mpc$^3$ is essential
  to overcome cosmic variance.
  Since the current data have not reached the full depth expected for
  the HSC-SSP, the color cut in $i-$NB816 or $z-$NB921 induces a bias
  towards star-forming galaxies with large equivalent widths,
  primarily seen in the stellar mass functions for the H$\alpha$
  emitters at $z \approx$ 0.25--0.40.
  Even so, the emission-line galaxies clearly cover a wide range of
  luminosity, stellar mass, and environment, thus demonstrating the
  usefulness of the narrowband data from the HSC-SSP to investigate
  star-forming galaxies at $z<1.5$.
\end{abstract}

%%%%%%%%%%%%%%%%%%%%%%%%%%%%%%%%%%%%%%%%%%%%%%%%%%%%%%%%%%%%%%%%
%%%%%%%%%%%%%%%%%%%%%%%%%%%%%%%%%%%%%%%%%%%%%%%%%%%%%%%%%%%%%%%%

\section{Introduction}
\label{sec:intro}

Emission lines from HII regions in galaxies are one of the important
spectral features to characterize galaxies. The intensity of the
nebular emission in the rest-frame optical wavelength such as
H$\alpha$ ($\lambda=6565$\AA\ in a vacuum),
[OIII] ($\lambda\lambda=4960,5008$\AA),
H$\beta$ ($\lambda=4863$\AA)
and [OII] ($\lambda\lambda=3727,3730$\AA) is sensitive to star
formation rate (SFR) of galaxies and thus widely used as an indicator
of SFR of galaxies (e.g., \cite{Kennicutt1998,Moustakas2006}).
While the luminosity of hydrogen lines such as H$\alpha$ and H$\beta$
is directly linked to the number of ionizing photons, the physics
associated with the emission lines from oxygen, caused by collisional
excitation, are more complicated than those from hydrogen and depend
on the physical condition of the nebular gas such as metallicity and
ionization state. However, it is known that the intensity of emission
lines from ionized oxygen, especially from [OII], is a usable
indicator of SFR in galaxies not only in the local Universe but also
at high redshifts (e.g.,
\cite{Kennicutt1998,Kewley2004,Moustakas2006,Hayashi2013,Hayashi2015}).
Recent studies also demonstrate that typical star forming galaxies at
high redshifts, especially $z>$ 2--3, are often identified as [OIII]
emission-line galaxies
(\cite{Ly2007,Drake2013,Khostovan2015,Suzuki2016}). 
Therefore, emission lines are a useful tool to sample and investigate
star-forming galaxies at various redshifts.     

Many surveys, targeting emission-line galaxies at $z\lesssim2$, have
been conducted so far (e.g.,  
\cite{Bunker1995,Thompson1996,Moorwood2000,vanderWerf2000,Fujita2003,Doherty2006,Shioya2008,Villar2008,Dale2010,Ly2007,Ly2011,Bayliss2011,Lee2012,Best2013,Colbert2013,Ciardullo2013,Drake2013,Pirzkal2013,Sobral2013,Sobral2015,An2014,Khostovan2015,Comparat2015,Comparat2016,Stroe2015,Stroe2017}).  
As a result, the cosmic SFR density has been established with a peak
at $z=$ 1--3 and then a gradually decline towards the local Universe
(e.g., \cite{HopkinsBeacom2006,MadauDickinson2014}). Star-forming
galaxies at each redshift show a tight correlation between SFR and
stellar mass irrespective of environment where galaxies are located,
which is called a main sequence of star-forming galaxies 
(e.g., \cite{Daddi2007,Elbaz2007,Noeske2007,Koyama2013,Koyama2014,Suzuki2016,Hayashi2016,Ramraj2017,Oteo2015,Sobral2014,Kashino2013,Shivaei2015}),
and star-forming galaxies at a fixed stellar mass have smaller SFRs
at lower redshifts (e.g., \cite{Speagle2014,Whitaker2014,Tomczak2016}).
The main sequence of star-forming galaxies indicates that only a few
percent of galaxies are in a starburst phase \citep{Rodighiero2011}.
The fact that few galaxies are outliers from the main sequence is
considered to be an evidence that a time scale of starburst phase and
quenching of star formation is short. However, physics governing the
evolution of individual galaxies is not yet well understood. What 
triggers the starburst in galaxies? What quenches the star formation
of galaxies? How have galaxies evolved along large-scale structures of
the cosmic web? As a first step to address these important issues for
understanding galaxy evolution, a more comprehensive sampling of
star-forming galaxies at the redshifts of $z\lesssim3$ is required
that covers a wide range in terms of star formation activity and
environment.  
Recent studies have indeed demonstrated the importance of
investigating galaxy properties along the cosmic web (e.g.,
\cite{Darvish2014,Darvish2015,Darvish2016,Darvish2017,Malavasi2017,Laigle2017,Kuutma2017}).

Surveys of emission-line galaxies are basically performed with either
spectroscopy or narrowband (NB) imaging.  
Spectroscopic surveys have the advantage of identifying emission lines
and measuring line luminosities. However, target selection for
spectroscopy can cause biases with the observed galaxy sample. Rather
than slit or fiber spectroscopy, an integral field unit spectrograph
such as VLT/MUSE \citep{Bacon2015,Meillier2016,Bina2016,Swinbank2017}
or grism such as HST/WFC3 (e.g.,
\cite{Atek2010,vanDokkum2011,Straughn2011,vanderWel2011,Brammer2012,Colbert2013,Pirzkal2013,Mehta2015,Morris2015,Momcheva2016}) 
can overcome selection biases by allowing us to get spectrum of all
galaxies over the field surveyed. One weakness is the small field of
view (FoV). Alternatively, NB imaging surveys of emission-line
galaxies allow us to cover a wide, homogeneous FoV without any bias
for target selection.
Although, the selection with NB imaging requires an equivalent 
width (EW) of emission lines larger than a certain limit, thus
sampling strong nebular emission lines.
As a result, this approach is able to effectively produce
comprehensive samples of star-forming galaxies.   

Indeed, previous studies with NBs
(e.g., \cite{Ly2007,Drake2013,Sobral2011,Sobral2013,Sobral2015,Khostovan2015,Matthee2017})
have succeeded in constructing large samples of emission-line galaxies
at $z<2$.  
\citet{Ly2007} have selected 200-900 emission-line galaxies with 11
individual redshift slices at $z<1.5$ from deep imaging with 4 NB
filters in 0.24 deg$^2$ of the Subaru Deep Field (SDF). 
\citet{Drake2013} have presented catalogs of
more than 5000 emission-line galaxies at $z\lesssim1.6$ from a survey
with 6 NB filters in 0.63 deg$^2$ of the Subaru/XMM-Newton Deep Survey
(SXDS) field. Samples of emission-line galaxies of similar size at
$z\lesssim2$ are also presented by the High-redshift(Z) Emission Line
Survey (HiZELS).
\citet{Sobral2013} have conducted deep surveys in $\sim2.0$ deg$^2$ of
the COSMOS and SXDS fields, while \citet{Sobral2015} have conducted
shallower but wider surveys in $\sim10$ deg$^2$ of the SA22 fields.
They map large-scale structures of H$\alpha$ emitters over these
fields (see also \cite{Sobral2011}).

%%%%%%%%%%%% Figure 1 %%%%%%%%%%%%%%%%%
\begin{figure}  
  \begin{center}
    \includegraphics[width=0.5\textwidth]{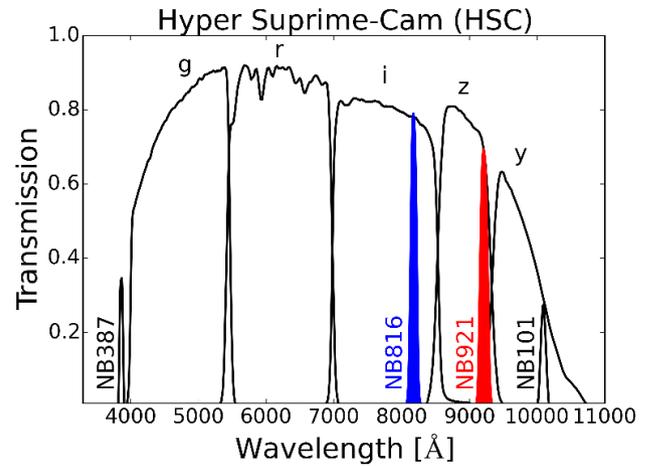}
  \end{center}
  \caption{The set of broadband (BB) and narrowband (NB) filters used
    in the HSC-SSP. Among the four NB filters, the data in two NB
    filters, NB816 (blue) and NB921 (red), are available with the PDR1
    data (Table\ref{tab:NBdata}). The response function of the filters
    and the quantum efficiency of CCD are taken into account for the
    transmission. The specification of the NB is shown in
    Table~\ref{tab:NBemitters}.   
  }\label{fig:filters}   
\end{figure}
%%%%%%%%%%%%%%%%%%%%%%%%%%%%%%%%%%%%%%%%

%%%%%%%%%%%% Table 1 %%%%%%%%%%%%%%%%%
\begin{table*}
  \tbl{Narrowband data available in HSC-SSP PDR1.}{%
    \begin{tabular}{lccccccc}
      \hline
      & \multicolumn{7}{c}{Narrowband filter} \\
      Field$^\ast$ & \multicolumn{3}{c}{NB816} && \multicolumn{3}{c}{NB921} \\
      \cline{2-4}\cline{6-8}
      & Area$^\dagger$ [deg$^2$] & Integration [hour] & Limit mag.$^\ddagger$ && Area$^\dagger$ [deg$^2$] & Integration [hour] & Limit mag.$^\ddagger$ \\
      \hline
      UD-COSMOS
      &$\cdot\cdot\cdot$&$\cdot\cdot\cdot$&$\cdot\cdot\cdot$&&
      1.47 (1.54) & 7.0 & 25.1\\
      UD-SXDS & 1.43 & 4.0 & 25.2&&
      1.49 & 4.8 & 25.1\\
      D-COSMOS
      &$\cdot\cdot\cdot$&$\cdot\cdot\cdot$&$\cdot\cdot\cdot$&&
      2.27 (2.91) & 2.0 & 24.8\\
      D-DEEP2-3 & 4.25 & 1.0 & 24.8 &&
      5.63 & 1.0 & 24.6\\
      D-ELAIS-N1
      &$\cdot\cdot\cdot$&$\cdot\cdot\cdot$&$\cdot\cdot\cdot$&&
      5.37 & 1.0 & 24.6\\
      \hline
      total area & 5.68 & & && 16.2 & & \\      
      \hline      
  \end{tabular}}
  \label{tab:NBdata}
  \begin{tabnote}
    $^\ast$ The coordinates of fiducial pointing in each field are
    defined in the survey design paper of \citet{HSCSSP}.\\   
    $^\dagger$ An effective area, i.e., the masked regions are
    excluded. The value in the parenthesis shows the effective area
    including the overlapping regions (0.72 deg$^2$) between the
    UD-COSMOS and D-COSMOS fields.\\   
    $^\ddagger$ A median value of $5\sigma$ limiting magnitudes in the
    individual {\tt patch}\footnotemark[\ref{foot:terms}] regions,
    which are estimated from the standard deviation of sky values
    measured with randomly-distributed 2\arcsec\ diameter apertures.
  \end{tabnote}
\end{table*}
%%%%%%%%%%%%%%%%%%%%%%%%%%%%%%%%%%%%%%%

%%%%%%%%%%%% Table 2 %%%%%%%%%%%%%%%%%
\begin{table*}
  \tbl{Narrowband filters.}{%
    \begin{tabular}{cccccccccccc}
      \hline
      & & &&
      \multicolumn{2}{c}{H$\alpha$} &&
      \multicolumn{2}{c}{[OIII]} &&
      \multicolumn{2}{c}{[OII]} \\
      & $\lambda_c$ $^\dagger$ [\AA] & $\Delta\lambda$ $^\dagger$ [\AA] &&
      \multicolumn{2}{c}{($\lambda=6564.6$)$^\ddagger$} &&
      \multicolumn{2}{c}{($\lambda=5008.2$)$^\ddagger$} &&
      \multicolumn{2}{c}{($\lambda=3727.1,3729.9$)$^\ddagger$} \\
      \cline{5-6}\cline{8-9}\cline{11-12}
      & & &&
      $z$ & $z$ range &&
      $z$ & $z$ range &&
      $z$ & $z$ range\\      
      \hline
      NB816 & 8177 & 113 &&      
      0.246 & 0.237 -- 0.254 &&
      0.633 & 0.621 -- 0.644 &&
      1.19 & 1.18 -- 1.21 \\ 
      NB921 & 9214 & 135 &&      
      0.404 & 0.393 -- 0.414 &&
      0.840 & 0.826 -- 0.853 &&
      1.47 & 1.45 -- 1.49 \\
      \hline      
  \end{tabular}}
  \label{tab:NBemitters}
  \begin{tabnote}
    $^\dagger$ The value is derived from the area-weighted mean
    response function \citep[in prep.]{HSCFILTERS}.\\ 
    $^\ddagger$ Vacuum wavelength in the rest-frame in units of \AA.
  \end{tabnote}
\end{table*}
%%%%%%%%%%%%%%%%%%%%%%%%%%%%%%%%%%%%%%%

Here, we use Hyper Suprime-Cam (HSC), an instrument on the Subaru
Telescope capable of delivering good image quality over a FoV of 1.77
deg$^2$ in a single pointing \citep{Miyazaki2012,HSC}. The Subaru
Strategic Program (SSP) with HSC is a three-layered (Wide, Deep
(D), and UltraDeep (UD)), multi-band (five broadband (BB) filters:
$grizy$ plus four NB filters: NB387, NB816, NB921, and NB101, 
Figure~\ref{fig:filters}) imaging survey \citep{HSCSSP}. The survey
started in March 2014 and is ongoing. The HSC-SSP program will be
conducted in 300 nights spread over 5--6 years. The observations with
the NB filters are conducted in the UD and D fields. When the program is
completed, the NB921 data will reach down to 26.5 (25.9) mag and the
$z$-band data down to 27.1 (26.6) mag over 3.5 (26) deg$^2$
of the UD (D) fields \citep{HSCSSP}. The set of four NB filters
installed in HSC enables us to select not only Lyman $\alpha$ emitters
at $z\sim2.2$, 5.7, 6.7, and 7.3, but also galaxies with nebular
emission such as H$\alpha$, [OIII]($\lambda\lambda=4960,5008$\AA), and
[OII]($\lambda\lambda=3727,3730$\AA) at $z\lesssim1.7$.    
Undoubtedly, the HSC-SSP survey provides us with one of the most
comprehensive samples of line-emitting galaxies which are useful to
study the evolution of galaxies at low and intermediate 
redshifts. Therefore, the goal of this paper is to construct catalogs
of the emission-line galaxies from the HSC-SSP PDR1 data and then
investigate basic global properties of the selected galaxies.  

The outline of this paper is as follows.
In \S~\ref{sec:data}, the NB data from HSC-SSP are described and the
quality is verified.
In \S~\ref{sec:ELGs}, emission-line galaxies at $z=$ 0.25--1.47 are
selected and then catalogs of the galaxies are created.
In \S~\ref{sec:results}, the spatial distribution and luminosity
functions are constructed.
In \S~\ref{sec:discussions}, we discuss cosmic variance, the bright end of
the luminosity function while taking advantage of the wide field data.
We also investigate the stellar mass function for the emission-line
galaxies to better understand the samples of emission-line galaxies.
Finally, our conclusions are presented in \S~\ref{sec:conclusions}.
Throughout this paper, a composite model magnitude named {\tt cmodel}
is used for the photometry of galaxies and magnitudes are presented in the AB
system \citep{OkeGunn1983}, unless otherwise mentioned.  
The {\tt cmodel} photometry measures fluxes of objects by
simultaneously fitting two components of a de Vaucouleur and an
exponential profile convolved with point spread function (PSF)
(see \cite{hscPipe} for the details).
The cosmological parameters of $H_0=70$ km s$^{-1}$ Mpc$^{-1}$,
$\Omega_m=0.3$ and $\Omega_\Lambda=0.7$, along with
\citet{Chabrier2003} initial mass function (IMF), are adopted.  

%%%%%%%%%%%%%%%%%%%%%%%%%%%%%%%%%%%%%%%%%%%%%%%%%%%%%%%%%%%%%%%%
%%%%%%%%%%%%%%%%%%%%%%%%%%%%%%%%%%%%%%%%%%%%%%%%%%%%%%%%%%%%%%%%

\section{DATA}
\label{sec:data}

\subsection{HSC-SSP PDR1}
\label{sec:data.hscssp}
This work is based on the first Public Data Release (PDR1) of the
HSC-SSP which were available on 2017 February 28 \citep{HSCSSPDR1}.
The processing of the HSC-SSP data, including data reduction, object
detection, and photometry, is conducted with the HSC software 
pipeline (hscPipe 4.0.1, \cite{hscPipe}), essentially equivalent to
that for the Large Synoptic Survey Telescope (LSST,
\cite{Ivezic2008,Axelrod2010,Juric2015}). 
The astrometry and photometry are calibrated with Pan-STARRS1 (PS1,
\cite{Tonry2012,Schlafly2012,Magnier2013}).
One should refer to \citet{HSCSSPDR1} for full details regarding the
data set of the HSC-SSP PDR1 and to \citet{hscPipe} for the data
processing. Hereafter, we briefly describe critical points with
respect to the data products including the photometry. 

In this release, NB data from two filters (NB816 and NB921;
Table~\ref{tab:NBdata}) are available. Of the two UD fields and four D
fields defined in the HSC-SSP survey, the NB816 data are taken in one
UD field and one D field, while the NB921 data are taken in both UD
fields and three D fields, which results in an effective coverage of
5.68 (16.2) deg$^2$ by 4 (12) FoVs of HSC using NB816 (NB921). 
Since the NB921 data have an 0.72 deg$^2$ of overlap between the
UD-COSMOS and D-COSMOS fields, we account for this when determining
the total effective area coverage. The median seeing is 0.62 (0.70)
arcsec in NB816 (NB921), respectively. The integration time of the NB
data are 4-7 hours in the UD fields and 1-2 hours in the D fields. The
$5\sigma$ limiting magnitudes of NB816 are 25.2 and 24.8 in the UD and
D fields, and those of NB921 are 25.1 and 24.6 in the UD and D fields.  
The limiting magnitudes provided here are estimated from the standard
deviation of sky values within randomly-distributed 2\arcsec\ diameter
apertures. Note that \citet{HSCSSPDR1} measure the $5\sigma$ depth for
point sources in a different manner.
The data from all of the five BBs, $grizy$, are also available in the
area covered by the NB data. The median seeing ranges from 0.61 --
0.83 arcsec. The $i$ and $z$-band data are $>0.5$ mag deeper than the
NB816 and NB921 data, thus indicating a depth sufficient for the
selection of emission-line galaxies.   

hscPipe conducts photometry in two ways for all individual sources
detected in any of the available bands. One is {\tt unforced}
photometry that is conducted in each band separately, while the other
is {\tt forced} photometry that is conducted in each band on a fixed
position with the fixed profile of objects determined in a single
reference band.
The reference band for the {\tt forced} photometry is selected
according to the priority order of $i,r,z,y,g$, NB921 and NB816 for
the individual objects based on which band and at what signal-to-noise
ratio (SNR) the objects are detected. For instance, even if an object
is detected in NB, as long as it has a detection in any of BBs at high
enough SNR, the BB (most likely $i$-band) is selected as a reference
band for the {\tt forced} photometry.  

\subsection{Catalogs of galaxies detected in narrowbands}
\label{sec:data.nb}

Catalogs of objects are retrieved from the Catalog Archive Server
(CAS) of the HSC-SSP data \citep[in prep.]{hscDatabase}.
We describe how to select the NB-detected galaxies in this section,
and the details of the flags we apply are shown in
Appendix~\ref{app:catalog}. 

\subsubsection{NB-detected objects}
\label{sec:data.nb.cas}
Detection in any of the NB filters is a necessary condition, because
we are interested in galaxies with a nebular emission line that enters
NB816 or NB921. We select deblended objects that are detected 
in NB816 or NB921 at a SNR,
{\tt flux\_cmodel}/{\tt flux\_cmodel\_err}, greater than 5 in the
{\tt unforced} photometry. 
We also require that the objects meet the criteria for a likely
significant detection of the {\tt cmodel} measurement in a region that
is not affected by saturation, cosmic rays, and bad pixels.
Since the edge regions where the number of co-added frames in NB816 or
NB921 is less than 3 (10) in the D (UD) fields are shallower than the
other regions of the D (UD) fields, objects in the shallow regions are
removed. 
With these requirements, we generate catalogs of NB-detected
objects with clean photometry from the catalog database. However, we
notice that there can be satellite trails, moving objects, and 
ghost-like features  left in the coadded NB images, which are the known
problems in the HSC-SSP PDR1 \citep{HSCSSPDR1}. Some regions around
very bright stars and near the edge of the FoV also have bad
quality. Therefore, we apply masks to exclude such objects and regions
from the catalogs. While hscPipe already provides bright object masks
and flags for masking \citep{BrightStarMask}, the masks that we
produce as described in the next section are complementary to those
provided by the HSC-SSP.  

%%%%%%%%%%%% Figure 2 %%%%%%%%%%%%%%%%%
\begin{figure}[t]  
  \begin{center}
    \includegraphics[width=0.5\textwidth]{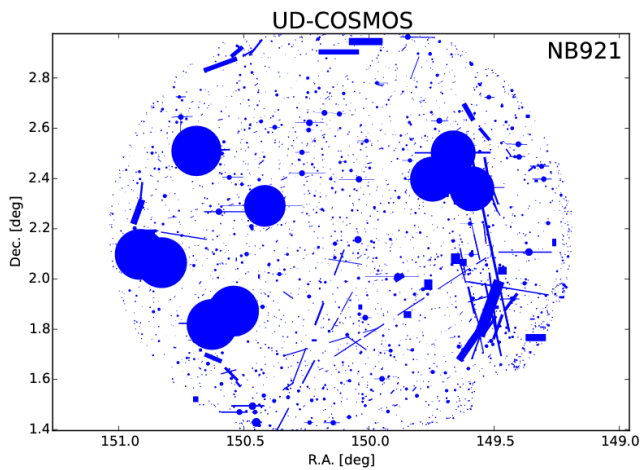}    
  \end{center}
  \caption{The mask regions for NB921 data in the UD-COSMOS field are
    shown in blue, which are used to exclude satellite trails, moving
    objects, and ghost-like features in the coadded NB
    images.}\label{fig:mask}   
\end{figure}
%%%%%%%%%%%%%%%%%%%%%%%%%%%%%%%%%%%%%%%%

\subsubsection{Masked regions}
\label{sec:data.nb.mask}

Because a single {\tt visit}%
\footnote{\label{foot:terms}
  The followings are HSC/LSST-specific terms seen in this paper \citep{HSCSSPDR1}.
  {\tt visit} is an ID number of the individual shots.
  {\tt tract} is a pre-defined region covering $\sim1.7\times1.7$ deg$^2$.  
  A single {\tt tract} is divided into $9\times9$ sub-regions called
  {\tt patch} that covers $\sim12\times12$ arcmin$^2$.
} %
with the NB filter is taken with a long exposure time (15 min),
satellite trails are seen in NB images. Note that the bright satellite
trails are removed from the {\tt coadd} image by combining the
{\tt visit} images while clipping pixels having significantly deviant
fluxes. If there are satellite trails and moving objects only in the
NB images, such objects can be regarded as objects that are much
brighter in NB than in BB and then misidentified as emission-line
galaxies. Therefore, it is important to exclude such objects by masking 
them.  

To find the satellite trails and moving objects in the {\tt coadd} NB
image, the individual warped {\tt visit} images are examined. Among
pixels with a detection flag on in the mask layer stored in the
{\tt coadd} fits image, we search for pixels that attribute the
detection to a large flux from only a single {\tt visit} image. Pixels
forming a line on the {\tt coadd} image are candidates of satellite
trails and moving objects. We set the mask regions to cover the pixels
of satellite trails and moving objects. We then make fine adjustments
to the masks manually. We also add additional mask regions for the
bleed trails of saturated stars based on the mask image layer of the
{\tt coadd} fits image. 

Figure~\ref{fig:mask} shows the mask regions for NB921 data in the
UD-COSMOS field. The masks for NB816 and NB921 data are available in
all of the other fields as well. The effective area shown in
Table~\ref{tab:NBdata} is an area of the unmasked region that is
calculated with the {\tt random} catalog including the randomly
distributed objects \citep{HSCSSPDR1}.   

\subsubsection{Removal of stars and junk objects}
\label{sec:data.nb.stars}
Furthermore, we need to exclude stellar objects as well as junk
objects that hscPipe cannot identify. 
In order to exclude stars, we use a probability of an object being a
star, {\tt pstar}, which is estimated by a supervised learning
star/galaxy classification code \citep{hscPipe}. The
information of {\tt pstar} is available in the S15B internal release
but not in the HSC-SSP PDR1 data as of the public release in 2017
February.
Note that in the UD and D fields the S15B internal release data are
equivalent to the PDR1 data \citep{HSCSSPDR1}.
We also use the flag of {\tt classification\_extendedness}, which can
identify stellar objects with few contaminants down to $i\sim23$
\citep{HSCSSPDR1}.
We find that the two flags are complementary and the combination of
{\tt pstar} and {\tt classification\_extendedness} is more effective
at identifying stellar objects. Objects with the flag of {\tt
  pstar$>$0.5} or {\tt classification\_extendedness$=$0} $\land$ {\tt
  mag\_psf$<$23.0} in $i$-band are rejected as stellar objects, which
amounts to $\sim$7-13\% of the NB-detected objects in each field. This
procedure can exclude not only stars but also point sources such as
active galactic nuclei (AGNs), which is discussed in
\S~\ref{sec:ELGs.AGN}.
We find that large objects with a minor axis greater than 3 arcsec or
small objects with a major axis less than the size of PSF are highly
likely to be flagged as a junk object. These objects are also removed from
the catalogs.

\subsection{Quality verification of NB data}
\label{sec:data.nbquality}

%%%%%%%%%%%% Figure 3 %%%%%%%%%%%%%%%%%
\begin{figure}
  \begin{center}
    \includegraphics[width=0.5\textwidth]{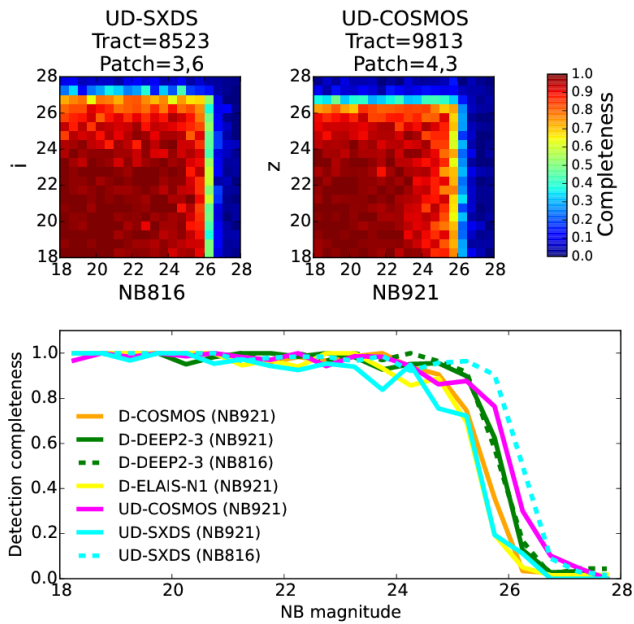}    
  \end{center}
  \caption{Detection completeness as a function of $i$ ($z$) and NB816
    (NB921) magnitudes. The upper left panel shows the completeness for
    NB816-detected objects in the UD-SXDS field, while the upper right
    panel shows for NB921-detected objects in the UD-COSMOS
    field. Synthetic objects with a profile of the PSF and various 
    combination of $i$ and NB816 magnitudes (or $z$ and NB921) are
    randomly embedded within images to measure the completeness. The
    lower panel shows the detection completeness in NBs for the
    individual fields as a function of NB magnitude.
  }\label{fig:completeness}
\end{figure}
%%%%%%%%%%%%%%%%%%%%%%%%%%%%%%%%%%%%%%%%

\subsubsection{Detection completeness}
\label{sec:data.nbquality.completeness}
Detection completeness is important to determine the faint end of
number counts (\S~\ref{sec:data.nbquality.numbercount}) and luminosity
functions (\S~\ref{sec:results.lf}). To assess the completeness for
individual fields, we use the software pipeline, SynPipe \citep{SynPipe}, 
which can randomly embed artificial objects in the individual CCD
images. Then, we run hscPipe on the CCD images with the synthetic
objects to conduct the same processes of coadd, object detection, and
photometry as for the original data.  
We compare the catalog of detected sources output by hscPipe with
the list of synthetic objects input to SynPipe, where we use an
aperture with a radius of 2 pixels (i.e., 0.34 arcsec) for matching
between the input and output catalogs without any constraint on
magnitude difference.
Since emission-line galaxies are selected based on a color of NB and
BB both of which cover the same wavelengths, the completeness of
emission-line galaxies should depend on the color of $i-$NB816
($z-$NB921) for NB816 (NB921) emitters. Therefore, we investigate the
completeness as a function of $i$ ($z$) and NB816 (NB921) as shown in
Figure~\ref{fig:completeness}.      

We construct objects with a profile of the PSF as synthetic objects
for the measurement of detection completeness.
The PSF is modeled on each CCD image by hscPipe based on the actual
stars (refer to \cite{hscPipe} for the details of the PSF modeling).
We note that galaxies do not have a profile of the PSF but rather have
an extended profile. However, star-forming galaxies at $z<2$ have
varying profiles depending on their intrinsic properties thus it is
reasonable to assume such a simplification of the profile of synthetic
objects. Our assumption of the PSF profile can result in the
overestimate of the detection completeness.  

To investigate the completeness of objects with a given $i$ or $z$
magnitude, we make an input list of 1,000 synthetic objects with NB816
or NB921 of 18.0--28.0 mag at 0.5 mag bin of BB in each 
{\tt patch}. Then, twenty input
lists are prepared to cover the same range of BB magnitudes as the NB
magnitudes (18.0--28.0 mag). In this procedure, we run SynPipe
in a single {\tt patch} in each field per each NB data. We
select the following representative {\tt patch}, where there
are not any very bright stars, satellite trails, and bad areas, for
each field:   
({\tt tract}\footnotemark[\ref{foot:terms}], {\tt patch}) =
(9813, `7,7') for D-COSMOS,
(9706, `1,3') for D-DEEP2-3,
(17130, `3,2') for D-ELAIS-N1,
(9813, `4,3') for UD-COSMOS,
(8523, `3,6') for UD-SXDS.
The detection completeness estimated in UD-SXDS (UD-COSMOS) for NB816
(NB921) is shown in Figure~\ref{fig:completeness}. 
As expected, the completeness is lower at fainter magnitude in both BB
and NB. 

\citet{HSCSSPDR1} also measure the detection completeness by
embedding synthetic objects with a PSF profile in coadded images
directly and then running hscPipe to retrieve the embedded
objects. Our measurement is consistent with theirs.

\subsubsection{Dependence of NB response function on radius}
\label{sec:data.nbquality.NBresponse}
Since the diameter of the HSC filters is as large as 60 cm, the filter
response curve is slightly dependent on the position on the filter.
The measurement of response curve at the various positions of the
filter shows that the response curve can change along the radial
direction and not along angular direction. The central wavelength and
full width at half maximum (FWHM) of the response curve can become
longer and wider by 14\AA\ (15\AA) and 4.0\AA\ (4.5\AA) at larger
radius for NB816 (NB921), respectively.
One should also refer to \citet[in prep.]{HSCFILTERS} for the details
of the HSC filters.

We compare the photometry between UD and D layer data for identical
objects in an overlapping {\tt patch} of the COSMOS field. Since the
pointing coordinates are different between the UD and D layers, the
objects in the region of the UD-COSMOS are observed in a position
closer to the center of HSC FoV than those in the D-COSMOS. We make
sure that no systematic difference is seen between the photometry. We
also investigate the dependence of the BB-NB colors in galaxies and
stars and the number counts of NB-detected galaxies on the {\tt
  patch}. All of the validation tests suggest that there is no
significant impact of the slight non-uniformity in the filter response
on the photometry of objects in the NBs, the detection of the objects
in the NBs, and selection of NB emission-line galaxies.    

%%%%%%%%%%%% Figure 4 %%%%%%%%%%%%%%%%%
\begin{figure}[t]
  \begin{center}
    \includegraphics[width=0.5\textwidth]{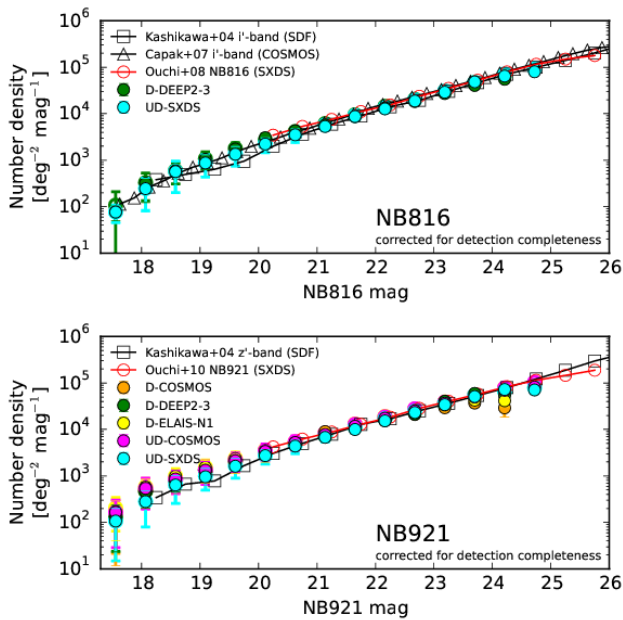}    
  \end{center}
  \caption{
  Number counts of NB-detected galaxies. Open circles
  show the number densities of the galaxies detected in NB816 or
  NB921 regardless of detection in any other bands. Among the
  galaxies, the number densities of the NB-detected galaxies with
  detection in all of the BBs are shown by the filled circles.
  Note that the number densities are corrected for the detection
  completeness estimated in \S~\ref{sec:data.nbquality.completeness}.
  The objects with magnitudes brighter than 17.5 mag are rejected due
  to the saturation.
  The upper panel shows the NB816 data and the lower panel shows the
  NB921 data. The number counts are compared with the previous studies
  \citep{Ouchi2008,Ouchi2010,Kashikawa2004,Capak2007}.  
  }\label{fig:NC_NBobjs}   
\end{figure}
%%%%%%%%%%%%%%%%%%%%%%%%%%%%%%%%%%%%%%%

\subsubsection{Number counts}
\label{sec:data.nbquality.numbercount}
Number counts of detected galaxies are a useful tool to verify
our detection and galaxy selection. Figure~\ref{fig:NC_NBobjs} shows
the number densities of the NB-detected galaxies that we have selected
in the NB816 and NB921 data. Both number densities in NB816 and NB921
show the similar behavior to each other. As expected, the number
densities of galaxies increase up to about 24.5 (25.0) magnitudes in
the D (UD) fields. The values are consistent with the limiting
magnitudes shown in Table~\ref{tab:NBdata}.
In the range of 20--24 mag, the number densities of galaxies detected
in individual D or UD fields are consistent with each other. This
suggests that there is no significant field-to-field variance for the
NB-detected galaxies within this magnitude range.
However, there is larger variance at brighter magnitudes of $<20$ mag.
This is expected since the number of
bright galaxies is small and thus the bright end of number densities
is more sensitive to a field variance than the fainter part.
The NB921 number densities in the UD fields show the slight
difference, while those in the D fields show better consistency. Thus,
it is important to survey area wider than $\sim2$ deg$^2$ (1 FoV of
the HSC) to overcome the field variance in bright galaxies.

As mentioned in \S~\ref{sec:data.nb.cas} (see also
Appendix~\ref{app:catalog}), we exclude bright objects when the
central 3$\times$3 pixels are saturated. \citet{Drake2013} point out
the possibility that bright emission-line galaxies can be missed due
to the magnitude cut at the bright end (see also
\cite{Stroe2014}). It is thus important to know the influence of the
saturation limit on the bright end of the number counts. To estimate
magnitudes where saturation occurs for extended sources, we compare
the HSC-SSP catalogs with the COSMOS2015 catalog \citep{Laigle2017}
for objects with {\tt flags\_pixel\_saturated\_center=True} in NB921,
and find that they have a 2'' aperture magnitude in $z'$ brighter than
17.3 mag in the COSMOS2015 catalog. The saturation magnitudes for
point sources are discussed in \citet{BrightStarMask}. Therefore, we
limit our samples to the objects with {\tt cmodel} magnitude fainter
than 17.5 mag conservatively.

We compare the number densities of the galaxies that we have selected
with previous studies in general deep fields
\citep{Ouchi2008,Ouchi2010,Kashikawa2004,Capak2007}.
We compare our results with the number densities of galaxies detected
in NB816 and NB921 of Subaru/Suprime-Cam in the SXDS field
\citep{Ouchi2008,Ouchi2010}. They surveyed $\sim$ 1 deg$^2$ with the 
similar NB filters to those of HSC down to 26 mag. 
We also compare with the number densities of galaxies detected in $i$
($z$) band for NB816 (NB921) detected galaxies, that is, galaxies
detected in bands at similar wavelength are compared.
\citet{Kashikawa2004} surveyed the SDF covering 0.25 deg$^2$, while
\citet{Capak2007} observed the COSMOS field covering 2 deg$^2$. Since
the number densities of emission-line galaxies showing an excess in
BB-NB is much smaller than those of galaxies without the color excess,
the comparison is meaningful in particular in the magnitude range of
$<20$ mag where the NB816 and NB921 number counts from the previous
studies are not available.
Figure~\ref{fig:NC_NBobjs} shows that the number densities of galaxies
selected based on HSC-SSP data are consistent with the previous
studies. This suggests that the object detection by hscPipe works well
and the magnitude cut of $17.5$ mag to avoid saturation is reasonable.

%%%%%%%%%%%%%%%%%%%%%%%%%%%%%%%%%%%%%%%%%%%%%%%%%%%%%%%%%%%%%%%%
%%%%%%%%%%%%%%%%%%%%%%%%%%%%%%%%%%%%%%%%%%%%%%%%%%%%%%%%%%%%%%%%

%%%%%%%%%%%% Figure 5 %%%%%%%%%%%%%%%%%
\begin{figure}[t]
  \begin{center}
    \includegraphics[width=0.5\textwidth]{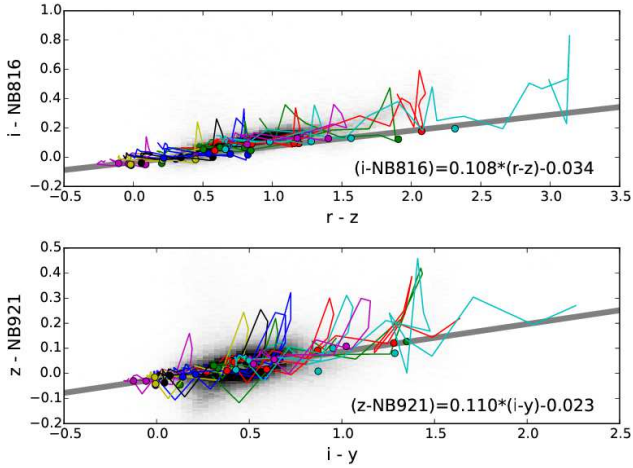}    
  \end{center}
  \caption{Colors of stellar continuum in $i-$NB816 ($z-$NB921) for
    galaxies as a function of $r-z$ ($i-y$). Since the effective
    wavelength of NBs is different from that of BBs, galaxies with
    redder colors tend to show a larger value in the BB-NB colors
    which can mimic the color excess in emission-line galaxies.  
    The thin lines show the color tracks of model spectra by
    \citet{BruzualCharlot2003} for galaxies at $z=$ 0--2. The details
    of the model spectra are described in the text. The circles show
    the colors of galaxies at the specific redshifts that NB816 or
    NB921 can probe. The gray thick lines show a line fitted to the
    circles. In background, color distribution for galaxies with NB =
    18--23 mag is shown in gray scale using HSC data, which is
    consistent with the model color tracks.
  }\label{fig:ColorTerm}
\end{figure}
%%%%%%%%%%%%%%%%%%%%%%%%%%%%%%%%%%%%%%%

%%%%%%%%%%%% Figure 6 %%%%%%%%%%%%%%%%%
\begin{figure}[t]
  \begin{center}
    \includegraphics[width=0.5\textwidth]{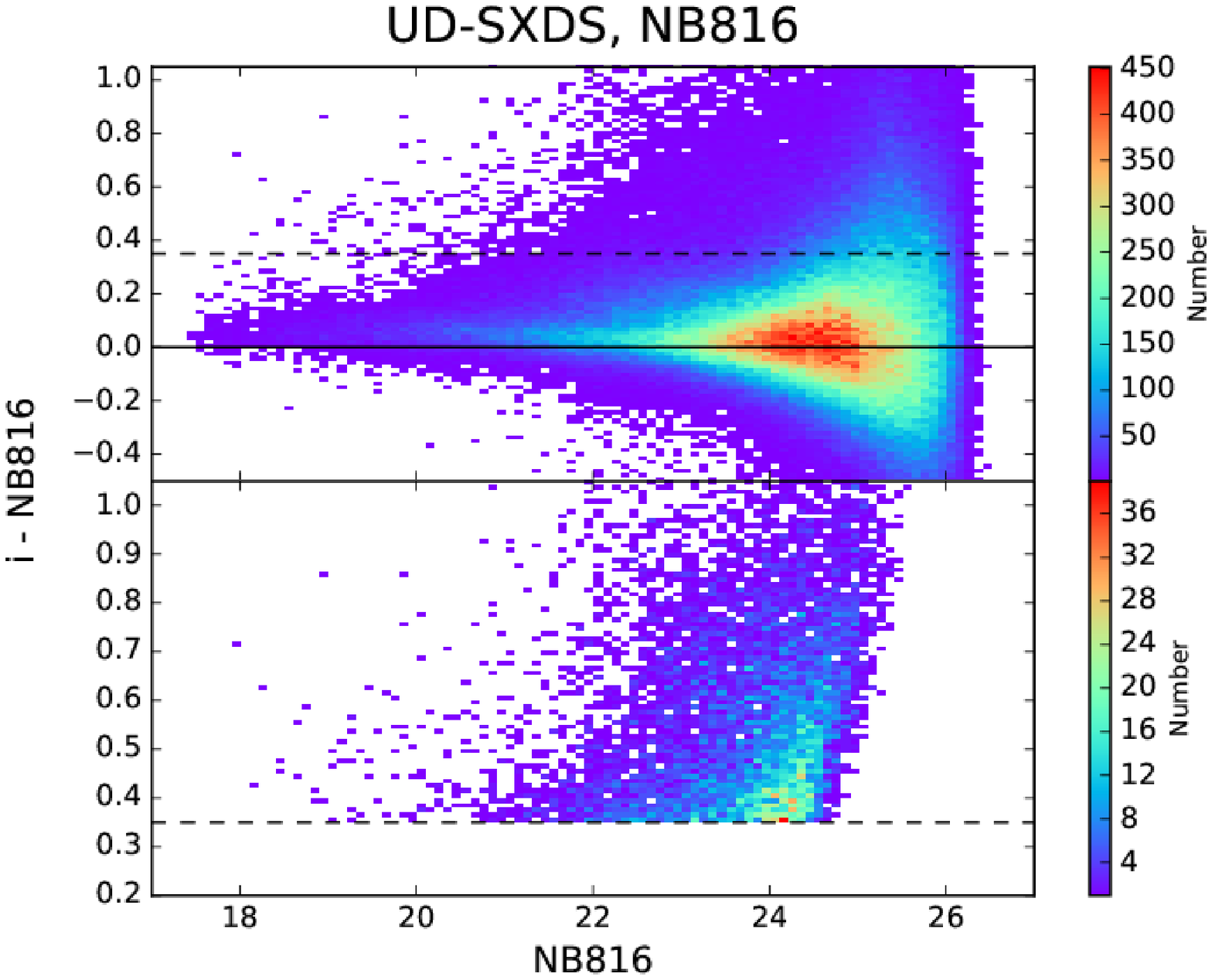}\\    
    \includegraphics[width=0.5\textwidth]{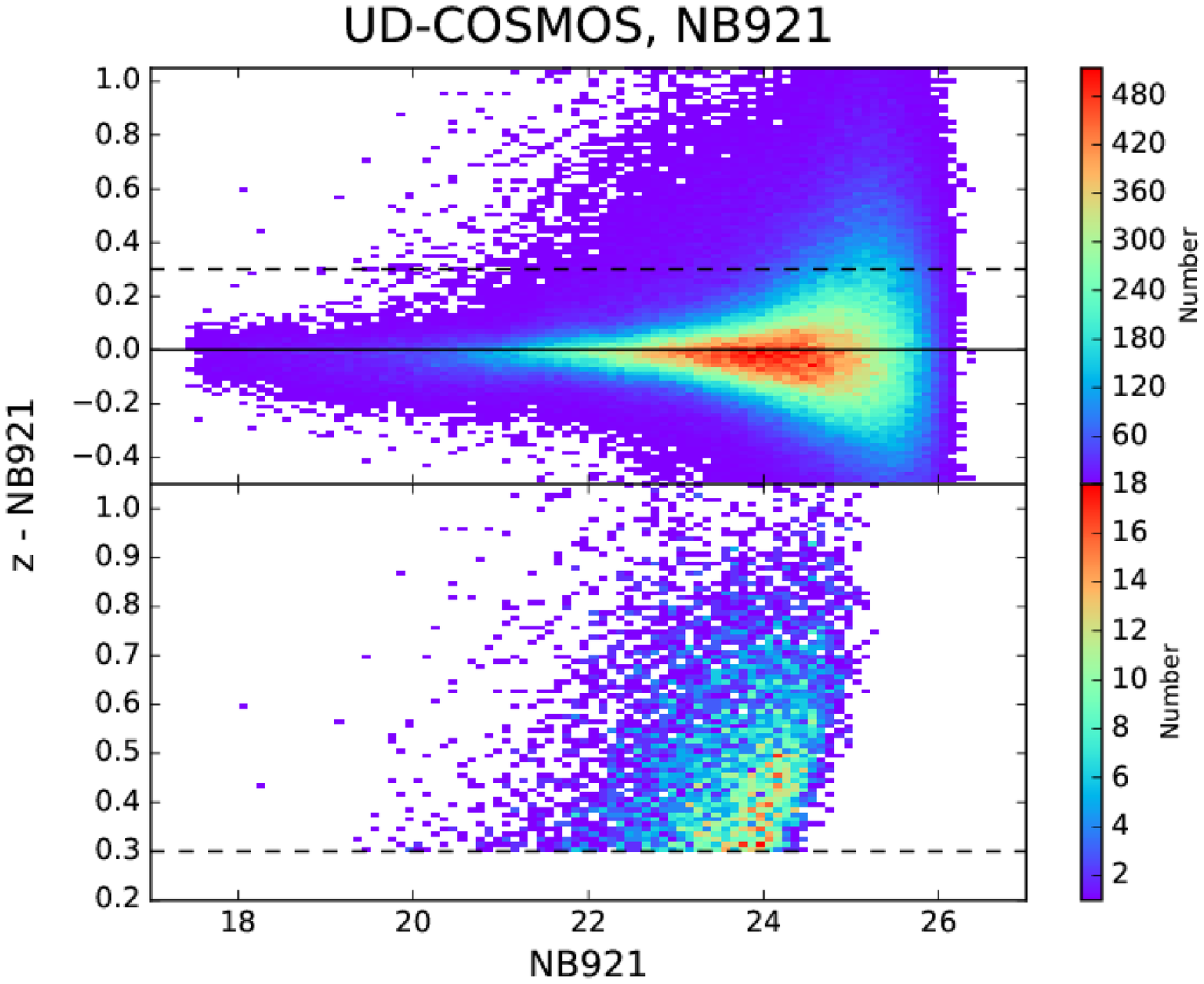}    
  \end{center}
 \caption{Colors of $i-$NB816 ($z-$NB921) as a function of NB816
   (NB921) magnitude in the UD-SXDS (UD-COSMOS) field. In each NB
   data, the upper panel shows the color distribution for galaxies,
   where the colors are corrected for the color term shown in
   Figure~\ref{fig:ColorTerm}. The color coding is based on the number 
   of galaxies in each bin. Although the sequence of galaxies without
   emission line is slightly shifted from $i-$NB816=0 for NB816 data,
   this shift is consistent with what is shown in
   Figure~\ref{fig:ColorTerm}. The correction of the color term is
   optimized for emission-line galaxies at the redshifts that NB816
   can probe. The correction by which galaxies are distributed around
   $i-$NB816=0 results in an overestimation of stellar continuum for
   the emission-line galaxies. The lower panel shows the color
   distribution for the emission-line galaxies selected.
   The dashed lines shows the color cut that we apply: $i$-NB816=0.35
   and $z$-NB921=0.30.  
 }\label{fig:BB-NB}
\end{figure}
%%%%%%%%%%%%%%%%%%%%%%%%%%%%%%%%%%%%%%%%

\section{Emission-line galaxies at $z<1.5$}
\label{sec:ELGs}

The flux density observed in a spectral band where an emission line enters is
expressed by the following:
$f_{\lambda, {\rm band}}=f_{\lambda, {\rm cont}} + f_{\rm el}/\Delta_{\rm band}$,
where $f_{\lambda}$ is the flux density, $f_{\lambda, {\rm cont}}$ is
for the stellar continuum, $f_{\rm el}$ is the emission-line flux, and
$\Delta_{\rm band}$ is the width of the filter. Magnitudes in NB are 
more sensitive to an emission line entering the filter than those in
BB by a difference in the width of filter.
Since the flux density of the stellar continuum also contributes to
magnitudes in NB, emission lines selected by NB imaging are required
to have not only flux but also EW large enough to make a difference
between NB and BB magnitudes.  
In practice, galaxies with such an emission line at specific redshifts
matched to a wavelength of the NB are observed significantly brighter in
the NB than the closest BB (e.g., NB816 against $i$ or NB921 against $z$).
In this section, we select H$\alpha$ emitters (HAEs) at $z\sim0.25$,
[OIII] emitters (O3Es) at $z\sim0.63$, and [OII] emitters (O2Es) at
$z\sim1.19$ from the NB816-detected galaxies, and HAEs at $z\sim0.40$,
O3Es at $z\sim0.84$, and O2Es at $z\sim1.47$ from the NB921-detected
galaxies (Table~\ref{tab:NBemitters}).

\subsection{Selection}
\label{sec:ELGs.selection}

%%%%%%%%%%%% Figure 7 %%%%%%%%%%%%%%%%%
\begin{figure*}  
  \begin{center}
    \begin{tabular}{cc}    
      \includegraphics[width=0.5\textwidth]{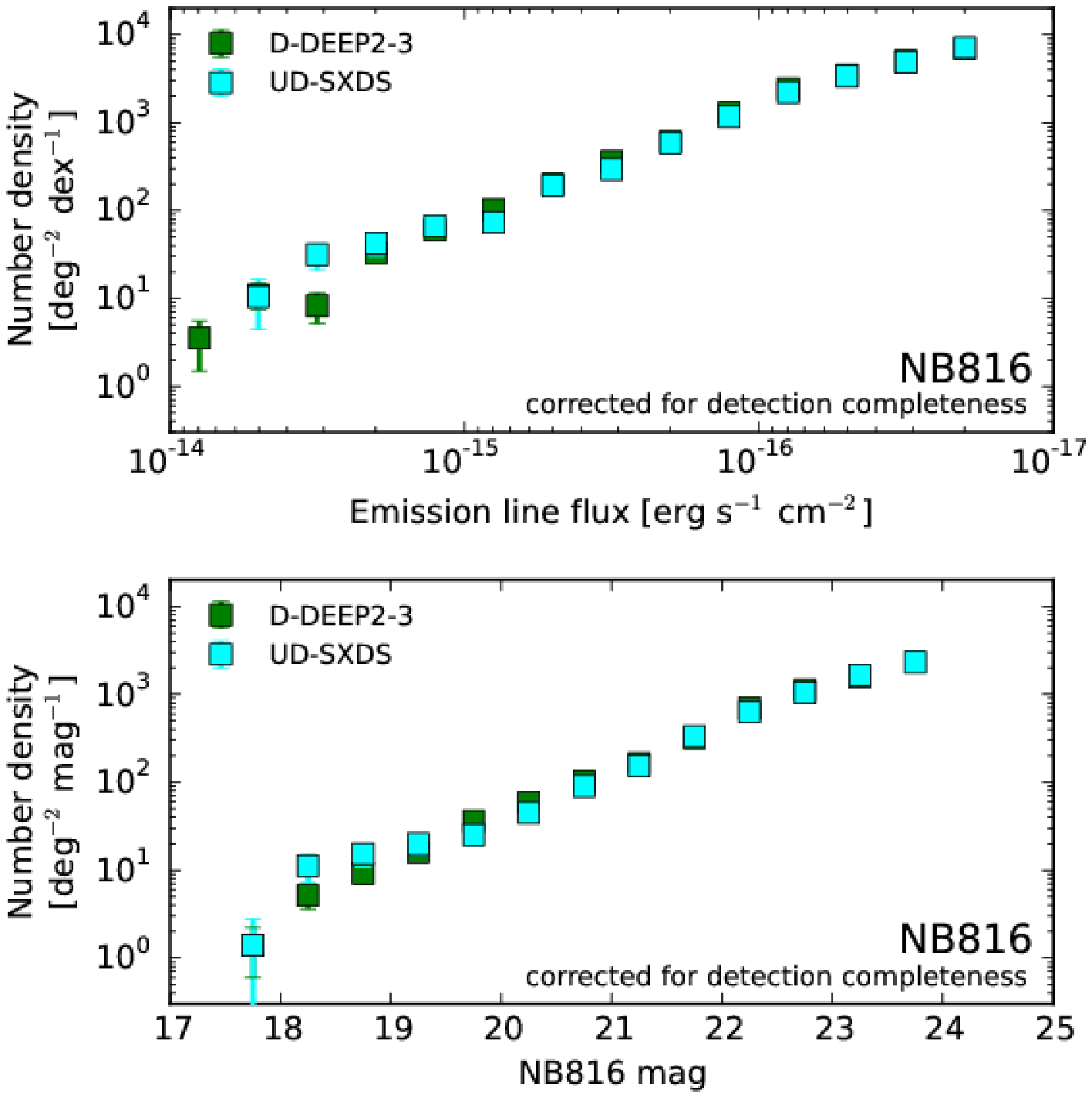}&
      \includegraphics[width=0.5\textwidth]{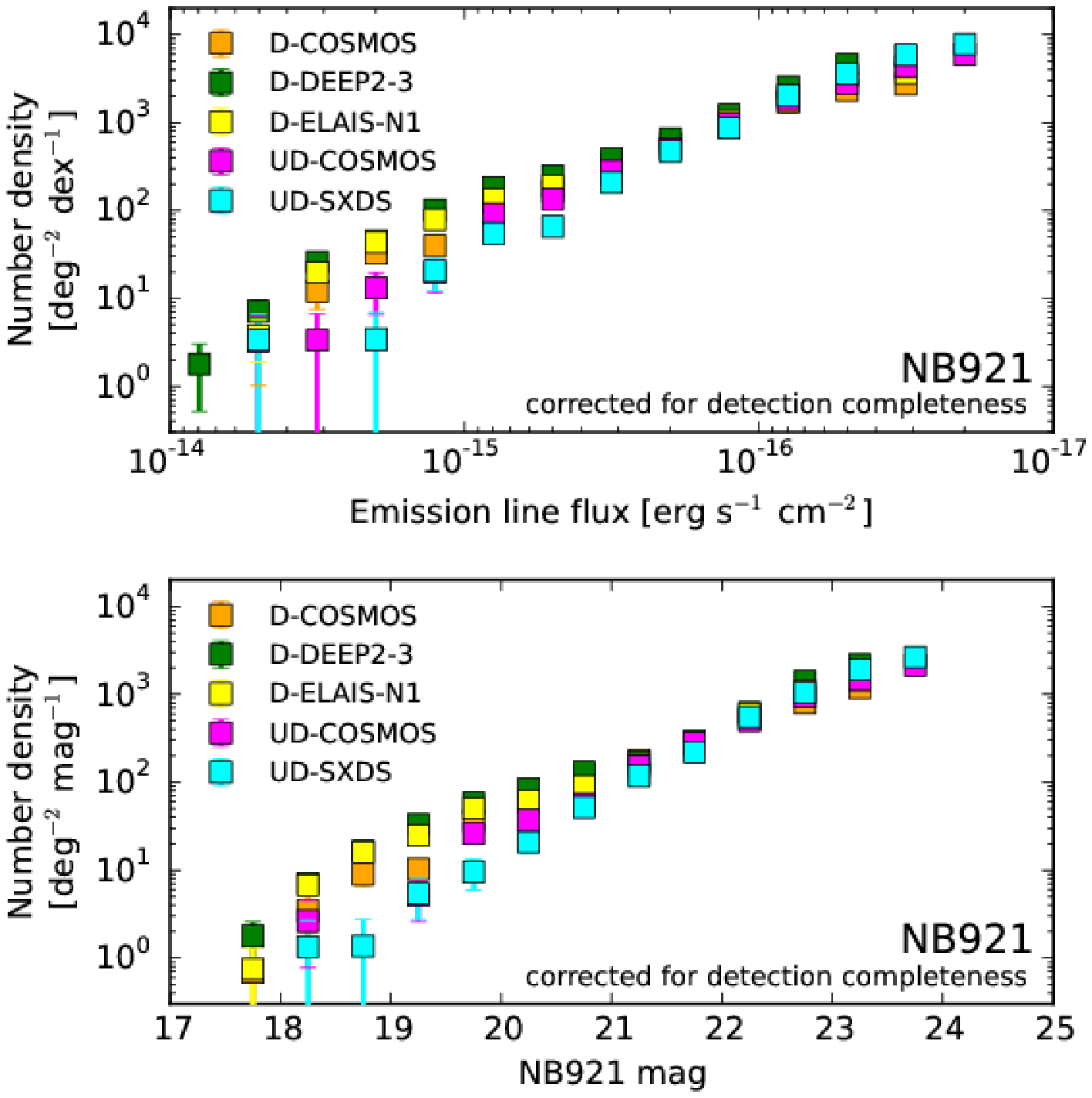}
    \end{tabular}        
  \end{center}
  \caption{Number counts of emission-line galaxies. The left two
    panels show the NB816 emitters and the right two panels show the
    NB921 emitters. The upper two panels are the number counts as a
    function of emission-line flux and the lower panels are the same
    but as a function of NB magnitude.
    The number densities are corrected for the detection completeness
    as estimated in \S~\ref{sec:data.nbquality.completeness}.
  }\label{fig:NC_emitters}  
\end{figure*}
%%%%%%%%%%%%%%%%%%%%%%%%%%%%%%%%%%%%%%%%%%

We compare the flux densities between $i$ and NB816 for the
NB816-detected galaxies ($z$ and NB921 for the NB921-detected
galaxies) to select the NB816 (NB921) emission-line galaxies.
The flux densities measured in {\tt unforced} photometry are used for
the comparison. The flux densities in BB of $i$ or $z$ are dominated
by the stellar continuum, while the flux densities in NB are dominated
by the emission line if the line with an EW large enough to be
detected enters the NB. We must keep in mind that the spatial
distribution of regions emitting the nebular emission lines is not
necessarily the same as that of the stellar component in a galaxy 
\citep{Nelson2012,Nelson2016}. 
The extended nebular emission galaxies such as [OII] blobs are one of
the extreme cases \citep{Brammer2013,Yuma2013,Yuma2017}.
As described in \S\ref{sec:data.hscssp}, one of the BBs is selected in
a higher priority than NB as a reference band for the {\tt forced}
photometry by hscPipe in the case of $z<1.5$ galaxies with stellar
continuum detected, although the situation may change for Ly$\alpha$
emitters that have no detection in some of bluer BBs. 
This means that the {\tt forced} photometry is conducted based on the
light profile of the stellar component. As a result, the comparison
between the {\tt unforced} fluxes in BB and NB allows us to properly
estimate the contribution of the emission line to NB.

There is a slight difference in effective wavelength between $i$
($z$)-band and NB816 (NB921) (see Fig.~\ref{fig:filters}), implying
that the photometry in NB and BB results in the measurement of the
flux densities at slightly different wavelengths. It is thus important
to correct for the color term to properly estimate the stellar continuum
underlying a emission line. Since the redder stellar continuum is more
sensitive to the difference of the effective wavelength of the filter
response function, the correction of color term should depend on the
galaxy color. We use the stellar population synthesis models of
\citet{BruzualCharlot2003} to estimate the intrinsic colors of stellar
continuum for galaxies at redshifts that the NB filters can probe.
The color of $i-$NB816 ($z-$NB921) is investigated as a function of
$r-z$ ($i-y$) color. Figure~\ref{fig:ColorTerm} shows the color tracks
for the galaxies with simple stellar population (SSP) or constant star
formation (CSF) with ages of 100Myr, 500Myr, 1Gyr and 3Gyr at
$z=$ 0--2. We assume a dust extinction of $E(B-V)=0.0$ for the galaxies
with SSP and $E(B-V)=0.0$ or 0.4 for the galaxies with CSF under the
\citet{Calzetti2000} extinction curve. No emission line is taken into
account in the model spectra, because we focus on the colors of
stellar continuum. The color tracks indicate that there is a color
variation at a given $r-z$ or $i-y$ color and some galaxies can have
red BB-NB colors of more than 0.2. Since we focus on the emission-line
galaxies at specific redshifts and prefer to avoid an overcorrection
of the color term for the galaxies, we derive the relation between
$i-NB816$ and $r-z$ ($z-NB921$ and $i-y$) by fitting the colors of the
model galaxies at redshifts that the NB filters can probe (the gray
line in Figure~\ref{fig:ColorTerm}):  
\begin{eqnarray}
  i-NB816 = 0.108(r-z)-0.034
\end{eqnarray}
for NB816-detected galaxies,
\begin{eqnarray}
  z-NB921 = 0.110(i-y)-0.023 
\end{eqnarray}
for NB921-detected galaxies.
The color distribution of the actual galaxies is consistent with the
color tracks of the model galaxies. Therefore, we apply the color term
to correct for the BB colors as a function of $r-z$ ($i-y$) for NB816
(NB921) detected galaxies as shown in Figure~\ref{fig:ColorTerm}. The
validity of this correction is supported by the result that the
corrected BB-NB colors of galaxies are more tightly distributed around
BB-NB=0 than before the correction (Figure~\ref{fig:BB-NB}).     

We select galaxies with a difference between BB and NB magnitudes
larger than 5$\sigma$ uncertainty as shown in Figure~\ref{fig:BB-NB}.
This 5$\sigma$ cut is relevant at faint magnitudes. At bright
magnitudes, Figure~\ref{fig:ColorTerm} suggests that bright galaxies
with intrinsic red colors in the stellar continuum can be included as
contaminants. We thus apply an additional cuts of $i$-NB816 $>$ 0.35
and $z$-NB921 $>$ 0.30. The color cuts are determined to exclude
galaxies with spectroscopic redshifts located outside of the redshift
ranges of possible NB emission-line galaxies. The criterion of the
color cut is supported by the color tracks shown in
Figure~\ref{fig:ColorTerm}.  
The BB -- NB of 0.35 (0.30) mag corresponds to an observed EW of
48\AA\ (56\AA) for NB816 (NB921) emitters. We further apply a
requirement that objects are detected in all five broadbands (i.e.,
{\tt merge\_peak=True} and {\tt detected\_notjunk=True}). This is
because photometric redshifts or galaxy colors are required to
distinguish between different emission lines, which is discussed in
\S~\ref{sec:ELGs.identification}.
As a result, 8,597 and 18,310 candidates of NB816 emission-line
galaxies are found in the UD-SXDS and D-DEEP2-3 fields, while 6,416,
6,074, 5,570, 15,909, and 11,856 candidates of NB921 emission-line
galaxies are found in the UD-COSMOS, UD-SXDS, D-COSMOS, D-DEEP2-3, and
D-ELAIS-N1 fields, respectively.

\subsection{Number counts of emission-line galaxies}
\label{sec:ELGs.NC_emitters}

Figure~\ref{fig:NC_emitters} shows the number counts of emission-line
galaxies. Based on these plots, we choose to apply magnitude cuts to
ensure complete samples. We place a cut at 23.5 and 24.0 in the D and
UD fields. The line fluxes are 2--3$\times10^{-17}$ erg s$^{-1}$
cm$^{-2}$ in the D fields, and 1.5--2$\times10^{-17}$ erg s$^{-1}$
cm$^{-2}$ in the UD fields. 
Note that these magnitudes are not the limiting magnitudes in the NB
data itself but the magnitudes at the peak of the number counts of the
emitters which ensures a high level of completeness. Indeed, the
magnitudes are $\sim$1 mag shallower than the $5\sigma$ limiting
magnitudes of NB816 and NB921. The imposition of detection in all BBs
on the emitters also contributes to the difference with the limiting
magnitudes of the NB detections themselves. For the NB816 emitters,
the number counts are consistent with each other between the two
fields. For the NB921 emitters, the number counts are consistent with
each other between the fields at the faint end of 
$<2\times10^{-16}$ erg s$^{-1}$ cm$^{-2}$ or $>21$ mag. However, there
is a large variation between the fields at the bright end (right
panels of Figure~\ref{fig:NC_emitters}), suggesting that the wide-area
survey is essential to search for bright galaxies whose number density
is small. 
Since we limit the samples to galaxies with NB magnitudes fainter than
17.5 mag, it is probable that we miss very bright NB816 (NB921)
emitters (see also the discussion in \cite{Drake2013} and
\cite{Stroe2014}). The cut of 17.5 mag and BB-NB colors of 0.35 
(0.30) that we apply in the selection of emission-line galaxies
implies that most of NB816 (NB921) emitters with line fluxes larger
than 5.5 (5.0) $\times$ 10$^{-15}$ erg s$^{-1}$ cm$^{-1}$ cannot be
selected. 

\subsection{Line identification}
\label{sec:ELGs.identification}

The NB816 and NB921 emission-line galaxies selected in
\S~\ref{sec:ELGs.selection} could be identified with one of several
strong nebular emission lines. Possible candidates are typical
emission lines of H$\alpha$, [OIII], H$\beta$, and [OII] in
rest-frame optical wavelengths from star-forming galaxies at the
specific redshifts of $z=$ 0.2--1.5 (Table~\ref{tab:NBemitters}).

We take three steps to identify emission lines.
First, we use public spectroscopic redshifts available in the D and UD
fields from the literature\footnote{%
  zCOSMOS DR3 \citep{Lilly2009},
  UDSz \citep{Bradshaw2013,McLure2013},
  3D-HST \citep{Skelton2014,Momcheva2016},
  FMOS-COSMOS \citep{Silverman2015},
  VVDS \citep{LeFevre2013},
  VIPERS PDR1 \citep{Garilli2014},
  SDSS DR12 \citep{Alam2015},
  WiggleZ DR1 \citep{Drinkwater2010},
  DEEP2 DR4 \citep{Davis2003,Newman2013},
  and PRIMUS DR1 \citep{Coil2011,Cool2013}.
} which are incorporated into the HSC-SSP database
\citep[in prep.]{hscDatabase}.
We can identify a specific emission line as contributing to the excess
flux in the NB data if a spectroscopic redshift is available and it
agrees with the observed wavelength coverage of the NB filter. As
shown in Table~\ref{tab:NBemitters}, the NB can probe a specific range
of redshifts. If the spectroscopic redshift of the galaxy is within the
range, the emission line is identified. Otherwise, if a redshift
exists but is in disagreement with the NB line detection, the galaxy
is excluded from the sample and considered a contaminant.   

%%%%%%%%%%%% Figure 8 %%%%%%%%%%%%%%%%%
\begin{figure}
  \begin{center}
    \includegraphics[width=0.5\textwidth]{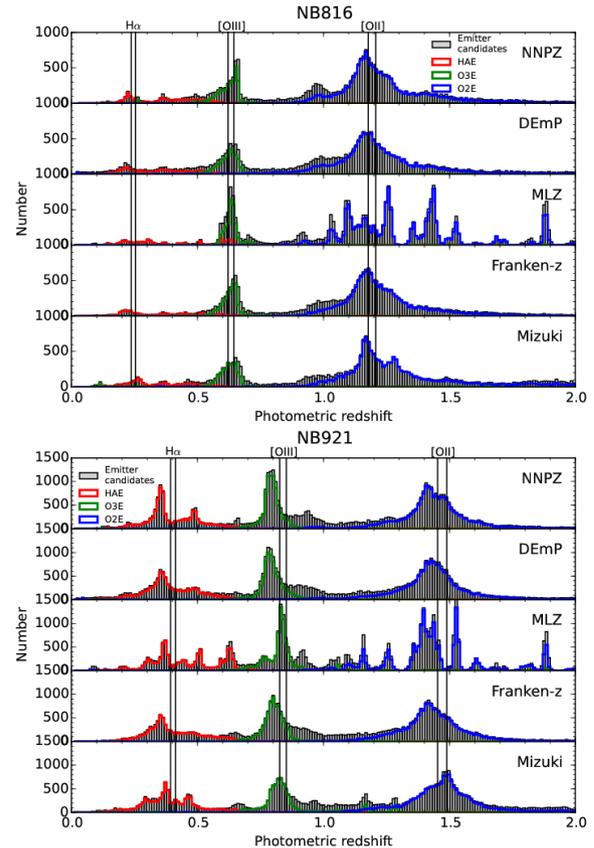}
  \end{center}
  \caption{Distribution of photometric redshift of galaxies selected
    as NB emission-line galaxy candidates in all of the D and UD
    fields (gray histograms), where five photometric redshifts
    calculated by different codes (MNPZ, DEmP, MLZ, Franken-z, and
    Mizuki; \cite{hscPhotoZ}) are available for individual
    galaxies. The upper panels are for the NB816 emitters and the
    lower panels are for the NB921 emitters. Among them, the galaxies
    with H$\alpha$, [OIII], [OII] emissions identified in
    \S~\ref{sec:ELGs.identification} are shown in red, green, and blue
    histograms, respectively.   
  }
  \label{fig:zphot_emitters}
\end{figure}
%%%%%%%%%%%%%%%%%%%%%%%%%%%%%%%%%%%%%%%%

%
In addition, we use photometric redshifts and their errors to identify
emission lines. We calculate five independent sets of photometric
redshifts by the different codes for the S15B internal release data
\citep{hscPhotoZ}. Among them, we use four sets for the line
identification. One photometric redshift catalog (DEmP,
\cite{Hsieh2014}) is not used because the uncertainties on the
photometric redshifts are largely over-estimated%
\footnote{
  This issue is only for the S15B internal data release and the DEmP
  photo-z's in PDR1 do not have such an issue.
}.
However, note that there is no problem with the estimated photometric
redshifts themselves. We thus use a combination of the four
photometric redshifts in order to identify the emission line. 
First, we apply a photo-z quality cut appropriate for the individual
photo-z sets, e.g., reduced $\chi^2 < 5$ for the {\tt MIZUKI} code, to
ensure that uncertain photo-z's are eliminated and only more likely
photo-z's are used for the identification. If the galaxy meets the
quality cut, we look for possible emission-line redshifts. If one of
the emission-line redshifts is consistent within the photo-z 95\%
confidence interval, then we assign the galaxy to that emission
line. If an object is consistent with multiple emission-line
redshifts, we assign the object to the emission line at the redshift
closest to the photo-z. Finally, we combine the four photo-z
estimates; when three estimates or more agree on the emission-line
redshift, we assign the redshift to that object. Otherwise, the
galaxies are excluded from the sample.  
Figure~\ref{fig:zphot_emitters} show the redshift distributions of
emission-line galaxies for individual photometric redshifts. There are
several peaks around the emission-line redshifts that we expect from
the wavelength of NB (Table~\ref{tab:NBemitters}), which indicates that
our method to select emission-line galaxies works well. However, we
also note that the galaxies selected as a NB816 emitter show the peak
at $z\sim1.0$, which is not expected. The peak is likely due to red
galaxies with strong Balmer/4000\AA\ break being incorrectly selected
as an emission-line galaxy (Figure~\ref{fig:ColorTerm}).
The contamination of the strong Balmer/4000\AA\ break galaxies is also
reported by \citet{Sobral2012}. Because the effective wavelength of NB
is longer than that of BB, if galaxies without emission line have the
strong continuum break between the effective wavelengths of BB and NB,
the NB magnitude can be brighter than the BB magnitude
significantly. Although we correct the BB magnitudes for the color
term to not select such galaxies with red continuum colors as discussed in
\S \ref{sec:ELGs.selection}, the color term correction is optimized
for galaxies at the specific redshifts that the NBs can probe.
Figure~\ref{fig:ColorTerm} suggests that the correction can be
underestimated for galaxies at other redshifts. Furthermore, since 
the $i$-band filter is wider than the $z$-band and the difference of
effective wavelength between BB and NB is larger for the pair of $i$
and NB816, the $i-$NB816 colors are easier to select such red galaxies
than the $z-$NB921 colors. It is difficult to distinguish the NB816
O2Es at $z\sim1.2$ from the red galaxies at $z\sim1$ based on the two
color diagrams shown below. There is another small peak at the
redshift next to the [OIII] line, i.e., $z\sim0.7$ for NB816 and
$z\sim0.9$ for NB921. This is likely that for H$\beta$ emission-line
galaxies. Therefore, Figure~\ref{fig:zphot_emitters} indicates that 
the line identification based on the photometric redshift is effective
in reducing contamination, as discussed in more detail in
\S~\ref{sec:ELGs.validation}.

%%%%%%%%%%%% Figure 9 %%%%%%%%%%%%%%%%%
\begin{figure}
  \begin{center}
    \hspace*{-7mm}
    \includegraphics[width=0.55\textwidth]{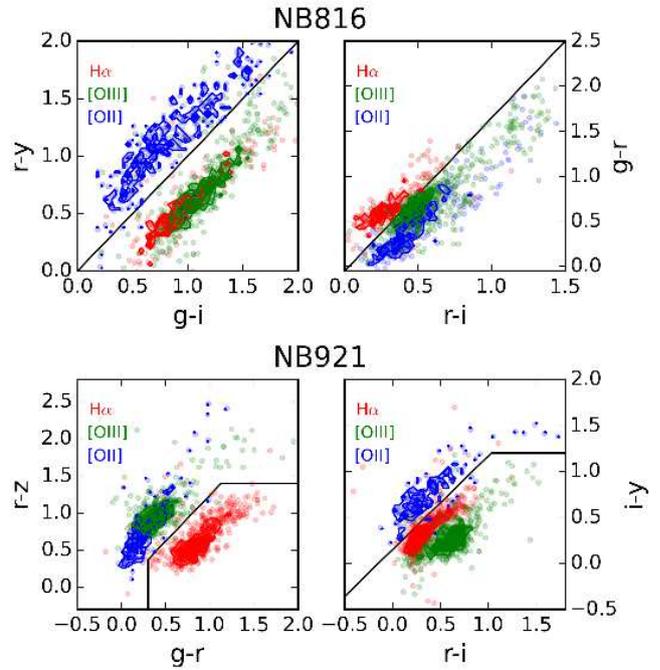}
  \end{center}
  \caption{Color--color diagrams used for line identification. The
    upper panels are the diagrams for NB816 emitters, and the lower
    panels are for NB921 emitters. In each panel, emission-line
    galaxies with spectroscopic redshifts in all of the fields are
    plotted. Individual galaxies are shown as fainter symbols and the
    contours are shown in solid lines for H$\alpha$ in red, 
    [OIII] in green, and [OII] in blue. For NB816 emitters, O2Es at
    $z \approx 1.19$ are distinguished from the others based on $g-i$
    vs.~$r-y$, and then HAEs at $z \approx 0.25$ are distinguished
    from O3Es at $z \approx 0.63$ based on $r-i$ vs.~$g-r$. For NB921
    emitters, HAEs at $z \approx 0.40$ are distinguished from the
    others based on $g-r$ vs.~$r-z$, and then O2Es at
    $z \approx 1.47$ are distinguished from O3Es at $z \approx 0.84$
    based on $r-i$ vs.~$i-y$. The boundaries are specified in the text.} 
  \label{fig:colorcolor}
\end{figure}
%%%%%%%%%%%%%%%%%%%%%%%%%%%%%%%%%%%%%%%%

%%%%%%%%%%%% Figure 10 %%%%%%%%%%%%%%%%%
\begin{figure}[t]
  \begin{center}
    \includegraphics[width=0.5\textwidth]{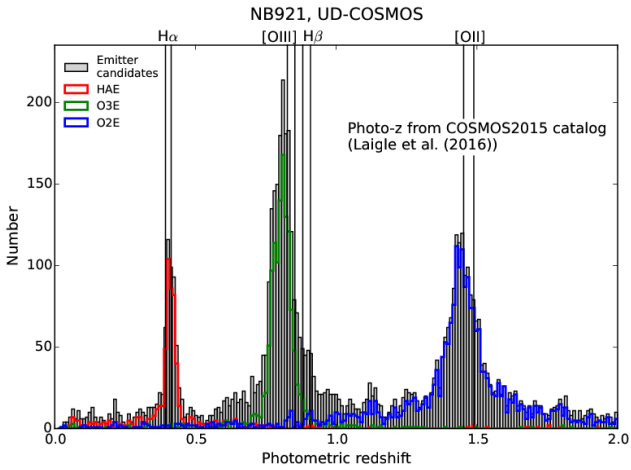}
  \end{center}
  \caption{
    Same as Figure~\ref{fig:zphot_emitters}, but for NB921 emitters in
    the UD-COSMOS field. The photometric redshifts are from the
    COSMOS2015 catalog \citep{Laigle2016} instead of those estimated
    from HSC-SSP data.
  }
  \label{fig:zphot_emitters_UD-COSMOS}
\end{figure}
%%%%%%%%%%%%%%%%%%%%%%%%%%%%%%%%%%%%%%%%

%
Finally, if the galaxy is not identified yet based on the information
on spectroscopic and photometric redshifts, we use colors to identify
emission line as shown in Figure~\ref{fig:colorcolor}. This case can
apply to galaxies with faint stellar continuum or galaxies with strong
emission lines that impact on the BB magnitudes. 
The colors of the galaxies are measured with magnitudes in
{\tt forced} photometry, because we are now interested in the colors
of stellar continuum. For NB816 emitters, we use $g-i$ versus $r-y$
colors to distinguish [OII] at $z\sim1.19$ from the others:  
\begin{eqnarray}
(r-y)>(g-i) ,
\end{eqnarray}
and then $r-i$ versus $g-r$ colors to distinguish H$\alpha$ at
$z\sim0.25$ from [OIII] at $z\sim0.63$:
\begin{eqnarray}
(g-r)>1.7(r-i)-0.05 .
\end{eqnarray}
For NB921 emitters, we use $g-r$ versus $r-z$ colors to distinguish
H$\alpha$ at $z\sim0.40$ from the others:
\begin{eqnarray}
(g-r)>0.3\ \land\ (r-z)<1.4\ \land\ (r-z)<1.24(g-r) ,
\end{eqnarray}
and then $r-i$ versus $i-y$ colors to distinguish [OII] at $z\sim1.47$
from [OIII] at $z\sim0.84$:
\begin{eqnarray}
(i-y)<1.2\ \land\ (i-y)<1.02(r-i)+0.15 .
\end{eqnarray}
The color criteria are determined based on the colors for
emission-line galaxies with spectroscopic redshifts.  

\subsection{Validation of the line identification}
\label{sec:ELGs.validation}

To validate our selection and line identification, we use the galaxies
that are selected as NB emission-line galaxies in
\S~\ref{sec:ELGs.selection} and have spectroscopic redshifts.
We find that among 84 NB816-HAEs with a spectroscopic redshift at
$z\approx0.25$, 
343 NB921-HAE at $z\approx0.40$,
203 NB816-O3Es at $z\approx0.63$,
274 NB921-O3E at $z\approx0.84$,
199 NB816-O2Es at $z\approx1.19$ and
109 NB921-O2E at $z\approx1.47$, 
our identification with photometric redshifts and BB colors
selects 
94\% of the NB816-HAEs at $z\approx0.25$,
85\% of the NB921-HAEs at $z\approx0.40$,
77\% of the NB816-O3Es at $z\approx0.63$,
84\% of the NB921-O3Es at $z\approx0.84$,
93\% of the NB816-O2Es at $z\approx1.19$ and
88\% of the NB921-O2Es at $z\approx1.47$.
The results indicate that the high fraction of emission-line galaxies
without spectroscopic redshift can be properly identified. 
On the other hand, the identification with photo-$z$'s or colors can
contain 33\% of the contaminants with an inappropriate spectroscopic 
redshift in the NB816-HAEs sample, 
17\% in the NB921-HAE sample,
13\% in the NB816-O3E sample,
15\% in the NB921-O3E sample,
40\% in the NB816-O2E sample and
43\% in the NB921-O2E sample.
The O2E samples can contain the relatively large fraction of
contamination compared with the samples of the other lines. This is
because since the [OII] doublet at $\lambda\lambda=3727,3730$\AA\ are
located near the Balmer/4000\AA\ break, it is hard to estimate the
stellar continuum level underlying the line only with the photometric
data and the BB-NB colors of strong Balmer/4000\AA\ break galaxies are
easy to mimic those of emission-line galaxies as discussed in the
previous section. 

Figure~\ref{fig:zphot_emitters_UD-COSMOS} shows the photometric
redshift distribution of emission-line galaxies in the UD-COSMOS.
As a further test, we use the many-band photometric redshifts from the
COSMOS2015 catalog \citep{Laigle2017}, which are available for
essentially all the NB emitters. Because the photometric redshifts in
the COSMOS2015 catalog are estimated from 30-band photometry covering
wide wavelengths from NUV to MIR, they are more precise
($\sigma(\Delta z/(1+z))\sim0.007$: \cite{Laigle2017}) than the
photometric redshifts from 5-band photometry of the HSC-SSP
($\sigma(\Delta z/(1+z))\sim0.05$: \cite{hscPhotoZ}). Therefore, the
distribution with the sharp peaks at the expected redshifts also
indicate the validity of our selection and line identification of
emission-line galaxies (see also Figure~\ref{fig:zphot_emitters}).     

In our line identification, we distinguish [OIII] from H$\beta$ based
on the photometric redshifts, because it is impossible to do that
based on the color--color diagrams (Figure~\ref{fig:colorcolor}).
We estimate an amount of H$\beta$ emitters that can be included in the
O3E sample. Among 14 H$\beta$ emitters at $z\sim0.68$ and 39 H$\beta$
emitters at $z\sim0.89$, which are classified by the spectroscopic
redshifts, 43\% of H$\beta$ emitters at $z\sim0.68$ and 21\% of
H$\beta$ emitters at $z\sim0.89$ are identified as O3Es based on the
photometric redshifts and colors. However, among the contamination in
the O3E samples, the fraction of H$\beta$ emitters are less than 
25\%. Therefore, we conclude that the contamination of H$\beta$
emitters in the O3Es sample is not so severe.
Figure~\ref{fig:zphot_emitters} also support our line identification
between [OIII] and H$\beta$.

%%%%%%%%%%%% Table 3 %%%%%%%%%%%%%%%%%
\begin{table*}
  \tbl{
    Samples of emission-line galaxies. The number of emission-line
    galaxies with a NB magnitude and an emission-line flux
    brighter/larger than the limit above which the sample is complete
    (\S~\ref{sec:ELGs.NC_emitters}) is shown. Magnitudes are given in
    units of AB mag, fluxes are in units of erg s$^{-1}$ cm$^{-2}$,
    and the survey volumes are in units of comoving Mpc$^3$
    $h_{70}^{-3}$. The value within the parenthesis shows the number
    and volume for the area excluding the overlapping regions between
    UD-COSMOS and D-COSMOS fields.  
  }{%
    \begin{tabular}{lccccccccc}
      %%%%%%%%%%%%%%%%%%%%%%%%%%%%%%%%%%%%%%%%%%%%%%%
      \multicolumn{10}{c}{H$\alpha$ emitters (HAEs)} \\
      \hline      
      & \multicolumn{4}{c}{NB816 ($z=0.25$)} && \multicolumn{4}{c}{NB921 ($z=0.40$)} \\
      \cline{2-5}\cline{7-10}
      Field & \# of objects & mag cut & flux cut & Volume && \# of objects &
      mag cut & flux cut & Volume \\
      \hline
      UD-COSMOS
      &$\cdot\cdot\cdot$&$\cdot\cdot\cdot$&$\cdot\cdot\cdot$&$\cdot\cdot\cdot$&&
      471 & 24.0 & $2.0\times10^{-17}$ & $1.5\times10^5$\\
      &&&&&& (441) & & & ($1.5\times10^5$)\\
      UD-SXDS & 304 & 24.0 & $1.5\times10^{-17}$ & $5.0\times10^4$ &&
      422 & 24.0 & $2.0\times10^{-17}$ & $1.2\times10^5$\\
      D-COSMOS
      &$\cdot\cdot\cdot$&$\cdot\cdot\cdot$&$\cdot\cdot\cdot$&$\cdot\cdot\cdot$&&
      974 & 23.5 & $3.0\times10^{-17}$ & $5.3\times10^5$\\
      &&&&&& (772) & & & ($4.2\times10^5$)\\
      D-DEEP2-3 & 889 & 23.5 & $2.0\times10^{-17}$ & $4.5\times10^5$ &&
      2,915 & 23.5 & $3.0\times10^{-17}$ & $2.0\times10^6$\\
      D-ELAIS-N1
      &$\cdot\cdot\cdot$&$\cdot\cdot\cdot$&$\cdot\cdot\cdot$&$\cdot\cdot\cdot$&&
      2,311 & 23.5 & $3.0\times10^{-17}$ & $1.6\times10^6$\\
      \hline\\
      %%%%%%%%%%%%%%%%%%%%%%%%%%%%%%%%%%%%%%%%%%%%%%%
      %%%%%%%%%%%%%%%%%%%%%%%%%%%%%%%%%%%%%%%%%%%%%%%
      \multicolumn{10}{c}{[OIII] emitters (O3Es)} \\
      \hline      
      & \multicolumn{4}{c}{NB816 ($z=0.63$)} && \multicolumn{4}{c}{NB921 ($z=0.84$)} \\
      \cline{2-5}\cline{7-10}
      Field & \# of objects & mag cut & flux cut & Volume && \# of objects &
      mag cut & flux cut & Volume \\      
      \hline
      UD-COSMOS
      &$\cdot\cdot\cdot$&$\cdot\cdot\cdot$&$\cdot\cdot\cdot$&$\cdot\cdot\cdot$&&
      1,127 & 24.0 & $2.0\times10^{-17}$ & $5.5\times10^5$\\
      &&&&&& (1,074) & & & ($5.2\times10^5$)\\
      UD-SXDS & 894 & 24.0 & $1.5\times10^{-17}$ & $2.9\times10^5$ &&
      762 & 24.0 & $2.0\times10^{-17}$ & $4.6\times10^5$\\
      D-COSMOS
      &$\cdot\cdot\cdot$&$\cdot\cdot\cdot$&$\cdot\cdot\cdot$&$\cdot\cdot\cdot$&&
      851 & 23.5 & $3.0\times10^{-17}$ & $1.6\times10^6$\\
      &&&&&& (609) & & & ($1.3\times10^6$)\\
      D-DEEP2-3 & 1,334 & 23.5 & $2.0\times10^{-17}$ & $2.5\times10^6$ &&
      2,409 & 23.5 & $3.0\times10^{-17}$ & $6.4\times10^6$\\
      D-ELAIS-N1
      &$\cdot\cdot\cdot$&$\cdot\cdot\cdot$&$\cdot\cdot\cdot$&$\cdot\cdot\cdot$&&
      1,574 & 23.5 & $3.0\times10^{-17}$ & $5.3\times10^6$\\
      \hline\\
      %%%%%%%%%%%%%%%%%%%%%%%%%%%%%%%%%%%%%%%%%%%%%%%
      %%%%%%%%%%%%%%%%%%%%%%%%%%%%%%%%%%%%%%%%%%%%%%%
      \multicolumn{10}{c}{[OII] emitters (O2Es)} \\
      \hline      
      & \multicolumn{4}{c}{NB816 ($z=1.19$)} && \multicolumn{4}{c}{NB921 ($z=1.47$)} \\
      \cline{2-5}\cline{7-10}
      Field & \# of objects & mag cut & flux cut & Volume && \# of objects &
      mag cut & flux cut & Volume \\      
      \hline
      UD-COSMOS
      &$\cdot\cdot\cdot$&$\cdot\cdot\cdot$&$\cdot\cdot\cdot$&$\cdot\cdot\cdot$&&
      1,309 & 24.0 & $2.0\times10^{-17}$ & $1.0\times10^6$\\
      &&&&&& (1,246) & & & ($9.9\times10^5$)\\
      UD-SXDS & 1,868 & 24.0 & $1.5\times10^{-17}$ & $6.9\times10^5$ &&
      2,230 & 24.0 & $2.0\times10^{-17}$ & $9.0\times10^5$\\
      D-COSMOS
      &$\cdot\cdot\cdot$&$\cdot\cdot\cdot$&$\cdot\cdot\cdot$&$\cdot\cdot\cdot$&&
      1,447 & 23.5 & $3.0\times10^{-17}$ & $3.3\times10^6$\\
      &&&&&& (1,222) & & & ($2.6\times10^6$)\\
      D-DEEP2-3 & 3,993 & 23.5 & $2.0\times10^{-17}$ & $5.9\times10^6$ &&
      3,055 & 23.5 & $3.0\times10^{-17}$ & $1.2\times10^7$\\
      D-ELAIS-N1
      &$\cdot\cdot\cdot$&$\cdot\cdot\cdot$&$\cdot\cdot\cdot$&$\cdot\cdot\cdot$&&
      3,263 & 23.5 & $3.0\times10^{-17}$ & $1.1\times10^7$\\
      \hline
      %%%%%%%%%%%%%%%%%%%%%%%%%%%%%%%%%%%%%%%%%%%%%%%
  \end{tabular}}
  \label{tab:Samples}
\end{table*}
%%%%%%%%%%%%%%%%%%%%%%%%%%%%%%%%%%%%%%%

\subsection{AGN contamination}
\label{sec:ELGs.AGN}

Since an AGN can contribute to the strong nebular emission lines from
galaxies, objects with an emission line detected by the NB imaging can
be AGNs, not star-forming galaxies.  
\citet{Sobral2016} find by spectroscopy that 30\% of luminous HAEs at
$z=$ 0.8--2.2 can be AGN and the fraction of AGN increases with
H$\alpha$ luminosity at $L({\rm H}\alpha)>10^{42}$ erg s$^{-1}$. The
trend is not dependent on the redshifts of HAEs. Other previous
studies have reported several percent of X-ray detected AGN
contamination in the NB emitter samples (e.g.,
\cite{Garn2010a,Calhau2017,Matthee2017}). It is thus important to
estimate the fraction of AGNs in the samples of emission-line galaxies
that we have selected.

We use the 4.6 Ms X-ray data by the Chandra COSMOS-Legacy Survey
\citep{Civano2016} which almost fully covers the UD-COSMOS field. The
D-COSMOS field is partially covered. \citet{Marchesi2016} present the
catalog of 4016 X-ray sources, 97\% of which have an optical/IR
counterpart. We match our catalogs to the X-ray sources and then
find that few emission-line galaxies ($\sim$ 0.1\%) have a counterpart
of X-ray source. The luminosities of emission line of galaxies with an
X-ray counterpart is not so large, which is against our expectation
that AGNs can have a large contribution to the bright end of
luminosity functions (discussed more in
\S~\ref{sec:discussions.brightendoflf}). 
Also, the fraction of emission-line galaxies with an X-ray counterpart
does not depend on the line, i.e., H$\alpha$, [OIII], or [OII]. 

As mentioned in \S~\ref{sec:data.nb.stars} and
\S~\ref{sec:data.nbquality.numbercount}, we remove the objects with NB
{\tt cmodel} magnitudes brighter than 17.5 or point sources from the
catalogs. Removing the objects may result in removing AGNs
unintentionally from the samples of NB emitters and that may explain
the small fraction of the emission-line galaxies with X-ray
counterpart. To investigate this possibility, we match the removed
objects to the X-ray source catalog in the UD-COSMOS field. 
We find that among 4016 X-ray sources, 369 NB921-detected point
sources (9.2\%) and 93 objects saturated in NB921 data (2.3\%) are
counterparts of the X-ray sources. Although five point sources
meet the criteria of the EW cut (i.e., $z$-NB921$>$0.3), only one
source has the spectroscopic or photometric redshifts consistent with
those of NB921 emitters. Therefore, we conclude that there is no
impact of the removal of the point sources and the saturated objects
on the discussion of AGNs in the NB emitter samples.

The COSMOS data indicates the fraction of X-ray detected AGN
contamination is small, which is $\sim0.1$\%. We assume that the
contamination rate in the other fields is the same as in the COSMOS
field. Even if there are 0.1\% of AGNs in the samples, our results
such as luminosity function do not change largely.      

\subsection{Stellar mass of emission-line galaxies}
\label{sec:ELGs.SEDFIT}

We calculate stellar mass for the emission-line galaxies by SED
fitting with five HSC BB data using the code with Bayesian priors
called `{\tt MIZUKI}' \citep{Tanaka2015}. 
If galaxies have a spectroscopic redshift, the SED fitting is
performed at the fixed redshift. Otherwise, we fix a redshift to that
estimated from the wavelength of NB (e.g., $z=0.40$ for NB921 HAEs,
see also Table~\ref{tab:NBemitters}). The model SED templates of
galaxies are generated by the code of \citet{BruzualCharlot2003}
and emission lines are taken into account. Metallicity is fixed to the
solar value and the extinction curve of \citet{Calzetti2000} is
adopted. Ages and optical depth in the $V$-band ranges 0.05--14 Gyr 
and 0--2, respectively. Star formation histories of exponentially
declining model with a decay timescale of 0.1 -- 11 Gyr, SSP model,
and constant SFR model are adopted. Note that the reddest HSC BB
(i.e., $y$-band whose effective wavelength is 9755\AA) corresponds to
the data at
$\sim4000$--4500 \AA\ in the rest frame for O2Es at $z=1.19$--1.47.
We significantly suffer from the degeneracy between age, metallicity,
and dust extinction. Thus, we impose no dust extinction only for the
O2Es at $z=1.19$--1.47. The assumption is not unreasonable, because
O2Es at $z\sim1.5$ are likely to be less dusty star-forming galaxies
\citep{Hayashi2013}.  

We compare the stellar masses estimated from the five HSC BBs with
those from the more multi-wavelength data in the COSMOS2015 catalog
\citep{Laigle2016}. Note that the same IMF \citep{Chabrier2003} is
assumed. We find that our stellar masses are in good agreement with
those of the COSMOS2015 catalog for the HAEs, while our stellar masses
are systematically larger by a factor of $\sim2$ at the most for the
O3Es and O2Es. This is a known bias of the code \citep{Tanaka2015}
and it will likely be reduced in our future runs. However, this may
also imply that multi-band data at longer wavelength such
near-infrared (NIR) data are required to estimate more accurate
stellar masses for the O3Es and O2Es at $z\gtrsim0.8$.
To keep the consistency in data set and photometry with hscPipe
between the fields, we use only the HSC data in this paper. The
studies with $u$-band and NIR data are planned in the next data
release \citep{HSCSSPDR1}.
On the other hand, the systematic difference by a factor of $\sim2$
can be caused by the different codes used for the SED fitting even if
the same data set is used \citep{Behroozi2010}. The stellar masses
that we estimate with the five BBs for the emission-line galaxies are
reliable for the HAEs and those for the O3Es and O2Es can still be
used with a caveat that there is a possible systematic difference by a
factor of $\sim2$. 

\subsection{Summary of catalogs of emission-line galaxies}
\label{sec:ELGs.summary}
We have identified
8,054 H$\alpha$ emitters at $z\approx$ 0.25 and 0.40,
8,656 [OIII] emitters at $z\approx$ 0.63 and 0.84, and
16,877 [OII] emitters at $z\approx$ 1.19 and 1.47 from
NB816 and NB921 data down to the limit of magnitude and line flux
where the samples are complete.  
The line fluxes above the completeness limit correspond to the
observed luminosities larger than
$\log$($L$/(erg s$^{-1}$)) $\gtrsim$ 39.4 and 40.1 in
H$\alpha$ at $z\approx$ 0.25 and 0.40,
$\log$($L$/(erg s$^{-1}$)) $\gtrsim$ 40.4 and 40.8 in [OIII] at
$z\approx$ 0.63 and 0.84, and 
$\log$($L$/(erg s$^{-1}$)) $\gtrsim$ 41.1 and 41.4 in [OII] at
$z\approx$ 1.19 and 1.47. 
These are among the largest samples ever constructed. The samples that
we have selected are summarized in Table \ref{tab:Samples}. Among the
emission-line galaxies, $\sim$3.9\% are identified by the
spectroscopic redshift, $\sim$71.5\% are by the
photometric redshift, and $\sim$24.6\% are by colors.
The catalogs are available at the HSC-SSP data release site%
\footnote{\label{foot:hscsspwebsite} https://hsc-release.mtk.nao.ac.jp/},
after the paper is published.

%%%%%%%%%%%%%%%%%%%%%%%%%%%%%%%%%%%%%%%%%%%%%%%%%%%%%%%%%%%%%%%%
%%%%%%%%%%%%%%%%%%%%%%%%%%%%%%%%%%%%%%%%%%%%%%%%%%%%%%%%%%%%%%%%

\section{Results}
\label{sec:results}

%%%%%%%%%%%% Figure 11 %%%%%%%%%%%%%%%%%
\begin{figure*}
  \begin{center}
    \begin{tabular}{cc}    
      \includegraphics[width=0.45\textwidth]{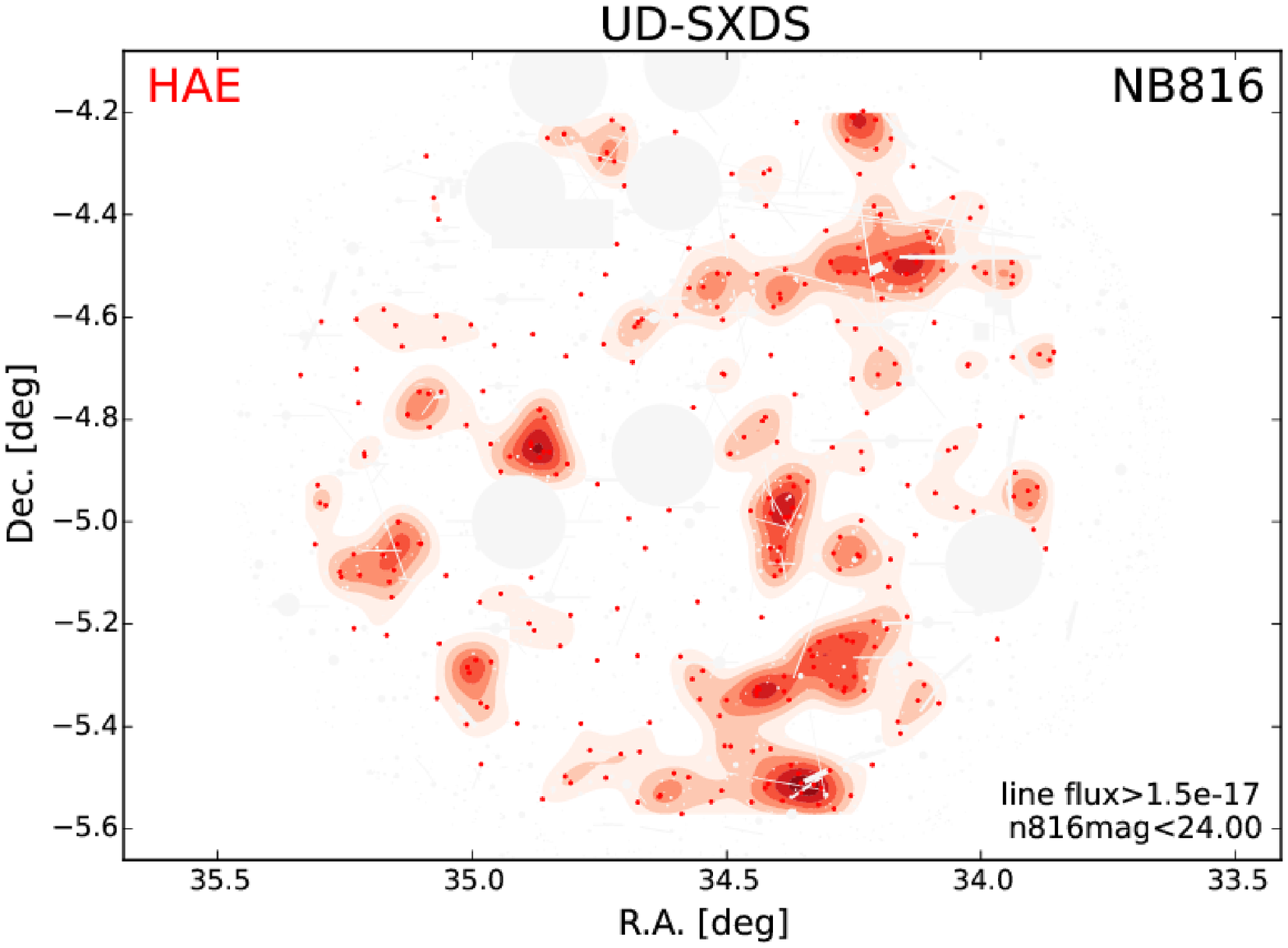}
      & 
      \includegraphics[width=0.45\textwidth]{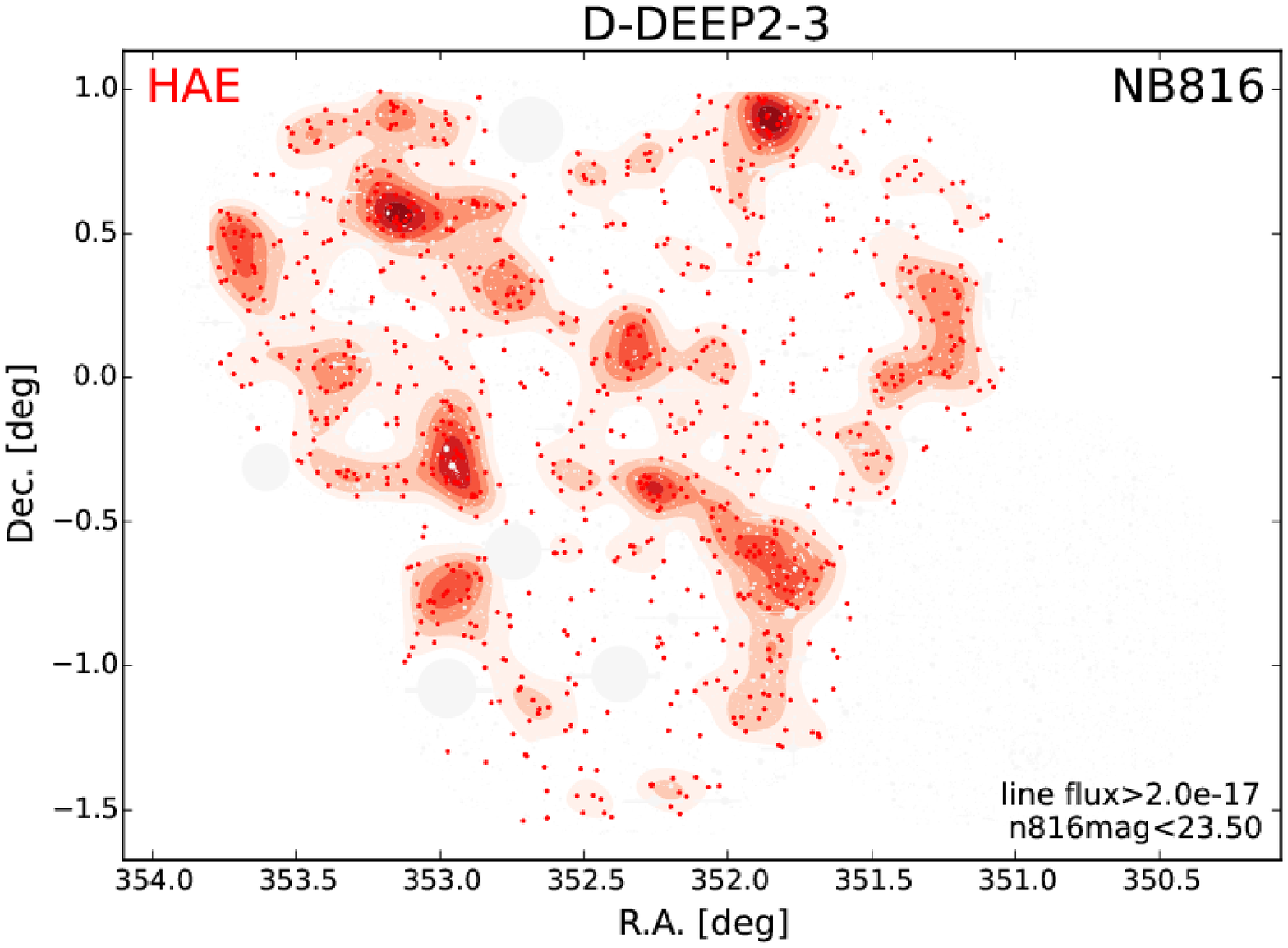}\\
      \includegraphics[width=0.45\textwidth]{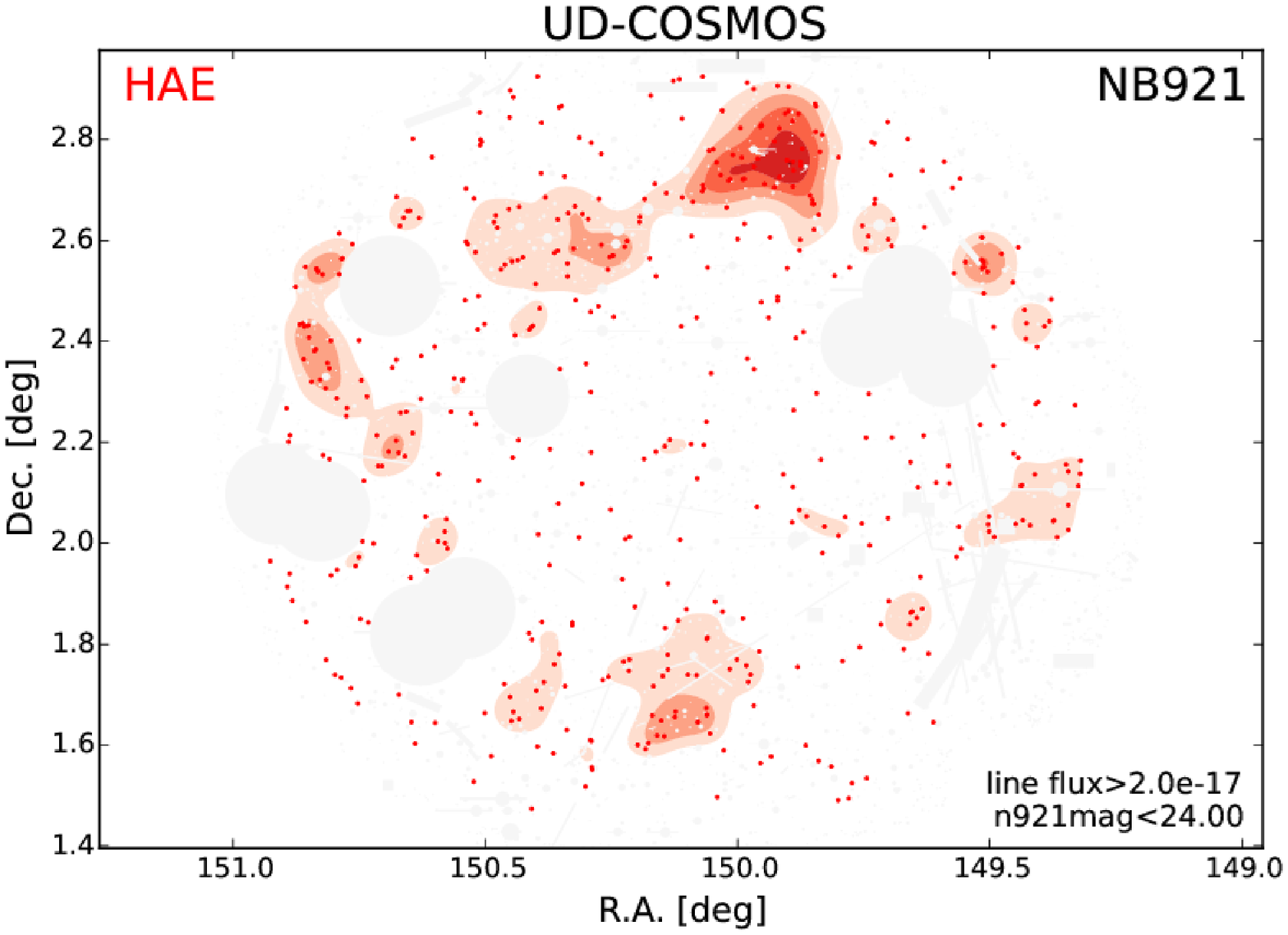}
      & 
      \includegraphics[width=0.45\textwidth]{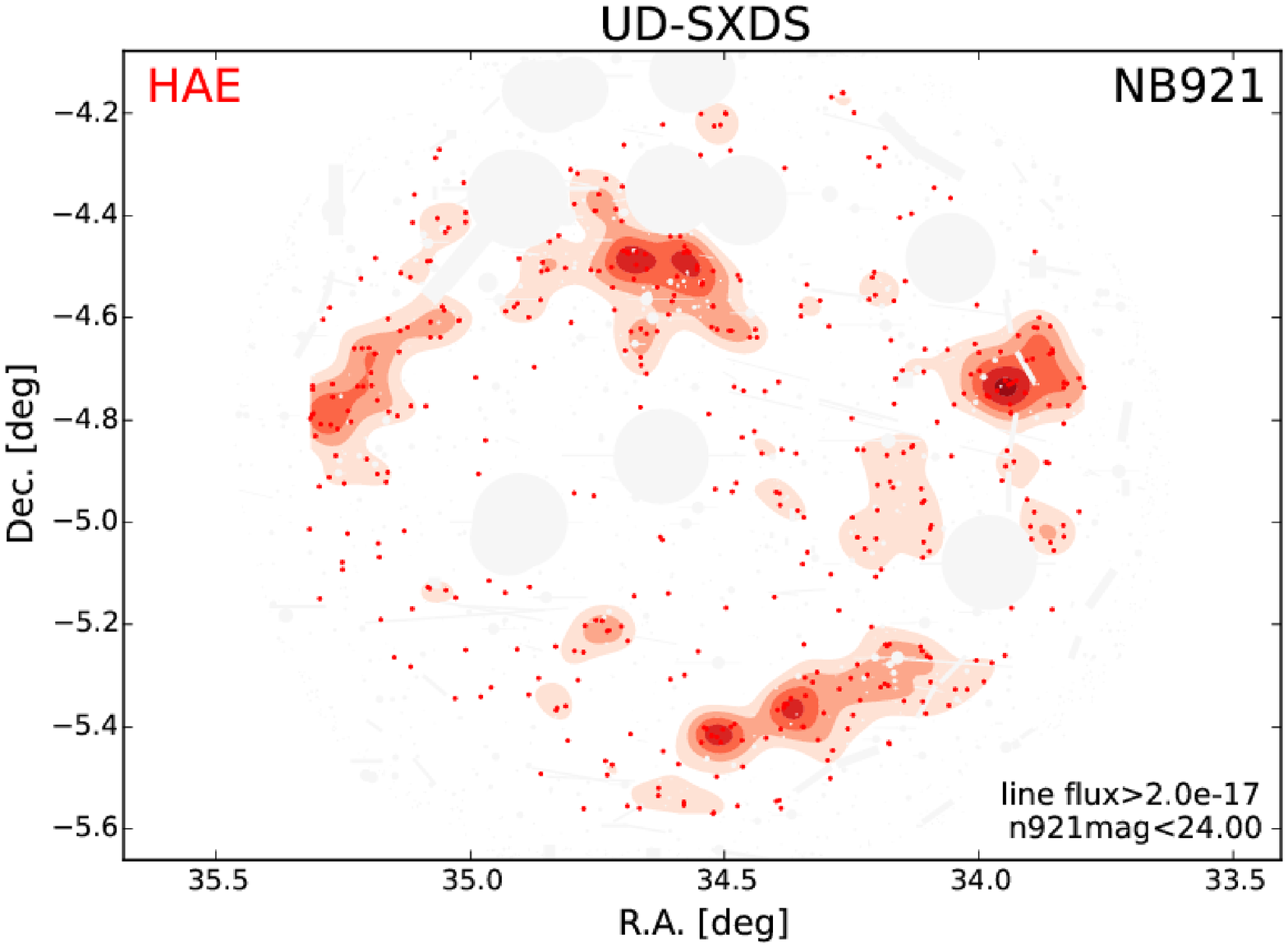}\\
      \includegraphics[width=0.45\textwidth]{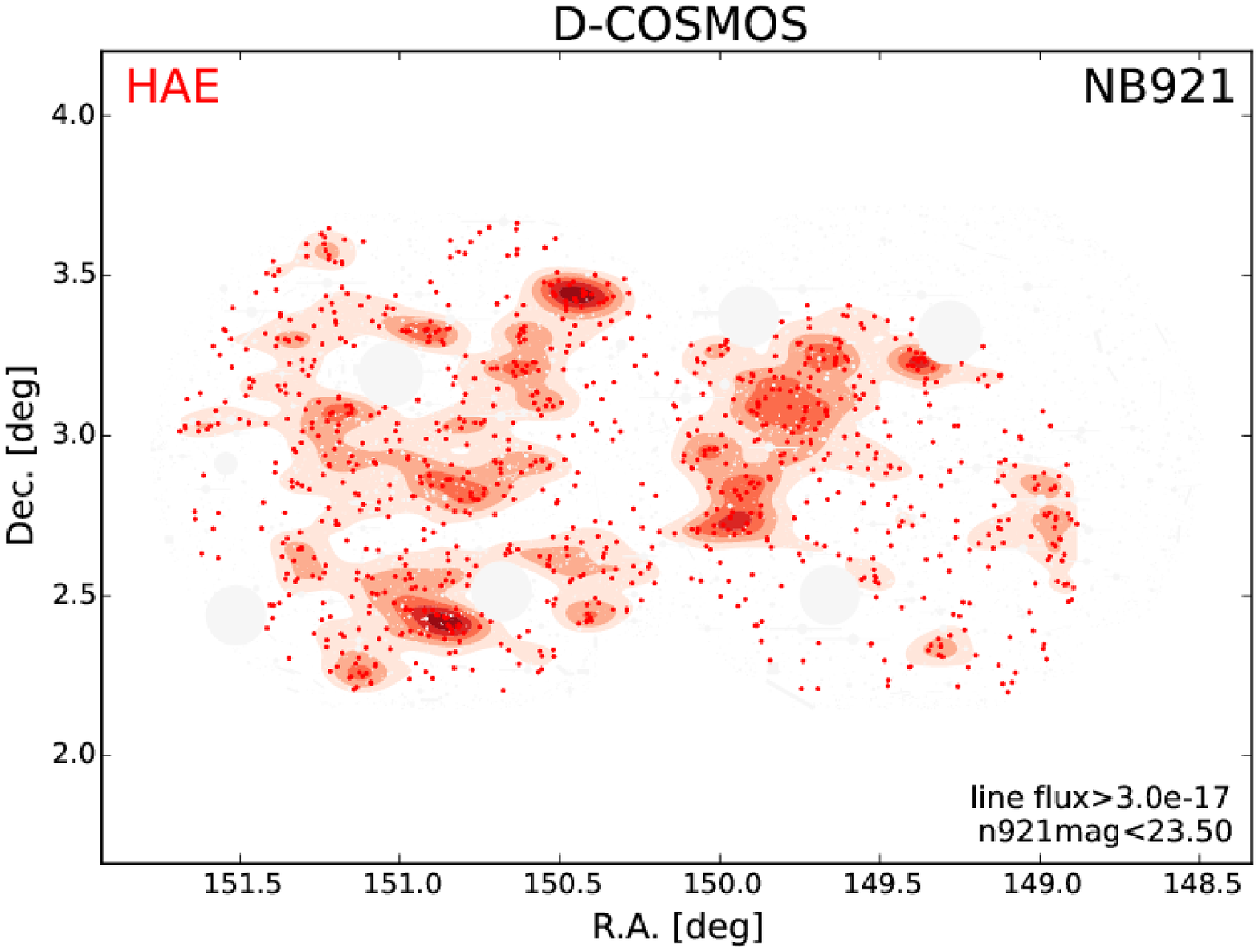}
      & 
      \includegraphics[width=0.45\textwidth]{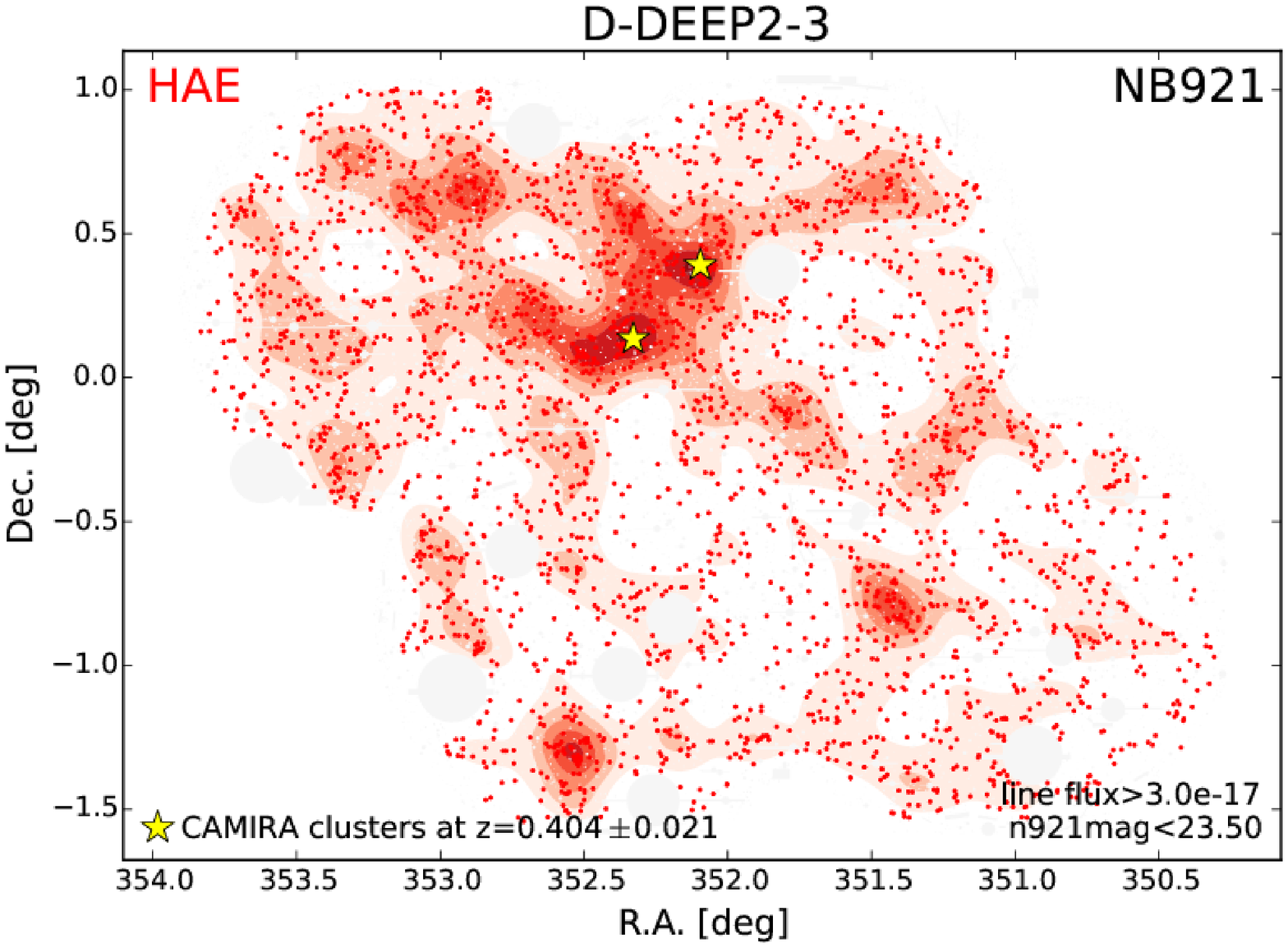}\\
      \includegraphics[width=0.45\textwidth]{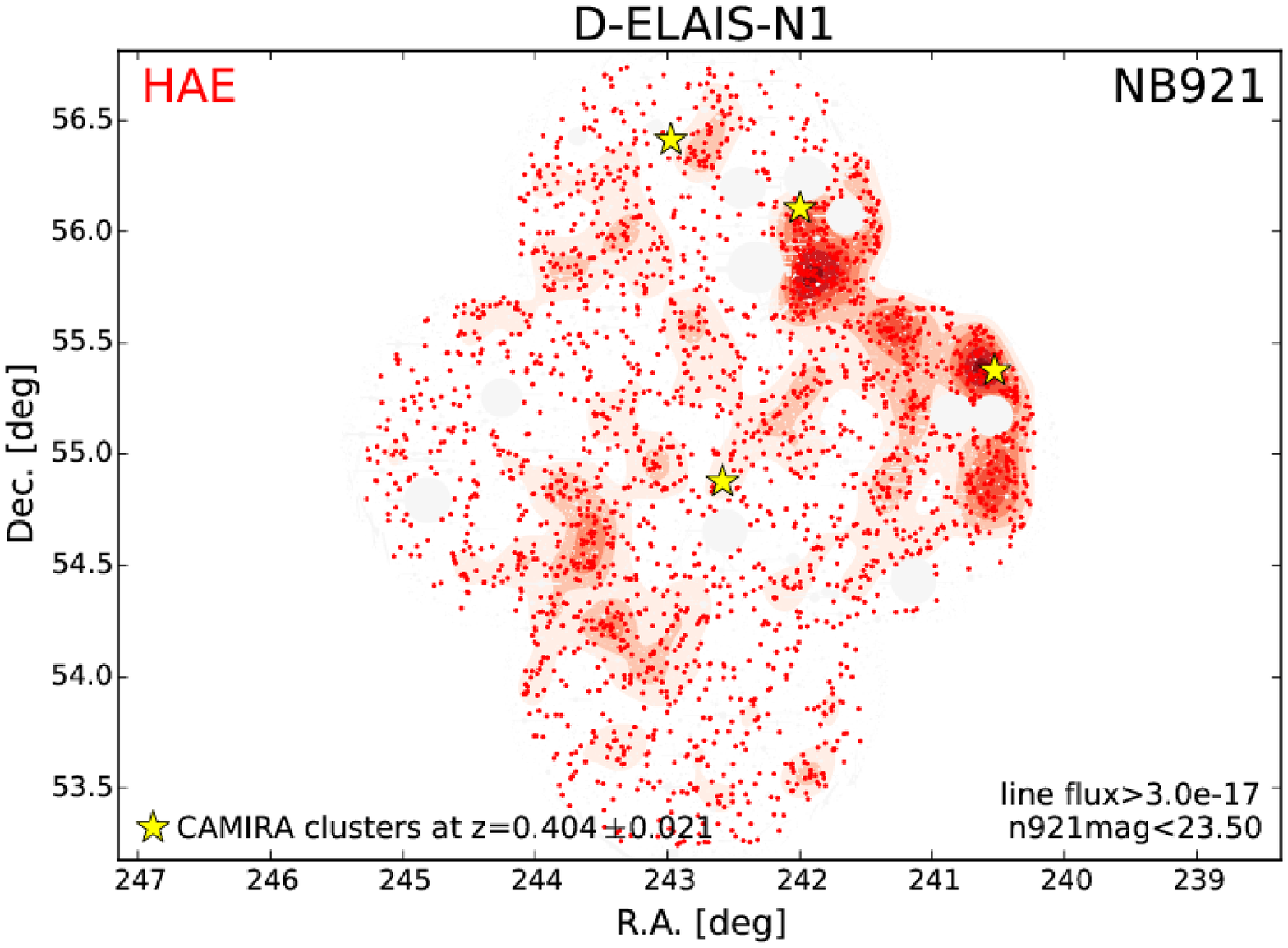}
      & \\
    \end{tabular}      
  \end{center}
 \caption{Spatial distribution of the HAEs at $z \approx$ 0.25 (for
   NB816) and 0.40 (for NB921) in each fields. The dots show the
   individual emission-line galaxies, whose number densities are shown
   by the contours. The gray regions are masked as described in \S
   \ref{sec:data.nb.mask}. 
   Star symbols show galaxy clusters at redshifts corresponding to
   $\lambda_c\pm\Delta\lambda$ of the NB, which are discovered from
   the red-sequence galaxies by the CAMIRA code
   \citep{Oguri2014,Oguri2017}.
 }\label{fig:map_HAEs}
\end{figure*}
%%%%%%%%%%%% Figure 12 %%%%%%%%%%%%%%%%%
\begin{figure*}
  \begin{center}
    \begin{tabular}{cc}    
      \includegraphics[width=0.45\textwidth]{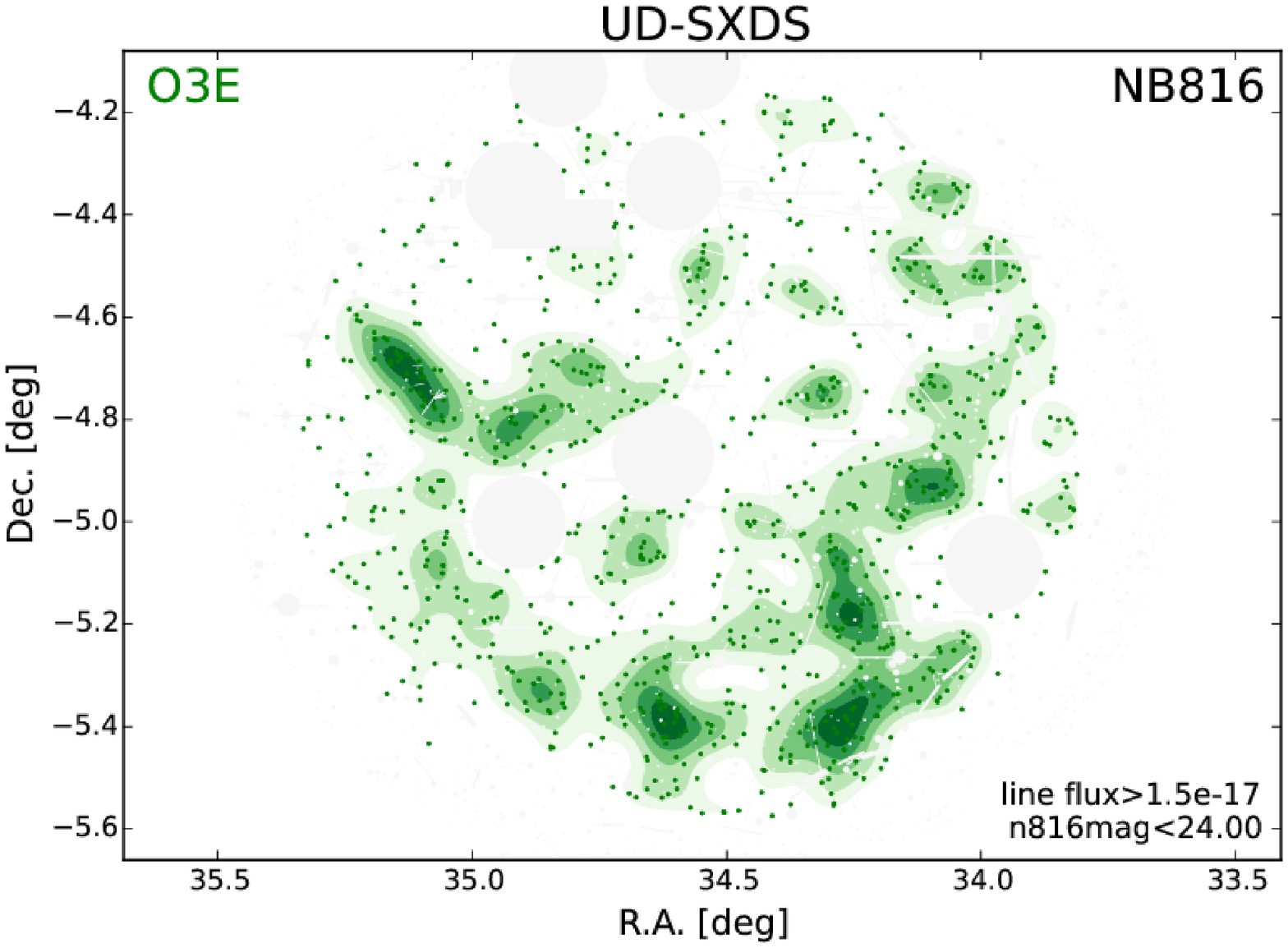}
      & 
      \includegraphics[width=0.45\textwidth]{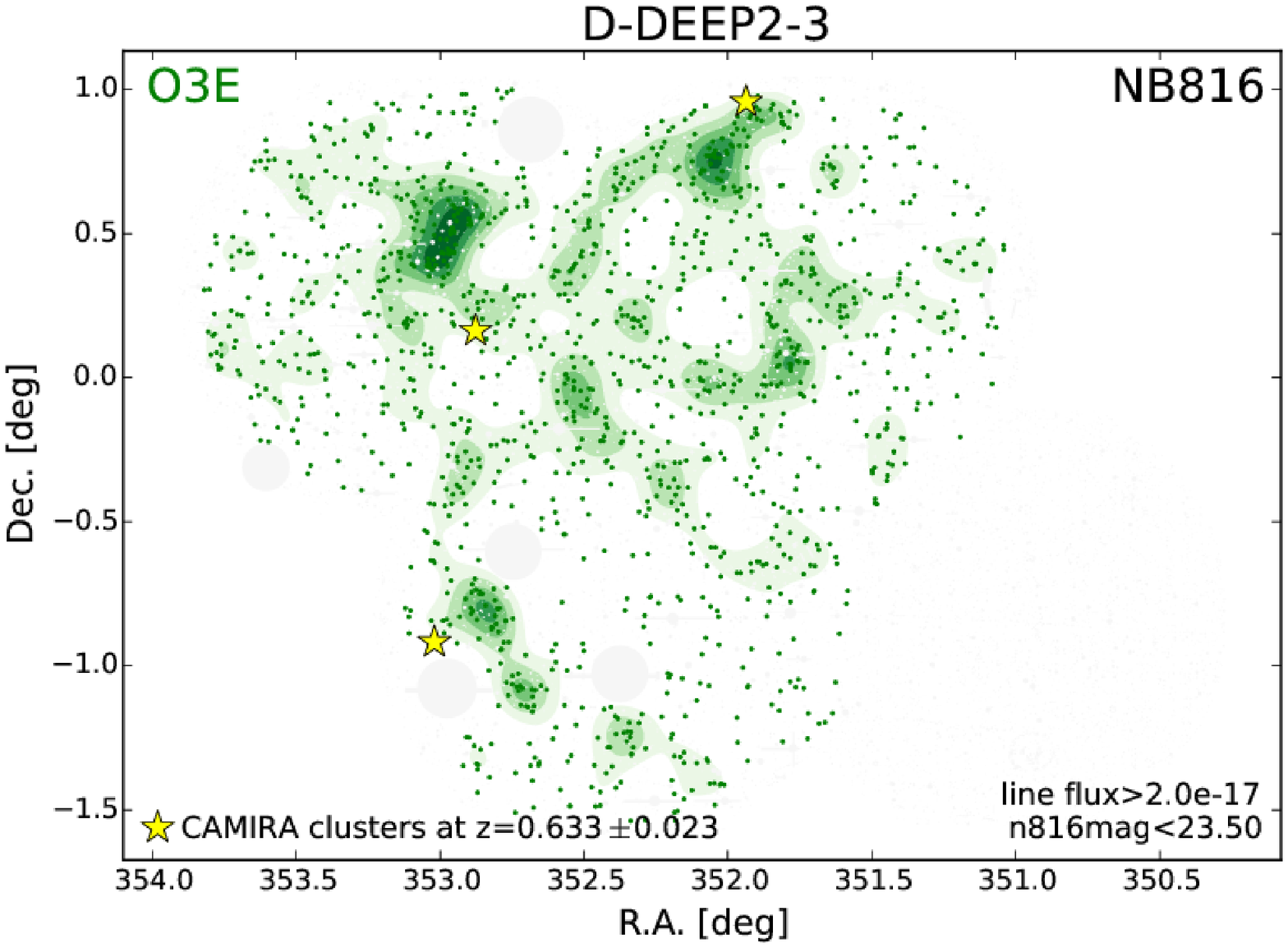}\\
      \includegraphics[width=0.45\textwidth]{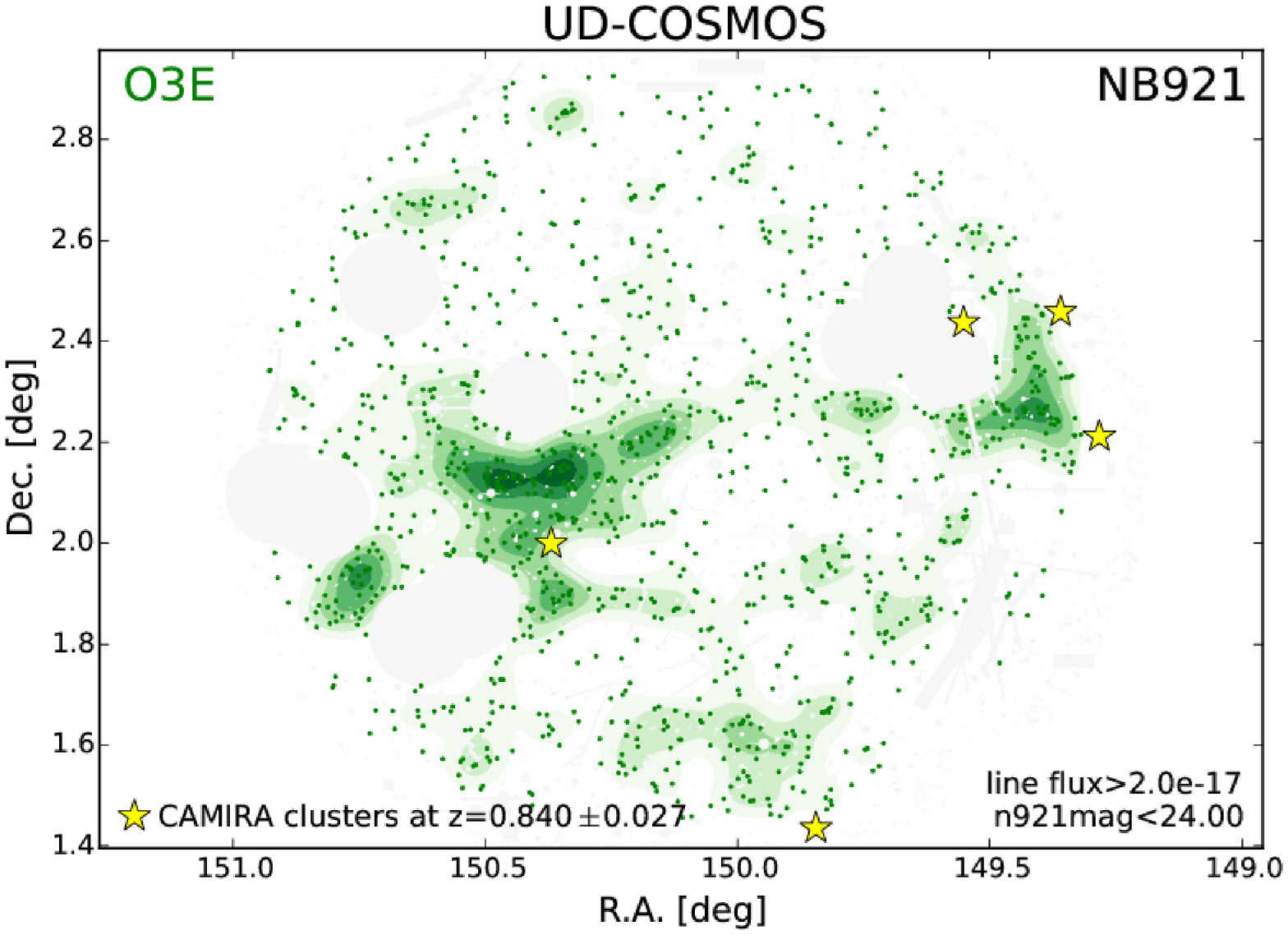}
      & 
      \includegraphics[width=0.45\textwidth]{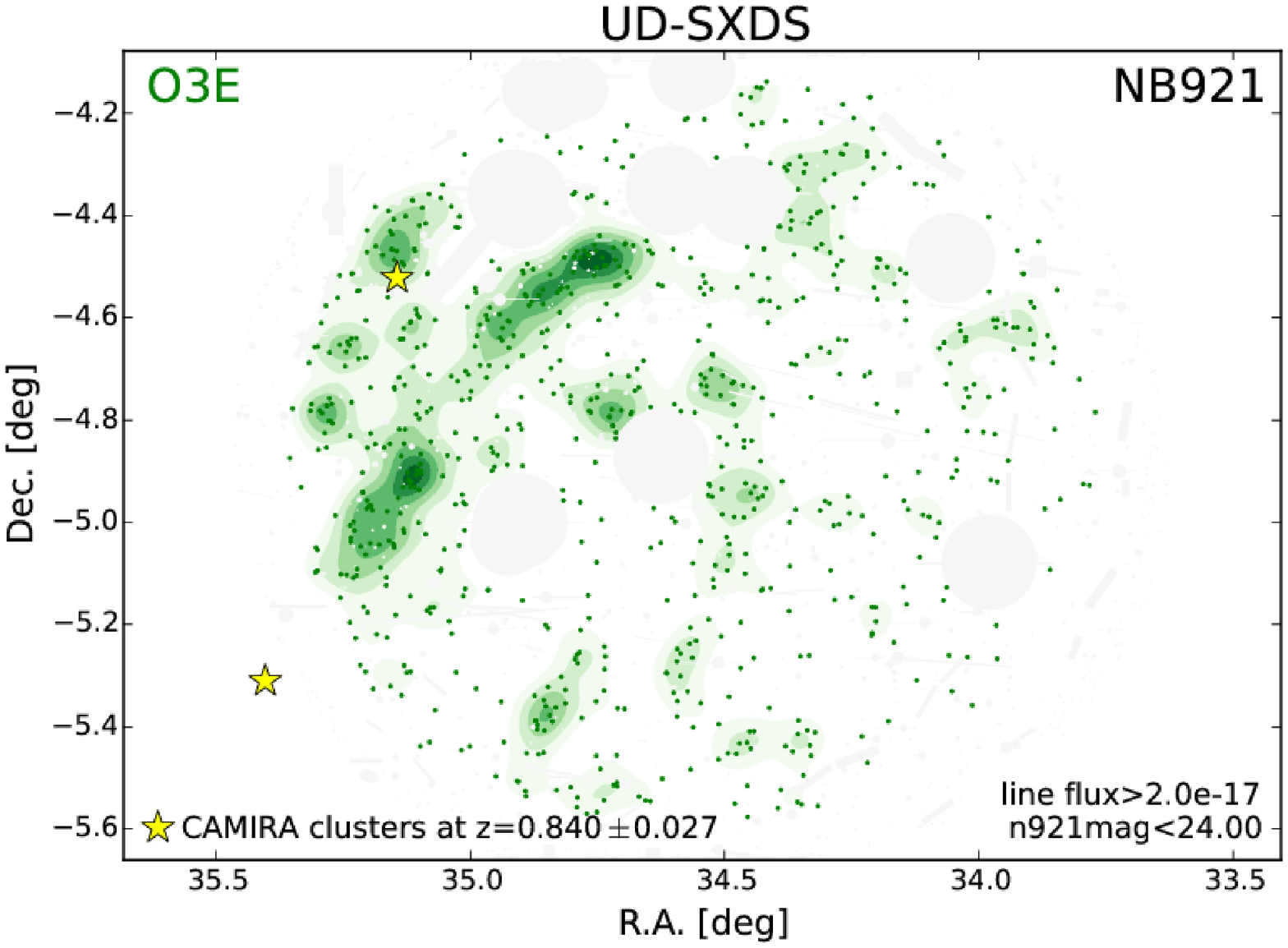}\\
      \includegraphics[width=0.45\textwidth]{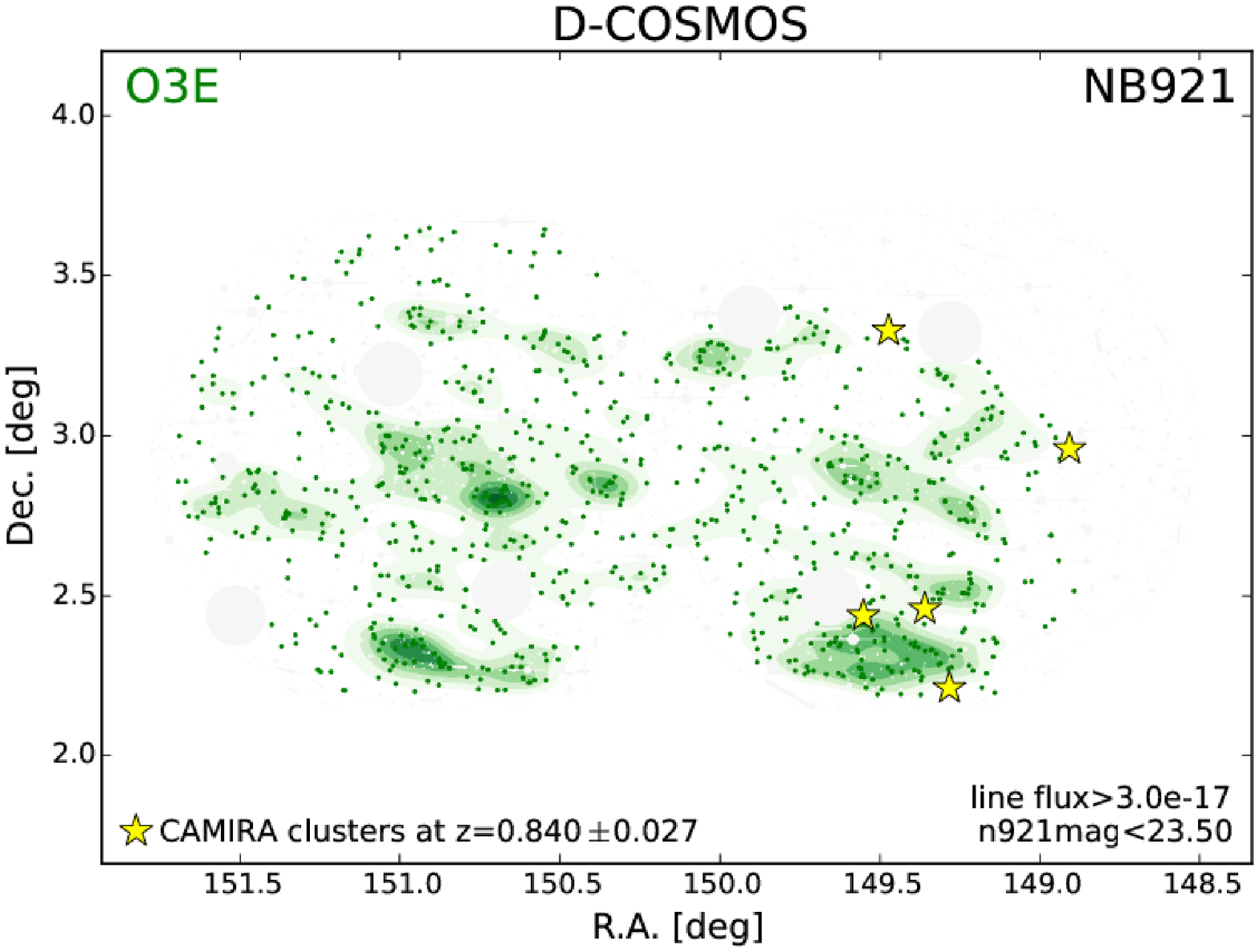}
      & 
      \includegraphics[width=0.45\textwidth]{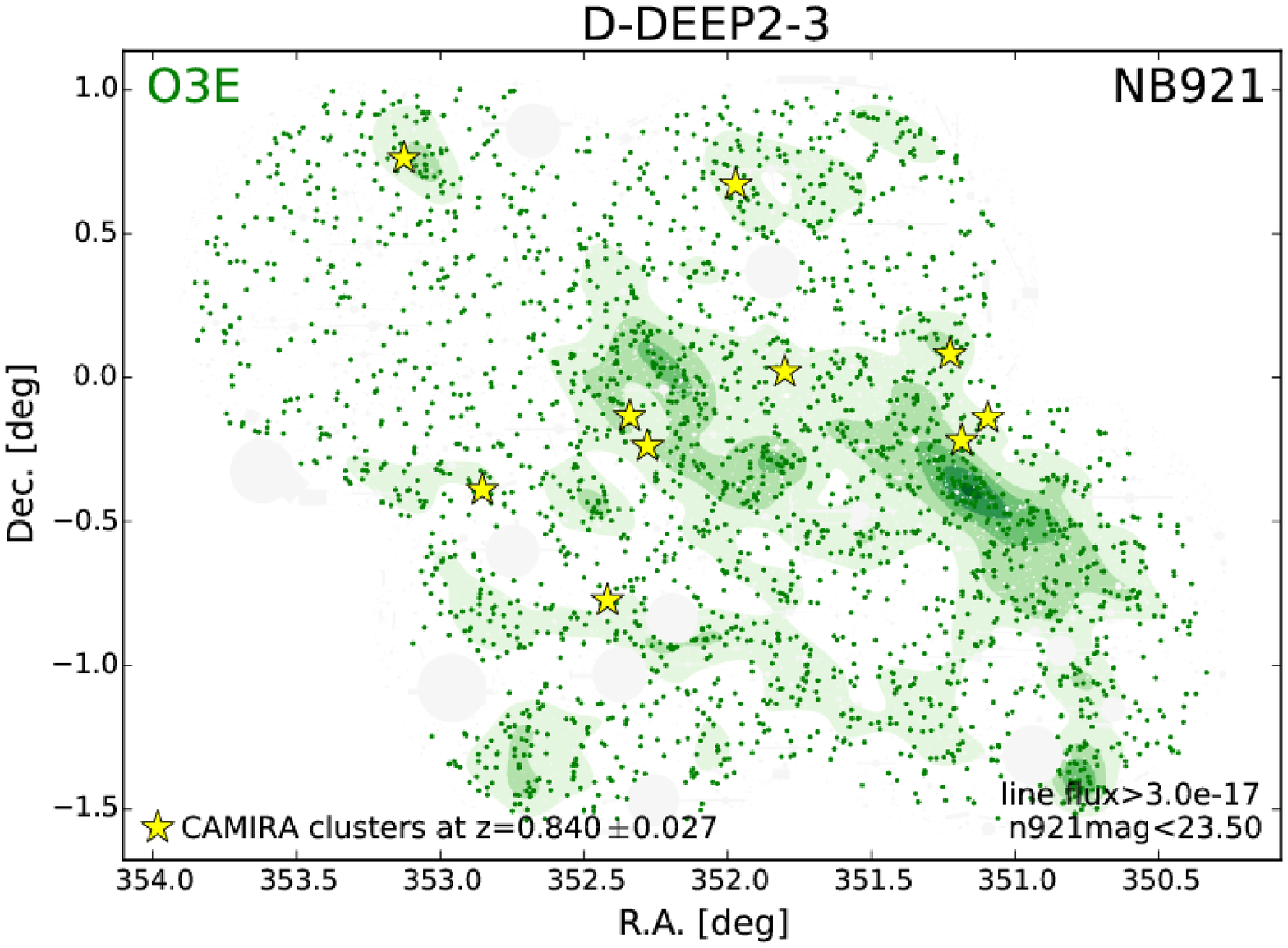}\\
      \includegraphics[width=0.45\textwidth]{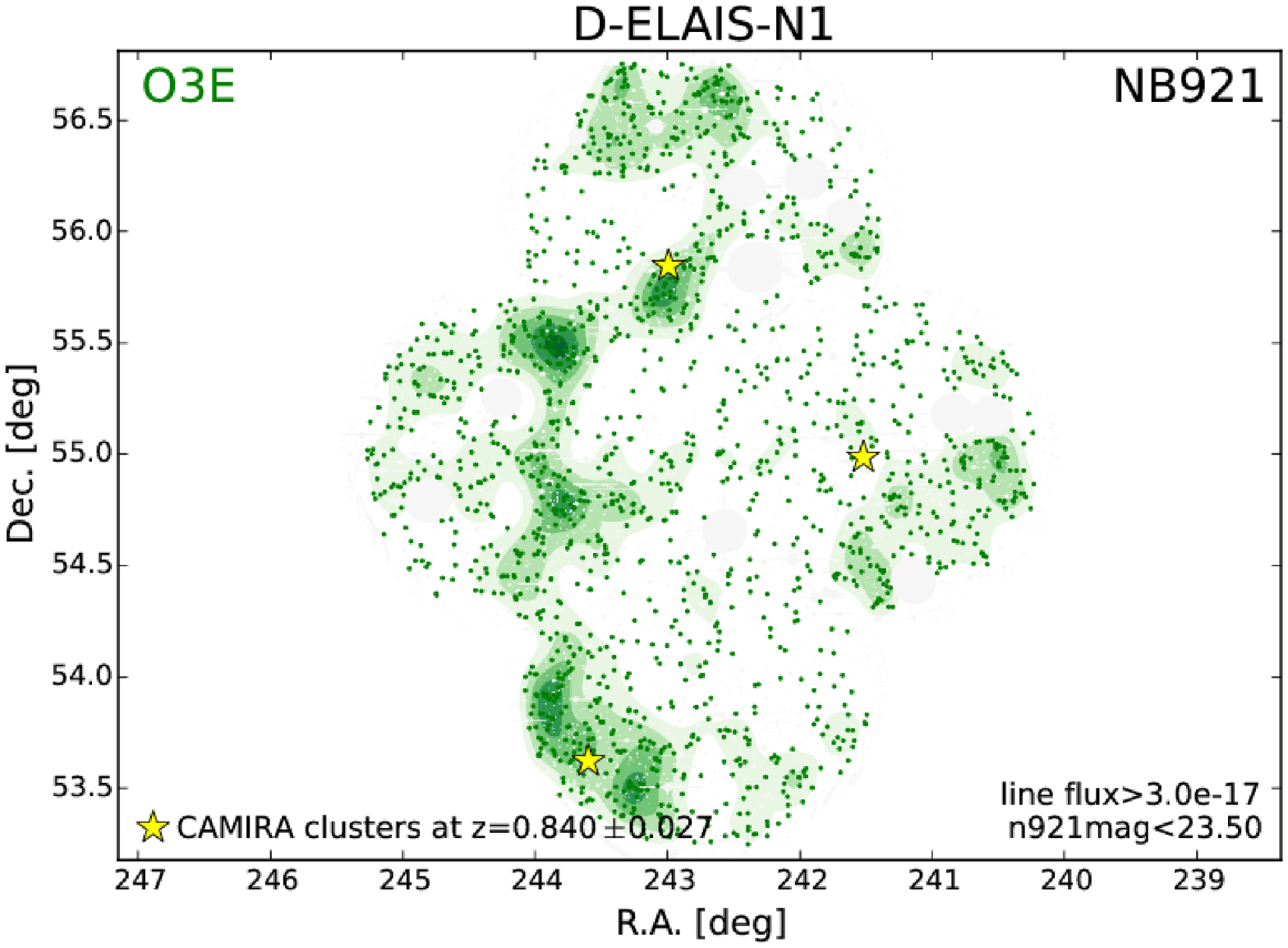}
      & \\
    \end{tabular}      
  \end{center}
 \caption{The same as Figure~\ref{fig:map_HAEs}, but for the O3Es at
   $z \approx$ 0.63 (for NB816) and 0.84 (for NB921).}\label{fig:map_O3Es}
\end{figure*}
%%%%%%%%%%%% Figure 13 %%%%%%%%%%%%%%%%%
\begin{figure*}
  \begin{center}
    \begin{tabular}{cc}    
      \includegraphics[width=0.45\textwidth]{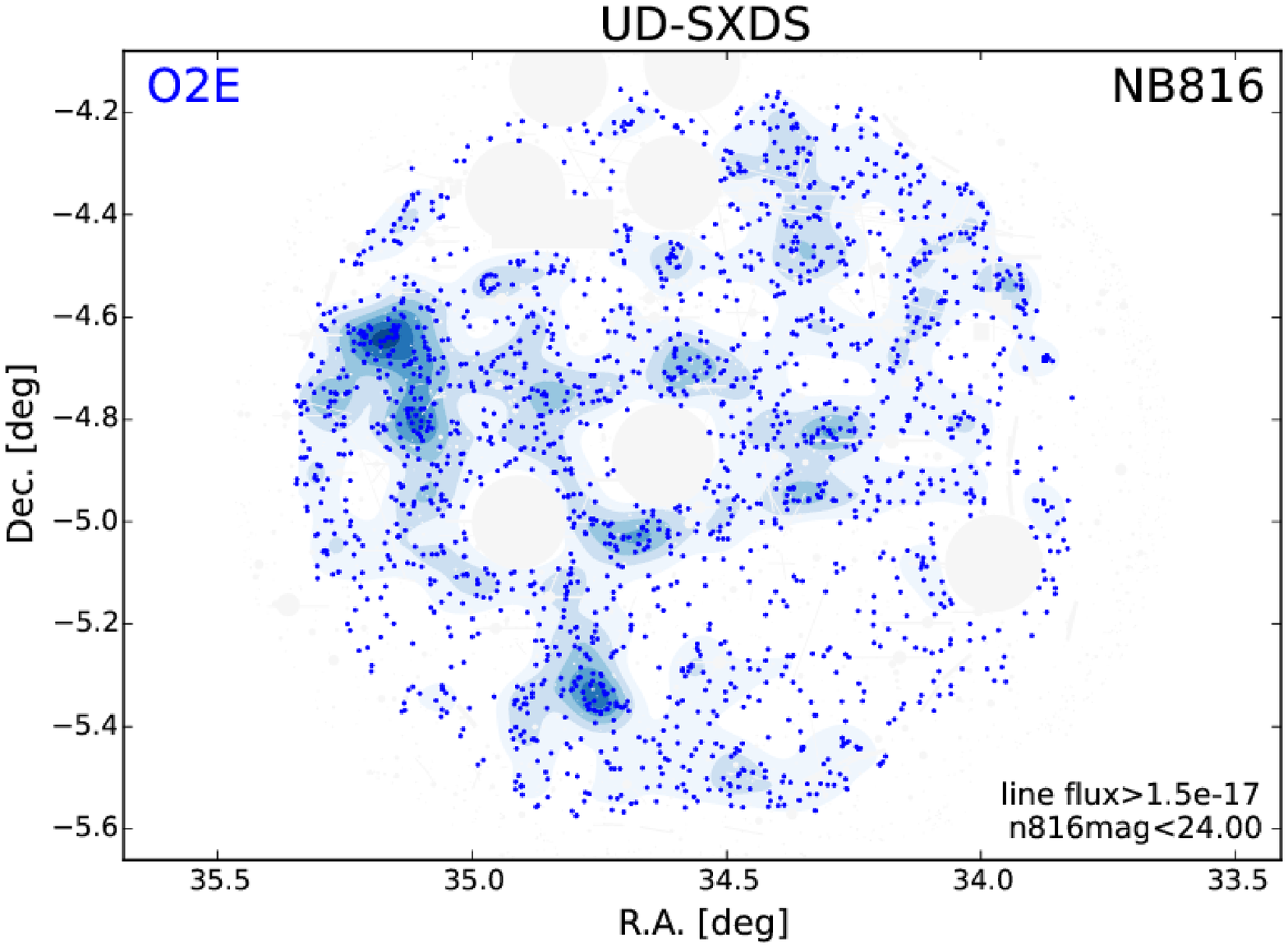}
      & 
      \includegraphics[width=0.45\textwidth]{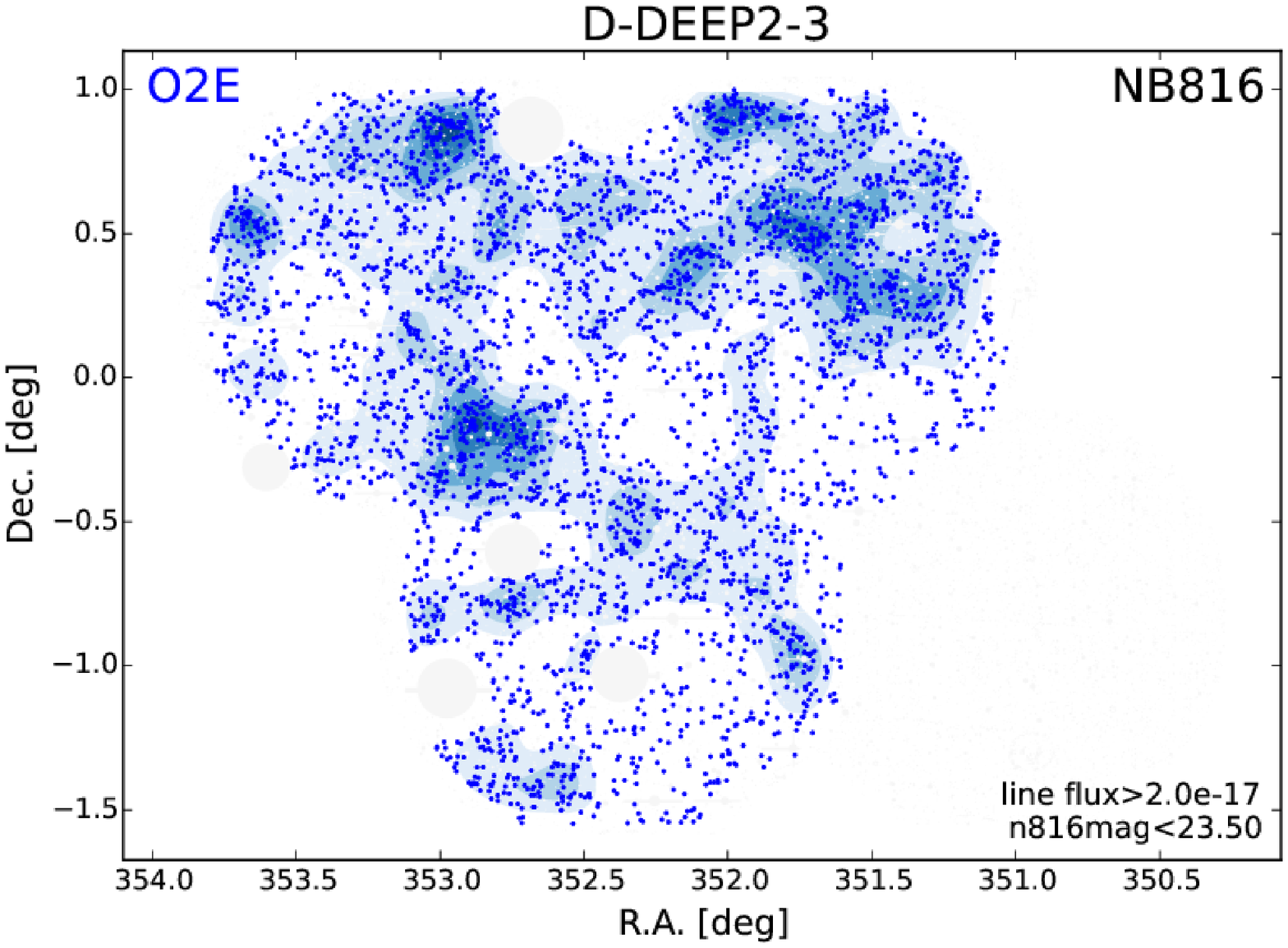}\\
      \includegraphics[width=0.45\textwidth]{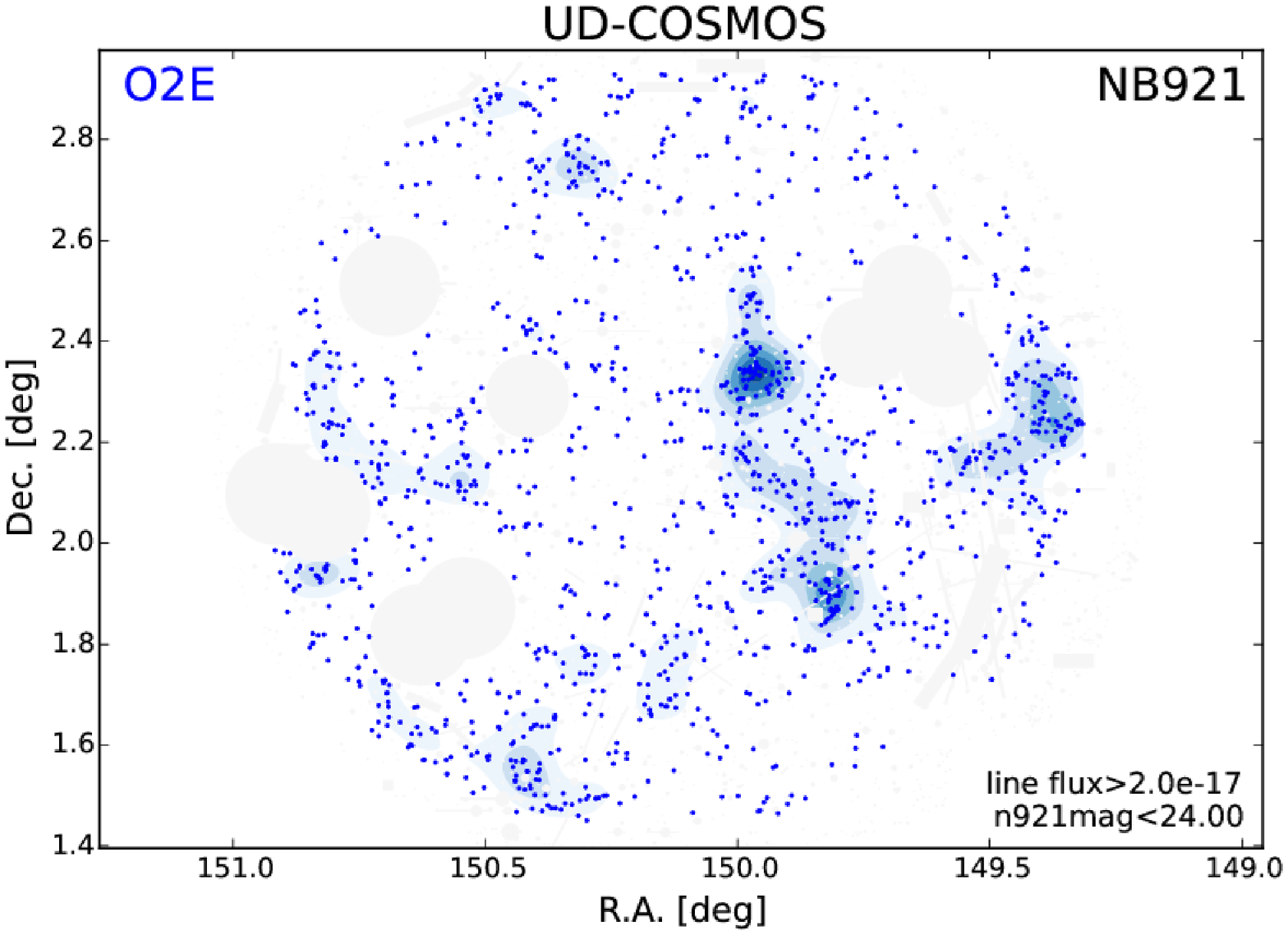}
      & 
      \includegraphics[width=0.45\textwidth]{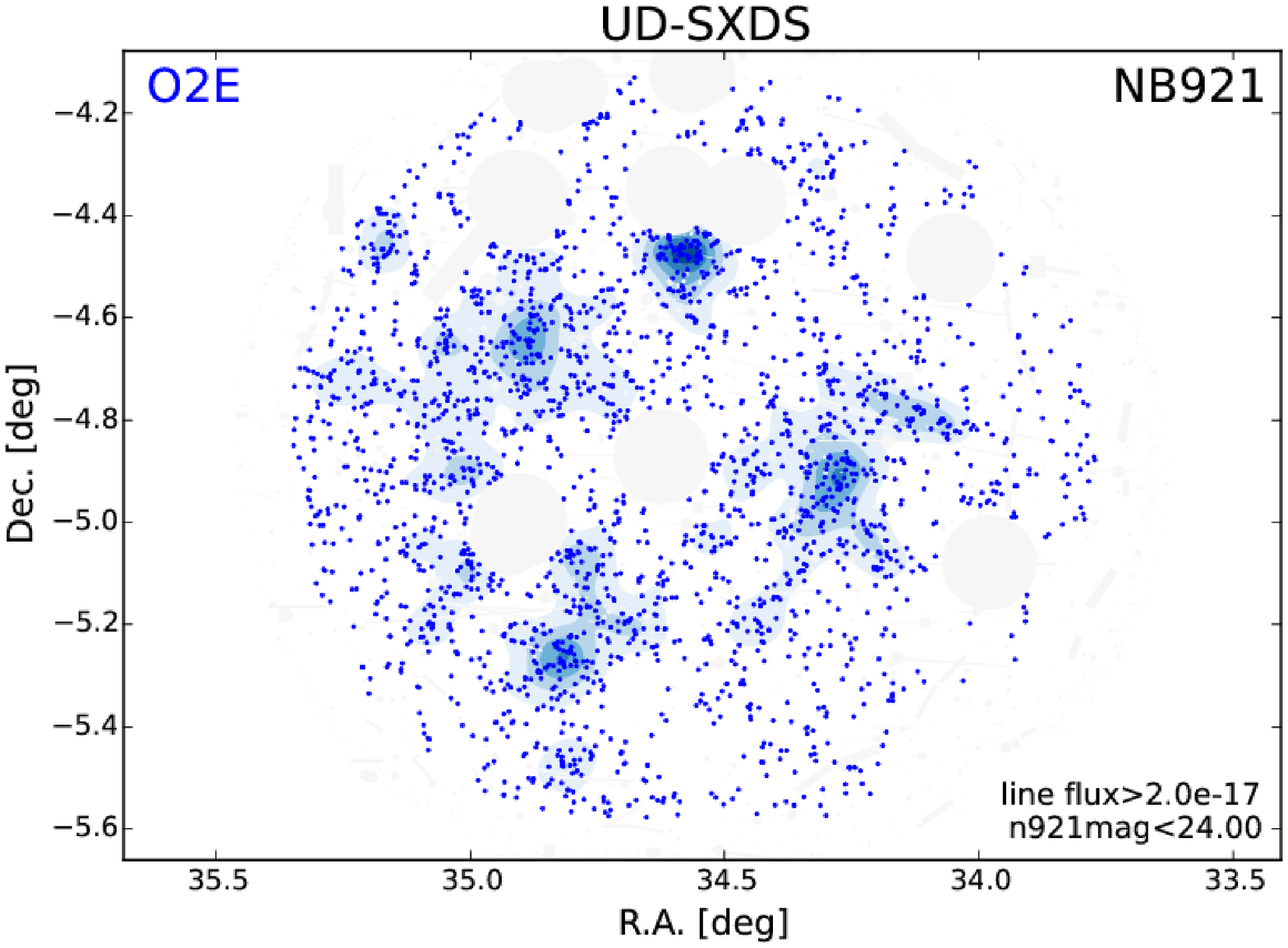}\\
      \includegraphics[width=0.45\textwidth]{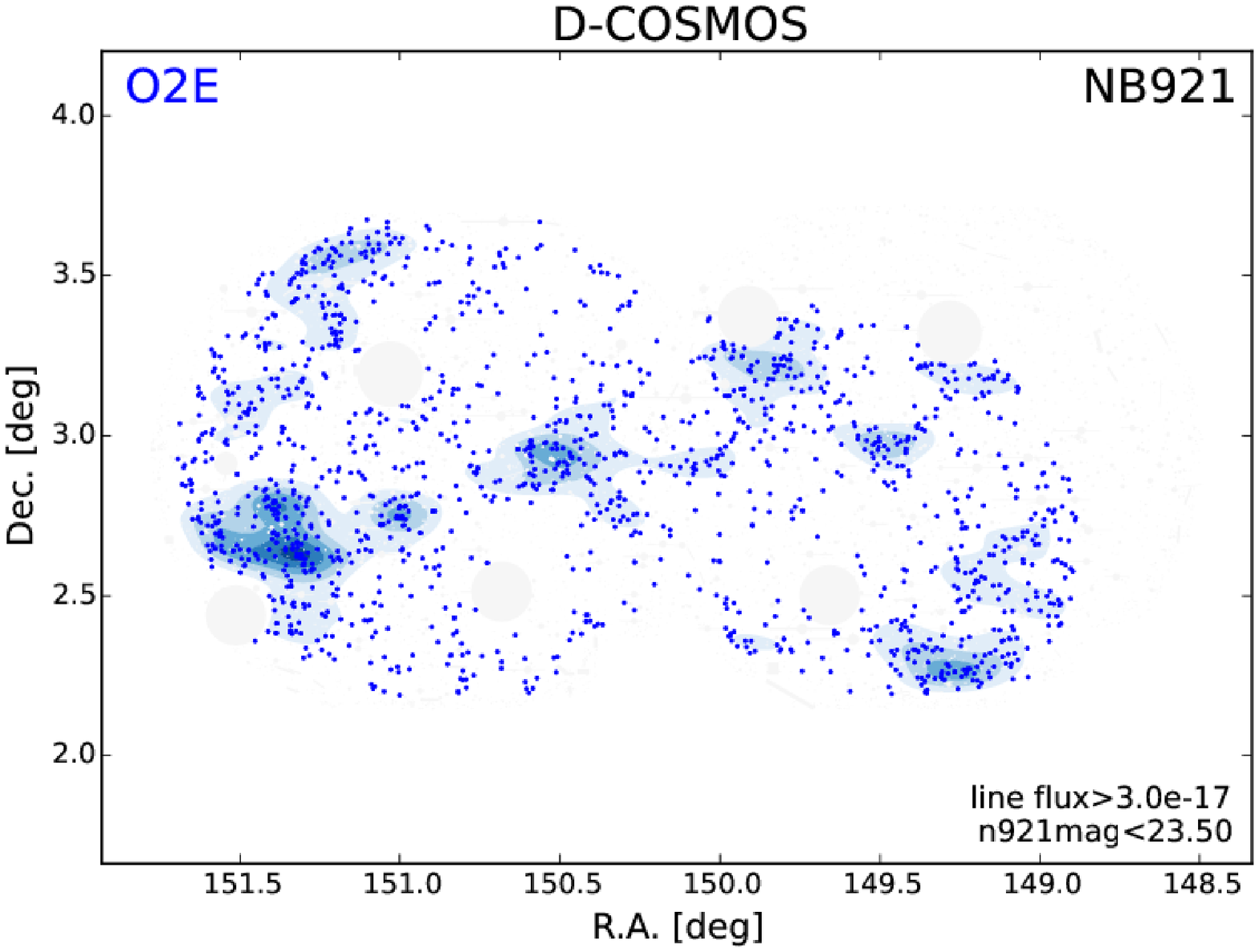}
      & 
      \includegraphics[width=0.45\textwidth]{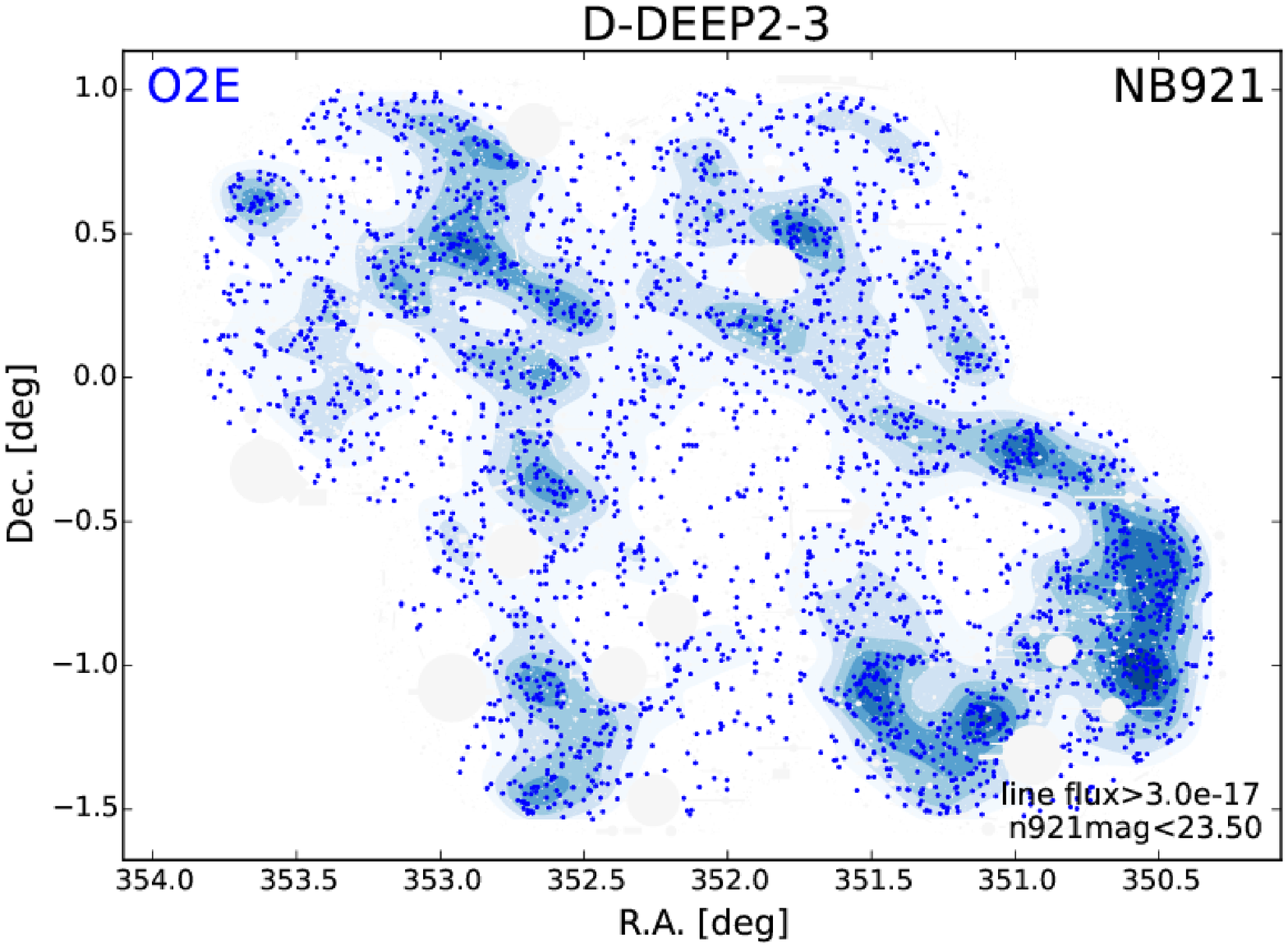}\\
      \includegraphics[width=0.45\textwidth]{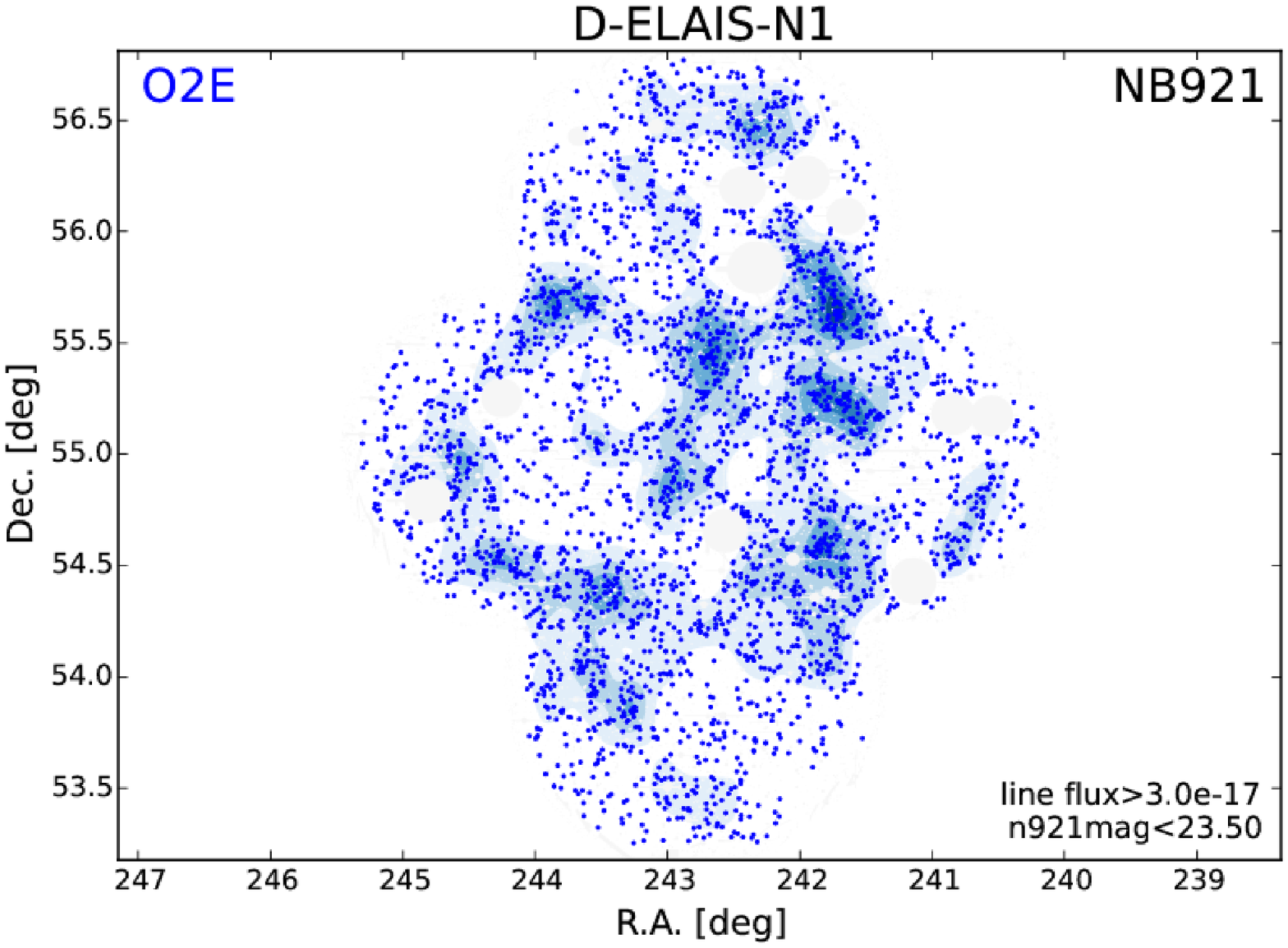}
      & \\
    \end{tabular}      
  \end{center}
  \caption{The same as Figure~\ref{fig:map_HAEs}, but for the O2Es at
    $z \approx$ 1.19 (for NB816) and 1.47 (for NB921). Note that the
    comparison with the CAMIRA clusters is not performed for these
    plots. This is because the CAMIRA code identifies the galaxy 
    clusters at only $z<1.1$ with the currently available
    data.}\label{fig:map_O2Es}  
\end{figure*}
%%%%%%%%%%%%%%%%%%%%%%%%%%%%%%%%%%%%%%%%%%%%%%%%%%%%%%%%%

\subsection{Spatial distribution of emission-line galaxies}
\label{sec:results..distribution}
Figures~\ref{fig:map_HAEs}--\ref{fig:map_O2Es} show the spatial
distribution of HAEs, O3Es, and O2Es at each redshift in each
field, where the emission-line galaxies with NB magnitude brighter
than and emission-line flux larger than the limits shown in Table
\ref{tab:Samples} are plotted to ensure homogeneity of the depth and
completeness over the fields surveyed. 
The figures clearly show that the individual populations of
emission-line galaxies are not distributed uniformly but
constitute large-scale structures over the fields surveyed.
The scale of some structures are larger than the $\sim1.5$ degree
diameter of the HSC FoV, which corresponds to $\sim$50 comoving Mpc at
$z\sim0.4$ (e.g., HAEs in the D-DEEP2-3 field shown in
Figure~\ref{fig:map_HAEs}).   
Evidently, HSC can allow us to reveal large-scale structures which
are hard to be fully identified even with Subaru/Suprime-Cam with a
smaller FoV ($0.57\times0.45$ deg$^2$).
In each field, there are several overdense regions which seem to be
located at the intersection of the filamentary structures of
emission-line galaxies. This is a similar picture to what the
hierarchical structure formation predicts (e.g.,
\cite{Vogelsberger2014,Schaye2015,Ishiyama2015}). 

It is interesting to compare the cosmic web of star-forming galaxies
with the location of massive galaxy clusters. \citet{Oguri2017} search
for galaxy clusters with the number of member galaxies larger than 15
up to $z\sim1.1$ in the HSC Wide and Deep fields by using a
red-sequence galaxy finder algorithm named CAMIRA (Cluster finding
Algorithm based on Multi-band Identification of Red-sequence gAlaxies:
\cite{Oguri2014}). We match our emitters to the catalog of the CAMIRA
clusters. The clusters, identified with photometric redshift estimates
matched to the NB selected galaxies, are plotted in
Figures~\ref{fig:map_HAEs}--\ref{fig:map_O2Es} with yellow star
symbols.   

Few CAMIRA clusters are found in the overdense regions of the HAEs at
$z\approx$ 0.25 and 0.40. It is known that the central region of
galaxy clusters at $z<1.0$ tends to be dominated by red quiescent
galaxies and the regions where active star formation occurs move to
the galaxy group environment in the outskirt of the clusters and/or
general fields \citep{Koyama2010,Koyama2011}. Star forming galaxies
may not be the best tracer of clusters at these redshifts, but in the
D-ELAIS-N1 field, there are four CAMIRA clusters at the edge of
redshifts that the NB921 can probe (Figure~\ref{fig:map_HAEs}). The
regions surrounding the clusters are overdense in HAEs, that may
indicate that the HAEs are located so that they connect the
clusters. Of equal interest, we have found a super cluster that
consists of two CAMIRA clusters at $z\approx$ 0.41 embedded in the
large-scale structures of HAEs in the D-DEEP2-3 field
(see Figure~\ref{fig:map_HAEs}). The separation between the clusters
of similar redshift suggests that they are likely merging clusters and
thus may have active star formation even in the central region of the
clusters. The filamentary structures are spread out from the
clusters. This structure including two galaxy clusters should be one
of the interesting regions to investigate the environmental dependence
of galaxy properties, which is discussed in another paper by 
\citet{Koyama2017}. 
HAEs in the merging clusters at $z\sim$ 0.15--0.3 are also
investigated by \citet{Stroe2017} (see also \cite{Stroe2015a}).

Figure~\ref{fig:map_O3Es} shows that the O3Es at $z\approx$ 0.63 and
0.84 do not solely trace galaxy clusters. However, the CAMIRA clusters
at $z\sim0.63$ and 0.84 are found along the large-scale distribution of 
the O3Es. [OIII] emission may not be the best indicator of SFR,
because there can be a contribution from an AGN and [OIII] emitting
galaxies can be biased towards less massive galaxies. However, the
O3Es seem to trace the cosmic web connecting the galaxy clusters. 

We have not searched for galaxy cluster at $z\gtrsim1.1$ with the
CAMIRA code due to lack of NIR data. At $z>1$, it is known that active
star formation occurs in the central region of galaxy clusters
(e.g., \cite{Hayashi2010,Hilton2010,Brodwin2013}), and we may expect
that overdense regions of NB emitters can indicate the sites of
clusters. Indeed, there are several overdensity regions of O2Es in
each field, all of which are strong candidates of galaxy clusters at
$z\approx$1.2--1.5. Follow-up spectroscopy is required to identify
galaxy clusters where the O2Es are concentrated.    

\subsection{Luminosity function}
\label{sec:results.lf}

\subsubsection{Luminosities of emission lines}
\label{sec:results.lf.lum}

Fluxes of the emission lines are derived from the NB and BB
photometry:
\begin{eqnarray}
F_{\rm line} = \frac{\Delta_{NB} \cdot \Delta_{BB}}{(\Delta_{BB} - \Delta_{NB})} \left(f_{\lambda,NB} - f_{\lambda,BB}\right) ,
\label{eq:NBflux}
\end{eqnarray}
where $F_{\rm line}$ is a line flux, $\Delta$ is a FWHM of the filter
in units of \AA\ (see Table~\ref{tab:NBemitters} for NBs, and
$\Delta_{BB}$ = 1555 and 782 for $i$ and $z$ bands), and $f_\lambda$
is a flux density in units of erg s$^{-1}$ cm$^{-2}$ \AA$^{-1}$.
As discussed in \S \ref{sec:ELGs.selection}, the {\tt cmodel}
magnitudes in {\tt unforced} photometry are used for the flux 
estimation. To measure the line fluxes according to the equation
(\ref{eq:NBflux}), it is assumed that the same stellar continuum level
underlying emission lines contributes to the flux densities of
$f_{\lambda,BB}$ and $f_{\lambda,NB}$. This implies that the
difference of effective wavelength between BB and NB can affect the
measurement of fluxes, in particular, for [OII]$\lambda3727$ doublet
being near the Balmer/4000\AA\ break or emission lines of red
galaxies. Thus, as in \S~\ref{sec:ELGs.selection}, we use the BB
magnitudes corrected for the color term to calculate the fluxes of the
emission lines. We use the best-fitting model spectra from the SED
fits with all the BB filters, and we find that the stellar continuum
underlying an emission line estimated from BB magnitudes are
consistent with the model spectra. In what follows, we use the
continuum estimated from the adjacent BB filters. 
The fluxes measured are converted to the luminosities using the
spectroscopic redshifts, if available. Otherwise, we use the
redshifts estimated from the NB to derive the luminosities.

There are [NII]($\lambda\lambda$=6550,6585\AA) doublet lines next to
the H$\alpha$ line.
While the width of NB is narrow, the targeted line and the adjacent
lines can be simultaneously observed with the NB. In general, the
fluxes of the contaminant emission lines are small compared with the
targeted line of H$\alpha$. However, it is important to estimate how much the
contamination contributes to the measurement of the fluxes of the
targeted line. The contribution of [NII] line is estimated from the
SDSS Data Release 7 (DR7) spectroscopic catalogs
\citep{Kauffmann2003,Salim2007,Abazajian2009}.
We select analogs of the HAEs by applying the same rest-frame EW cut
to the SDSS spectroscopic data and investigate the [NII]/H$\alpha$
line ratio as a function of stellar mass and observed H$\alpha$
luminosity for the analogs (see the details in
Appendix~\ref{app:lineratio}). We correct the line fluxes for the
contribution of [NII] lines in the HAEs based on the stellar mass and
the observed H$\alpha$ luminosity of the individual galaxies. The
median values of the fraction of [NII] over H$\alpha$+[NII] ranges
0.08--0.14 for NB816 and NB921 HAEs in each field.
Although we note that \citet{Villar2008} and \citet{Sobral2012} have
presented the correction of [NII] contribution based on the EW of
H$\alpha$+[NII], we apply the method, described above, to estimate the
line ratios such as [NII]/H$\alpha$ and H$\beta$/H$\alpha$
consistently throughout the paper (see also
\S~\ref{sec:results.lf.lfint} and Appendix~\ref{app:lineratio}).

\subsubsection{Survey volume}
\label{sec:results.lf.vol}

Since the transmission curve of HSC NB filters is not a perfect top-hat
(Figure~\ref{fig:filters}), emission lines at different redshifts go
through a different transmittance. Given that only emission lines with
fluxes above a limiting flux can be detected, emission lines with
an intrinsically large flux can be at lower or higher redshifts than
fainter emission lines. Therefore, the survey volume for an
emission-line galaxy depends on its luminosity. 

We estimate the survey volume for individual emission-line galaxies. For
a given luminosity, the minimum and maximum redshifts where the
emission line can be observed at more than 5$\sigma$ are calculated
based on the filter transmission curve and a limiting flux. Then, the
redshift range is converted to a comoving volume. Galaxies
with brighter lines have a larger survey volume. The median survey
volume in each field is shown in Table~\ref{tab:Samples}. 

Note that we assume that the emission line is observed at the central
wavelength of the NB filter. Given the fact that the NB filters do not
have a perfect top-hat shape of transmission curve, intrinsic
luminosity can be larger than observed luminosity. This results in
underestimating (overestimating) the number of emission-line galaxies
with bright (faint) intrinsic luminosities. Also, the survey volume
can be underestimated for galaxies with an intrinsically larger
luminosity. However, spectroscopic redshift is required to exactly
know at which wavelength within the NB filter the emission line is
observed. While some galaxies have the spectroscopic redshifts,
most of the emission line galaxies do not have. Therefore, we apply
the same survey volume to emission-line galaxies with a given observed
luminosity.

\subsubsection{Observed luminosity function}
\label{sec:results.lf.lfobs}

%%%%%%%%%%%% Figure 14 %%%%%%%%%%%%%%%%%
\begin{figure}[t]
  \begin{center}
    \includegraphics[width=0.5\textwidth]{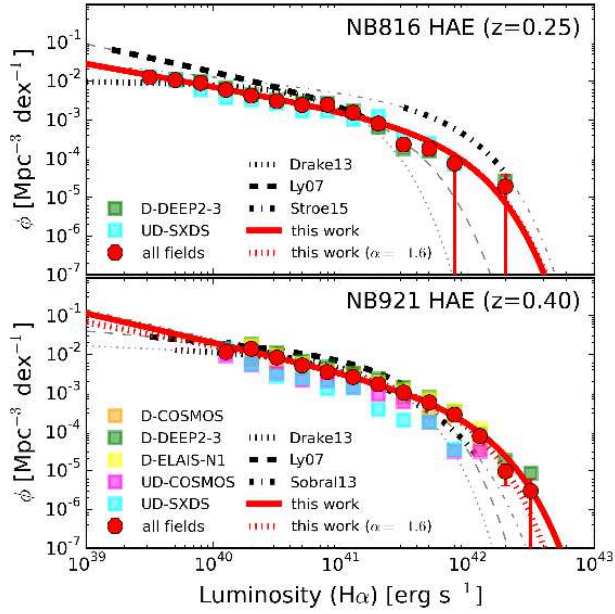}
  \end{center}
  \caption{
  Observed luminosity function of HAEs. The upper panel shows
  the luminosity function for NB816 HAEs at $z \approx 0.25$, while the
  lower panel shows that for NB921 HAEs at $z \approx 0.40$. The squares
  show the data in the individual fields with different colors.
  The red circles show the data in all of the fields surveyed and the
  error bars are estimated from the Poisson errors.
  The red curve is a best-fit Schechter function to the red data
  points, where the solid (dotted) curve is fitted without (with)
  $\alpha$ fixed. The black curves are the luminosity functions from
  the literature, where the  luminosity range covering the data in
  each study is shown by thick curve and the other range is shown by
  thin curve \citep{Ly2007,Drake2013,Stroe2015,Sobral2013}.   
  }\label{fig:LFobs_HAEs} 
\end{figure}
%%%%%%%%%%%% Figure 15 %%%%%%%%%%%%%%%%%
\begin{figure}[t]
  \begin{center}
    \includegraphics[width=0.5\textwidth]{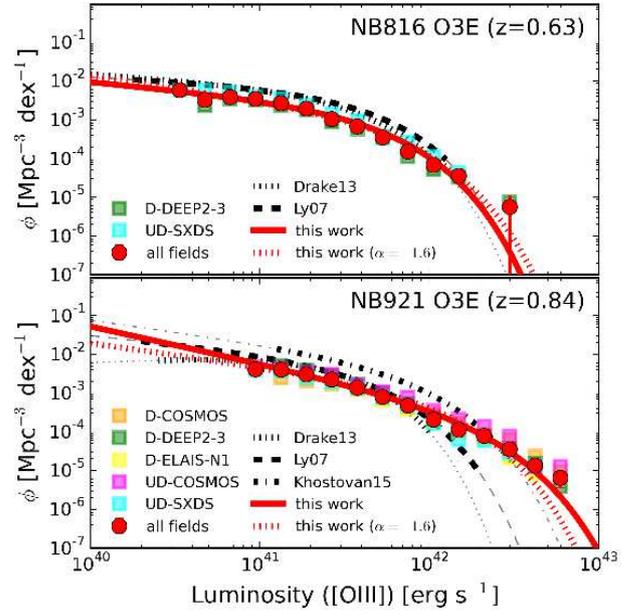}
  \end{center}
  \caption{
    The same as Figure~\ref{fig:LFobs_HAEs}, but for O3Es at
    $z \approx 0.63$ and 0.84. The luminosity functions from the
    literature \citep{Ly2007,Drake2013,Khostovan2015} are plotted
    for comparison. }\label{fig:LFobs_O3Es} 
\end{figure}
%%%%%%%%%%%% Figure 16 %%%%%%%%%%%%%%%%%
\begin{figure}[t]
  \begin{center}
    \includegraphics[width=0.5\textwidth]{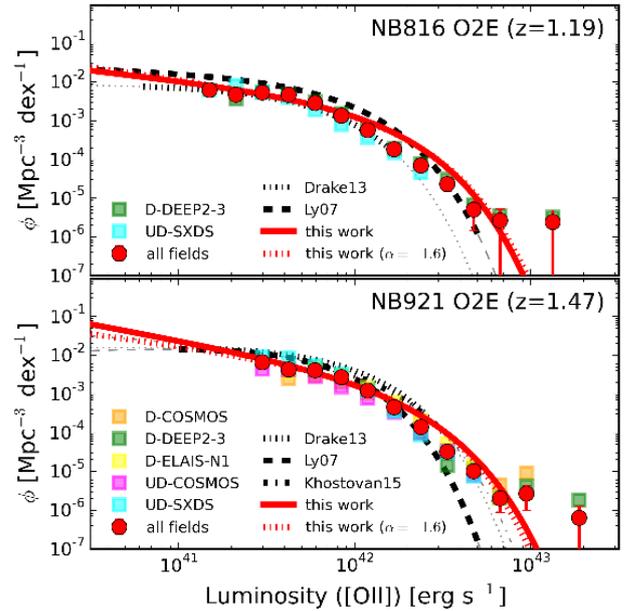}
  \end{center}
  \caption{
    The same as Figure~\ref{fig:LFobs_HAEs}, but for O2Es at
    $z \approx 1.19$ and 1.47. The luminosity functions from the
    literature \citep{Ly2007,Drake2013,Khostovan2015} are also plotted
    for comparison.}\label{fig:LFobs_O2Es}   
\end{figure}
%%%%%%%%%%%%%%%%%%%%%%%%%%%%%%%%%%%%%%%%

%%%%%%%%%%%% Table 4 %%%%%%%%%%%%%%%%%
\begin{table*}
  \tbl{Parameters of the Schechter function fitted to the luminosity
    function.}{%
    \begin{tabular}{lcccccccc}    
      %%%%%%%%%%%%%%%%%%%%%%%%%%%%%%%%%%%%%%%%%%%%%%%
      \hline
       && \multicolumn{3}{c}{No correction for dust extinction} && \multicolumn{3}{c}{Corrected for dust extinction} \\
      \cline{3-5}\cline{7-9}
      & redshift &
      $\log \phi^*$ & $\log L^*$ & $\alpha$ &&      
      $\log \phi^*$ & $\log L^*$ & $\alpha$ \\
      \hline
      HAE & 0.25 
      & -3.53$\pm$0.14 & 41.73$\pm$0.12 & -1.59$\pm$0.05 &&
      -3.55$\pm$0.11 & 41.95$\pm$0.09 & -1.53$\pm$0.04 \\
      HAE & 0.40
      & -3.45$\pm$0.16 & 41.86$\pm$0.11 & -1.75$\pm$0.06 && 
      -4.18$\pm$0.27 & 42.73$\pm$0.24 & -1.82$\pm$0.04 \\
      O3E & 0.63
      & -3.04$\pm$0.15 & 41.59$\pm$0.10 & -1.42$\pm$0.14 &&
      -3.36$\pm$0.16 & 41.97$\pm$0.10 & -1.66$\pm$0.11 \\
      O3E & 0.84
      & -3.71$\pm$0.21 & 42.17$\pm$0.11 & -1.95$\pm$0.11 &&
      -3.66$\pm$0.22 & 42.41$\pm$0.13 & -1.83$\pm$0.12 \\     
      O2E & 1.19
      & -2.81$\pm$0.25 & 41.99$\pm$0.15 & -1.51$\pm$0.28 &&
      -2.90$\pm$0.09 & 42.43$\pm$0.06 & -1.51$\pm$0.09 \\
      O2E & 1.47
      & -2.88$\pm$0.30 & 42.10$\pm$0.15 & -1.83$\pm$0.33 &&
      -2.74$\pm$0.10 & 42.48$\pm$0.06 & -1.41$\pm$0.14 \\
      \hline
      %%%%%%%%%%%%%%%%%%%%%%%%%%%%%%%%%%%%%%%%%%%%%%%
    \end{tabular}
  }\label{tab:Schechter}
  \begin{tabnote}
    Note: $\phi^*$ is in units of Mpc$^{-3}$ dex$^{-1}$ and $L^*$ is
    in units of erg s$^{-1}$.
  \end{tabnote}
\end{table*}
%%%%%%%%%%%%%%%%%%%%%%%%%%%%%%%%%%%%%%%

%%%%%%%%%%%% Table 5 %%%%%%%%%%%%%%%%%
\begin{table*}
  \tbl{Same as Table\ref{tab:Schechter}, but $\alpha$ is
    fixed to be -1.6 in fitting to the luminosity function.}{%
    \begin{tabular}{lcccccccc}    
      %%%%%%%%%%%%%%%%%%%%%%%%%%%%%%%%%%%%%%%%%%%%%%%
      \hline
       && \multicolumn{3}{c}{No correction for dust extinction} && \multicolumn{3}{c}{Corrected for dust extinction} \\
      \cline{3-5}\cline{7-9}
      & redshift &
      $\log \phi^*$ & $\log L^*$ & $\alpha$ &&      
      $\log \phi^*$ & $\log L^*$ & $\alpha$ \\
      \hline
      HAE & 0.25 
      & -3.55$\pm$0.07 & 41.74$\pm$0.09 & -1.6 &&
      -3.74$\pm$0.07 & 42.07$\pm$0.10 & -1.6 \\
      HAE & 0.40
      & -3.14$\pm$0.05 & 41.68$\pm$0.06 & -1.6 && 
      -3.43$\pm$0.07 & 42.21$\pm$0.09 & -1.6 \\
      O3E & 0.63
      & -3.25$\pm$0.07 & 41.70$\pm$0.08 & -1.6 &&
      -3.27$\pm$0.05 & 41.93$\pm$0.05 & -1.6 \\
      O3E & 0.84
      & -3.24$\pm$0.06 & 41.95$\pm$0.05 & -1.6 &&
      -3.32$\pm$0.06 & 42.24$\pm$0.06 & -1.6 \\     
      O2E & 1.19
      & -2.89$\pm$0.11 & 42.03$\pm$0.10 & -1.6 &&
      -2.99$\pm$0.04 & 42.48$\pm$0.03 & -1.6 \\
      O2E & 1.47
      & -2.72$\pm$0.10 & 42.02$\pm$0.07 & -1.6 &&
      -2.89$\pm$0.05 & 42.56$\pm$0.04 & -1.6 \\
      \hline
      %%%%%%%%%%%%%%%%%%%%%%%%%%%%%%%%%%%%%%%%%%%%%%%
    \end{tabular}
  }\label{tab:Schechter_alphafixed}
  \begin{tabnote}
    Note: $\phi^*$ is in units of Mpc$^{-3}$ dex$^{-1}$ and $L^*$ is
    in units of erg s$^{-1}$.
  \end{tabnote}
\end{table*}
%%%%%%%%%%%%%%%%%%%%%%%%%%%%%%%%%%%%%%%

The luminosity function of emission-line galaxies is derived according to
the $V_{max}$ method:
\begin{eqnarray}
\phi(\log L) = \sum_{i}\frac{1}{V_{max} \cdot f_c \cdot \Delta(\log L)},
\end{eqnarray}
where $i$ is for individual galaxies with $\log L\pm0.5\Delta(\log L)$
(\S \ref{sec:results.lf.lum}), 
$V_{max}$ is the survey volume (\S \ref{sec:results.lf.vol}) and $f_c$
is detection completeness (\S \ref{sec:data.nbquality.completeness}).
The detection completeness is taken into account based on both NB
(proxy of emission-line flux) and BB (proxy of stellar continuum
underlying the emission line) magnitudes (Figure~\ref{fig:completeness}).  

Figures~\ref{fig:LFobs_HAEs}--\ref{fig:LFobs_O2Es} show the observed
luminosity functions for HAEs, O3Es, and O2Es at each redshift, i.e.,
dust extinction in emission line is not corrected yet.
Because luminosity functions can be corrected for dust extinction in a
few different ways, the observed luminosity functions are the simplest
ones to compare with the results from the literature. 
The luminosity function is derived from galaxies in all of the
fields surveyed. Note that we do not double count the galaxies in the
overlapping regions between the UD-COSMOS and D-COSMOS fields to
derive the luminosity functions.
The luminosity functions in the individual UD and D fields are also
shown with different colors in the figures, which is useful to
indicate the field variance that illustrates how consistently the
luminosity functions are derived for the individual fields.
The sample variance is discussed in
\S~\ref{sec:discussions.samplevariance}. We fit a Schechter function
\citep{Schechter1976} to the luminosity functions, where the Schechter
function is:     
\begin{eqnarray}
\phi(L)dL = \phi^* \left(\frac{L}{L^*}\right)^{\alpha} \exp(-\frac{L}{L^*}) \frac{dL}{L^*}~,
\end{eqnarray}
or 
\begin{eqnarray}
\phi(L)d(\log L) = \phi^* \left(\frac{L}{L^*}\right)^{\alpha+1} \exp(-\frac{L}{L^*})~\ln10~ d(\log L)~.  
\end{eqnarray}
The Schechter parameters of the best-fit function are
shown in Table~\ref{tab:Schechter}.
Table~\ref{tab:Schechter_alphafixed} also show the Schechter
parameters of the best-fit function, but $\alpha$ is fixed to be -1.6
according to \citet{Sobral2013} and \citet{Sobral2015}.
The luminosity functions are well fit by the Schechter
function. However, we notice that some luminosity functions show the
excess of number density at the bright end from the best-fit Schechter
function (e.g., O2Es in Figure~\ref{fig:LFobs_O2Es}), which is
discussed in \S~\ref{sec:discussions.brightendoflf}.  

For comparison with our results, luminosity functions from previous
studies in wide fields are also plotted in 
Figures~\ref{fig:LFobs_HAEs}--\ref{fig:LFobs_O2Es}.
\citet{Ly2007} conducted NB imaging with NB816 and NB921 on the
Subaru/Suprime-Cam in the SDF of 0.24 deg$^2$.
They selected NB816 and NB921 emitters with the observed
EWs greater than 33 and 15 \AA, respectively.
The data are deeper than our HSC-SSP PDR1 data, but the area is much
smaller than our survey.
\citet{Drake2013} also conducted NB816 and NB921 imaging
with Subaru/Suprime-Cam in the SXDS field of 0.63 deg$^2$.
They applied the rest-frame EW cut of 100\AA\ to their samples to derive
luminosity functions. 
HiZELS conducted wider NB921 imaging with Subaru/Suprime-Cam in the
COSMOS and UDS fields of $\sim2$ deg$^2$ \citep{Sobral2013,Khostovan2015}.
They selected NB921 emitters with the rest-frame greater than 25 \AA.
The area covered by the previous studies are much smaller than HSC-SSP
PDR1 data, because the Suprime-Cam only has one sevenths the FoV of
HSC. However, these studies use the similar NB filters to those of HSC
and thus the H$\alpha$, [OIII], and [OII] luminosity functions at the
same redshifts are investigated. 

Although our luminosity functions for all of HAEs, O3Es, and O2Es
seem to be consistent with previous studies over the investigated
luminosity ranges, the number density of our emitters may be slightly
smaller than the previous studies in each luminosity bin. The small 
discrepancy can be caused by the relatively large EW cut in our
samples: an observed EW of 48\AA\ (56\AA) for NB816 (NB921) emitters.  
Since the current NB data of HSC-SSP is not so deep, we cannot 
give a strong constraint on the luminosity functions at less than
$\sim0.1L*$ ($0.3L*$) for O3Es (O2Es) at this time.
On the other hand, the faint end of luminosity functions for HAEs
reaches down to $\sim 0.01L*$. 

Our luminosity functions for NB816 (NB921)-selected emission-line
galaxies are based on 5.68 (16.2) deg$^2$ data, which
is one of the widest NB imaging data sets.
The number density of bright galaxies is low and survey volume for
galaxies at lower redshifts is also not large, which suggests that the
results for brighter galaxies at lower redshifts are more sensitive to
the area of the field surveyed. 
Indeed, the bright end of our luminosity functions is determined by
the emitters in the D fields. Thanks to the wide-field data, we can
give a constraint on the luminosity functions up to the larger
luminosity bins than investigated by the previous studies.

\subsubsection{Intrinsic luminosity function}
\label{sec:results.lf.lfint}

To derive the intrinsic luminosity functions for the emission-line
galaxies, it is important that the luminosities are corrected for dust
extinction. We estimate an amount of the dust extinction in the
NB-measured emission-line fluxes using the H$\alpha$/H$\beta$ line 
ratios of the SDSS analogs (Appendix~\ref{app:lineratio}). Based on
stellar mass and observed luminosity of the targeted emission line (for
example, [OII] for O2Es), we estimate the Balmer decrement for the
individual galaxies. Then, assuming the \citet{Cardelli1989}
extinction curve and the intrinsic H$\alpha$/H$\beta$ ratio of 2.86
for Case B recombination under an electron temperature of $T_e=10^4$ K
and electron density of $n_e=10^2$ cm$^{-3}$ \citep{Osterbrock1989},
the estimated H$\alpha$/H$\beta$ ratio is converted to the magnitude
of dust attenuation in the targeted line.   

The previous studies have presented the typical dust extinction in
emission line from a star-forming galaxy as a function of stellar
mass, using the SDSS data (e.g.,
\cite{Gilbank2010,Garn2010b}). Although the dust extinction estimated
by our method is comparable to the estimation by the previous studies,
our method allows us to correct the luminosity of emission lines for
dust extinction in the consistent way for HAEs, O3Es, and O2Es.

\begin{figure}[t]
  \begin{center}
    \includegraphics[width=0.5\textwidth]{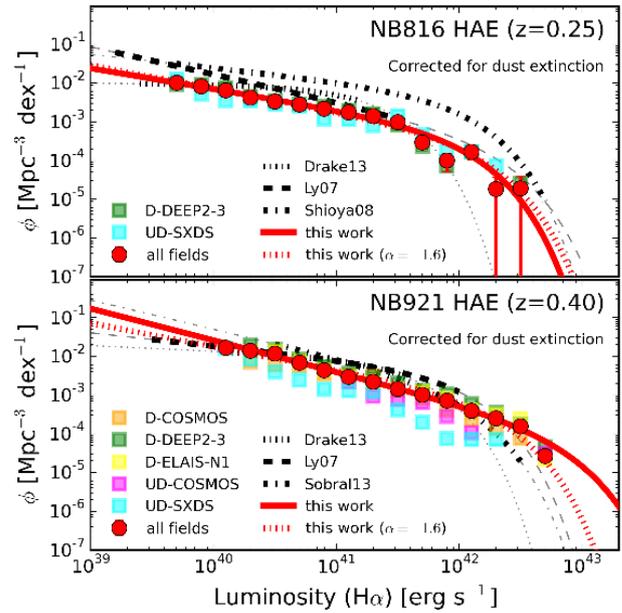}
  \end{center}
  \caption{
    The same as Figure~\ref{fig:LFobs_HAEs}, but the
    luminosities are corrected for dust extinction.
    The luminosity functions from the literature
    \citep{Ly2007,Drake2013,Sobral2013,Shioya2008} are plotted for
    comparison. 
  }\label{fig:LFcorr_HAEs}
\end{figure}

\begin{figure}[t]
  \begin{center}
    \includegraphics[width=0.5\textwidth]{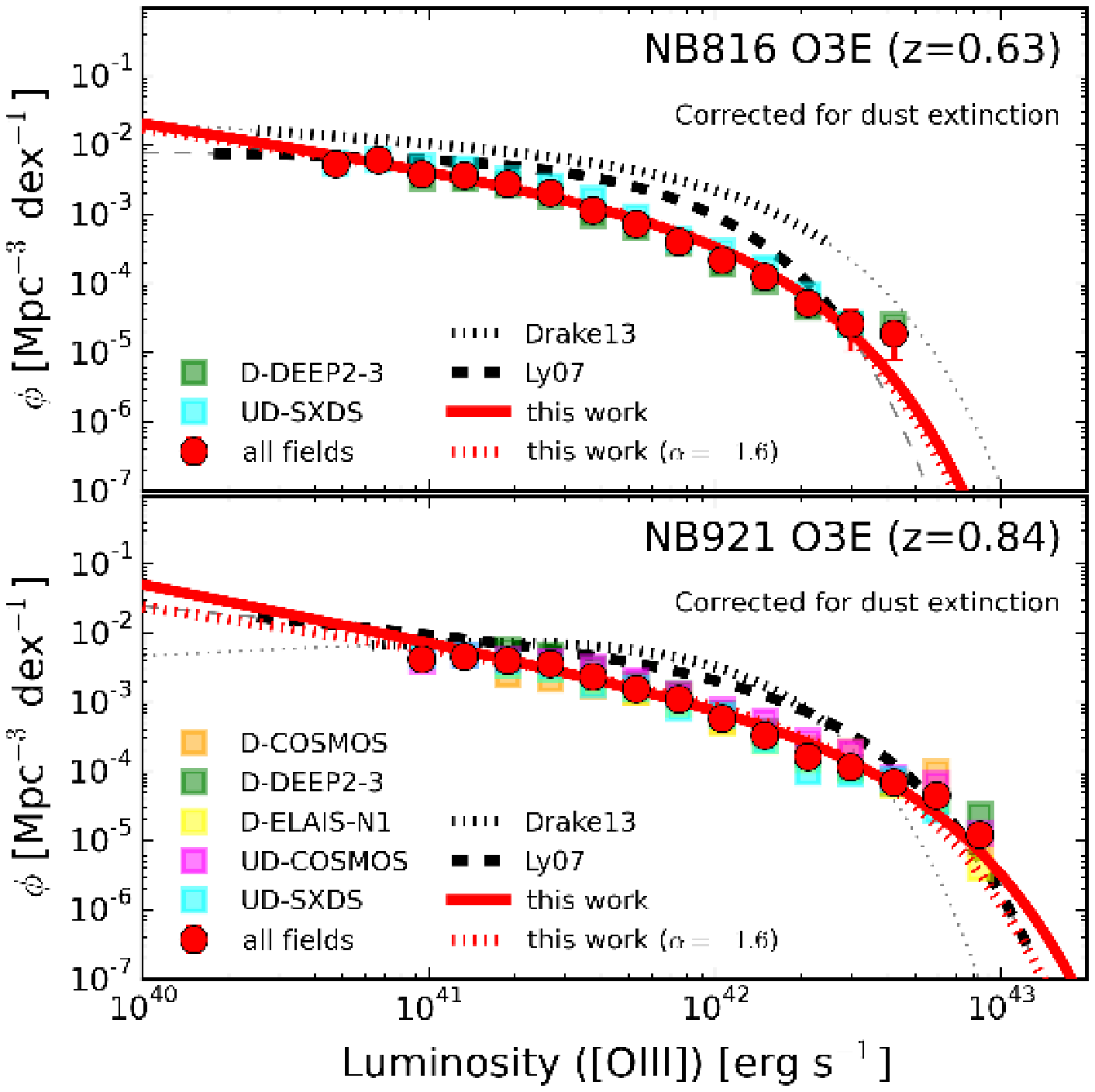}
  \end{center}
  \caption{
    The same as Figure~\ref{fig:LFobs_O3Es}, but the
    luminosities are corrected for dust extinction.
    The luminosity functions from the literature
    \citep{Ly2007,Drake2013} are plotted for
    comparison.
  }\label{fig:LFcorr_O3Es}
\end{figure}

\begin{figure}[t]
  \begin{center}
    \includegraphics[width=0.5\textwidth]{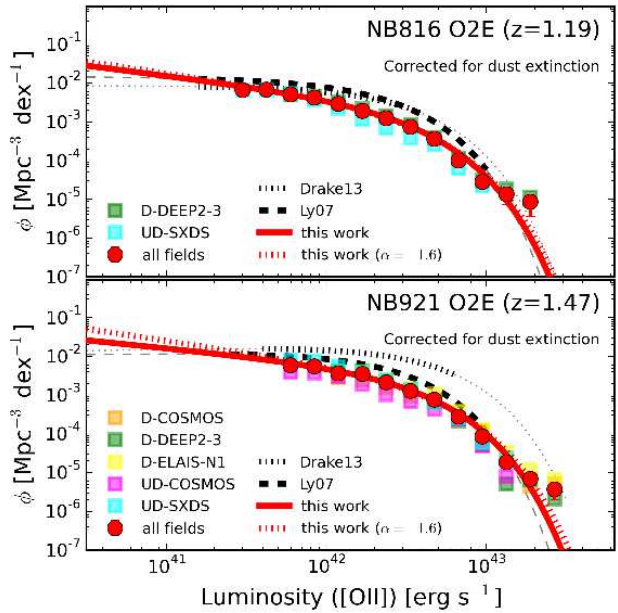}
  \end{center}
  \caption{
    The same as Figure~\ref{fig:LFobs_O2Es}, but the
    luminosities are corrected for dust extinction.
    The luminosity functions from the literature
    \citep{Ly2007,Drake2013} are plotted for
    comparison.    
  }\label{fig:LFcorr_O2Es}  
\end{figure}

Figures~\ref{fig:LFcorr_HAEs}--\ref{fig:LFcorr_O2Es} show the
luminosity functions corrected for dust extinction for HAEs, O3Es, and
O2Es at each redshift. The Schechter parameters of the best-fit
function are provided in Tables~\ref{tab:Schechter}--\ref{tab:Schechter_alphafixed}.
The luminosity functions from the previous studies are also plotted
for the comparison. The method of the dust correction is different
from ours, but our luminosity functions are consistent with the
previous studies.   
The number density of our emitters may be slightly smaller than the
previous studies at each luminosity bin, but that is likely due to the
relatively large EW cut in our samples.

%%%%%%%%%%%% Figure 17 %%%%%%%%%%%%%%%%%
%%%%%%%%%%%%%%%%%%%%%%%%%%%%%%%%%%%%%%%%%

%%%%%%%%%%%% Figure 18 %%%%%%%%%%%%%%%%%
%%%%%%%%%%%%%%%%%%%%%%%%%%%%%%%%%%%%%%%%%

%%%%%%%%%%%% Figure 19 %%%%%%%%%%%%%%%%%
%%%%%%%%%%%%%%%%%%%%%%%%%%%%%%%%%%%%%%%%

%%%%%%%%%%%%%%%%%%%%%%%%%%%%%%%%%%%%%%%%%%%%%%%%%%%%%%%%%%%%%%%%
%%%%%%%%%%%%%%%%%%%%%%%%%%%%%%%%%%%%%%%%%%%%%%%%%%%%%%%%%%%%%%%%

\section{Discussion}
\label{sec:discussions}

\subsection{Sample variance}
\label{sec:discussions.samplevariance}

Cosmic variance is a non-negligible problem to properly understand
representative properties of galaxy populations. Indeed,
Figures~\ref{fig:LFobs_HAEs}--\ref{fig:LFobs_O2Es} show that there is
a scatter in the luminosity functions for HAEs at $z=0.40$ between the
fields, although the other luminosity functions for HAEs at $z=0.25$,
O3Es and O2Es are mostly consistent between fields. 

The typical survey volume for HAEs at $z\sim0.40$ in the UD fields is
$\sim1\times10^{5}$ Mpc$^3$, which is a factor of $\sim5-10$
smaller than the survey volume of the other fields. 
The luminosity functions for the HAEs at $z\sim0.25$ in the two fields
and the HAEs at $z\sim0.40$ in the UD fields, all of which have a
survey volume of $<5\times10^{5}$ Mpc$^3$, are consistent with each other.
Also, the luminosity functions for the other emission-line galaxies in
larger survey volume are in agreement with each other between the
fields irrespective of the UD and D fields.

The slight discrepancy in luminosity function for the HAEs at
$z\sim0.4$ is between the UD and D fields. 
Given the luminosity functions shown in Figure~\ref{fig:LFobs_HAEs},
the survey volume of $\sim1\times10^{5}$ Mpc$^3$ in the UD fields
implies that there can be at most several galaxies detected in the
luminosity bins with $>10^{42}$ erg s$^{-1}$.
On the other hand, at $L\sim2\times10^{41}$ erg s$^{-1}$, where the
difference between the UD and D fields begins to become larger, there
should be dozens of galaxies detected in the luminosity bins even
in the UD fields.
According to \citet{Trenti2008}, we estimate the effect of cosmic
variance in the luminosity function of HAEs at $z\sim0.4$. The
estimated uncertainty in the number of HAEs is $\sim$40\%,
which suggests that the dispersion in the number density of HAEs with
$L>2\times10^{41}$ can be explained by the cosmic variance.
However, Figures~\ref{fig:LFobs_HAEs} and \ref{fig:LFcorr_HAEs} show
that the number density of HAEs in the UD fields are systematically
smaller than that in the D fields at any luminosity bins. This
indicates that the two UD fields may be an underdense region of HAEs
at $z\sim0.4$.  

Therefore, the comparison of the luminosity functions between the fields
shows that the luminosity functions in the individual fields are
consistent with each other and the survey volume of $>5\times10^{5}$
Mpc$^3$ is essential at least to overcome the cosmic variance.  
\citet{Sobral2015} also reach the similar conclusion that a survey
volume of $>5\times10^5$ Mpc$^3$ is required to derive the luminosity
functions with an error of less than 10\%\ irrespective of sample
variance.   

\subsection{Bright end of luminosity function}
\label{sec:discussions.brightendoflf}

It is known that hard radiation from AGNs can contribute to emission
lines from galaxies. The larger ionization energy required for [OIII]
than [OII] and H$\alpha$ implies that there is a possibility that the
fraction of AGN contamination in O3Es is higher than HAEs and O2Es.
However, contrary to the expectations, no strong excess of luminosity
function is seen at the bright end for HAEs and O3Es
(Figures~\ref{fig:LFobs_HAEs} and \ref{fig:LFobs_O3Es}), while a
slight excess is seen in the luminosity functions for O2Es
(Figure~\ref{fig:LFobs_O2Es}). Similar excess in bright end of
luminosity function is reported by the previous studies
\citep{Matthee2017}. 

As mentioned in \S~\ref{sec:data.nbquality.numbercount}, we limit the
samples to galaxies fainter than 17.5 mag. Taking account of the EW
cut as well that we apply in the selection of emission-line galaxies,
we can miss most of NB816 (NB921) emitters with line fluxes larger
than 5.5 (5.0) $\times$ 10$^{-15}$ erg s$^{-1}$ cm$^{-1}$. 
The fluxes corresponds to $L$/[erg s$^{-1}$] =
1.0$\times$10$^{42}$, 9.3$\times$10$^{42}$, and 4.4$\times$10$^{43}$ for
NB816 HAEs, O3Es, and O2Es,
and 2.9$\times$10$^{42}$, 1.7$\times$10$^{43}$, and 6.8$\times$10$^{43}$ for
NB921 HAEs, O3Es, and O2Es, respectively.
Therefore, the HSC-SSP PDR1 data are capable of selecting luminous
emitters in the luminosity ranges investigated,
although some emission-line galaxies at the bright end can be missed
if more luminous galaxies tend to have a larger EW (e.g., \cite{Sobral2014}).  
The limitation on the bright end confirms that there is no deviation
of luminosity function from the Schechter function in HAEs and O3Es,
while there is a slight deviation of O2E luminosity function from the
Schechter function at the bright end. 

\citet{Matthee2017} argue that the excess of the bright end of
luminosity function is caused by AGNs and for HAEs at $z\sim$1.47 and
2.2 the fraction of X-ray sources in the HAEs monotonically increases
with the H$\alpha$ luminosity at $L({\rm H}\alpha)>10^{42.5}$ erg
s$^{-1}$ irrespective of redshift (see also \cite{Sobral2016}).
On the other hand, as described in \S \ref{sec:ELGs.AGN}, matching our
samples of emission-line galaxies to the catalog of the Chandra
X-ray sources in the COSMOS field \citep{Marchesi2016} indicates that
at most 0.1\%\ of the emission-line galaxies are detected in X-ray as well.
The fraction of O2Es with an X-ray counterpart is similar to that for
HAEs and O3Es. Furthermore, since the X-ray counterpart of O2Es have
the observed [OII] luminosity of $<10^{42}$ erg s$^{-1}$, the O2Es
with X-ray detection do not have a large [OII] luminosity contributing
to the bright end of the luminosity function.
Indeed, the intrinsic luminosity functions corrected for dust
extinction for not only O2Es but also HAEs and O3Es show no excess
from the Schechter function at the bright end
(Figures~\ref{fig:LFcorr_HAEs}--\ref{fig:LFcorr_O2Es}). There are few
emission-line galaxies with the luminosities of $L>10^{42.5}$ erg 
s$^{-1}$, which may cause the discrepancy between our results and
those of \citet{Matthee2017}. Therefore, we conclude that a strong
excess at the bright end is not seen in our luminosity functions.      

\subsection{Stellar mass function}
\label{sec:discussions.SMF}

The stellar mass of galaxies is one of fundamental properties to
characterize galaxies. There are many properties such as SFR,
metallicity, dust extinction, and size that show a strong correlation
with stellar mass. Since our samples of emission-line galaxies are
based on the luminosity of nebular emissions which are indicators of
SFR, they are SFR-limited samples to a first order
approximation. Considering that there is a tight correlation between
stellar mass and SFR in star-forming galaxies at each redshift (e.g., 
\cite{Speagle2014}), it is expected that NB emission-line galaxy
samples can contain star-forming galaxies with stellar mass down to a
certain completeness limit. Therefore, it is interesting to
investigate the stellar mass function of emission-line galaxies to
better understand our samples.   

We discuss for only HAEs hereafter in this section. This is because
the stellar masses for HAEs are reliable, but those for O3Es and O2Es
at $z>0.8$ can be systematically overestimated (see
\S\ref{sec:ELGs.SEDFIT} and \cite{hscPhotoZ}).  

\begin{figure}[t]
  \begin{center}
    \includegraphics[width=0.5\textwidth]{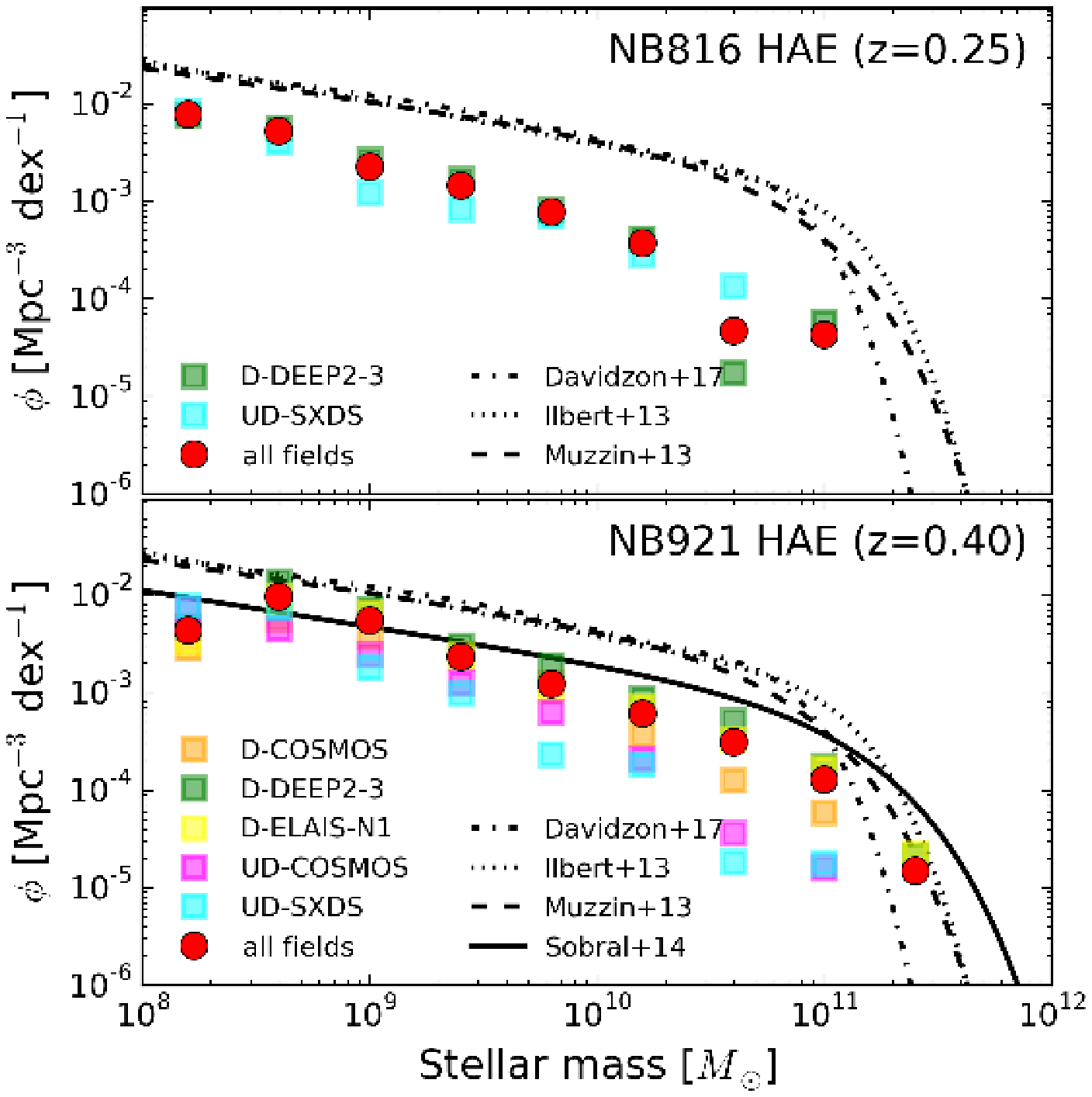}
  \end{center}
  \caption{
    Stellar mass function for HAEs. 
    The luminosity functions from the literature are plotted for
    comparison: HAEs at $z=0.4$ \citep{Sobral2014} and star-forming
    galaxies at $z =$ 0.2--0.5 \citep{Ilbert2013,Muzzin2013,Davidzon2017}
  }\label{fig:MassFunc_HAEs}  
\end{figure}

Figure~\ref{fig:MassFunc_HAEs} shows stellar mass function for HAEs.
The stellar mass functions of HAEs at $z\sim0.4$ show the difference
in mass bins of $>4\times10^{10}$ M$_\odot$ between the two fields of
D-DEEP2-3 and D-ELAIS-N1 and the others. Figure~\ref{fig:map_HAEs}
shows that the two fields have several galaxy clusters. Since the
galaxy clusters are identified by red-sequence galaxies, the
clustering of massive galaxies is reasonable. 

We compare our stellar mass function with that of \citet{Sobral2014}
(solid line) for HAEs at $z=0.4$. While our stellar mass function
is consistent with that of \citet{Sobral2014} at $M<2\times10^9$
M$_\odot$, the number density of our HAEs is smaller than the
\citet{Sobral2014} sample at $M>2\times10^9$ M$_\odot$. As mentioned
in \S~\ref{sec:ELGs.selection}, the HSC-SSP PDR1 data can select the
HAEs with the rest-frame EW larger than 40\AA, while the sample of
\citet{Sobral2014} includes the HAEs with the rest-frame EW larger
than 25\AA. Therefore, we miss the HAEs with small EWs that
\citet{Sobral2014} can select. Since more massive galaxies tend to
have a smaller EW on average (e.g., \cite{Sobral2014}), the larger
discrepancy between our stellar mass function and that of
\citet{Sobral2014} at higher stellar masses is likely due to the 
difference of the limiting EW in the HAE selection. 

Next, we compare our stellar mass function with those from
star-forming galaxies at $z=$ 0.2--0.5 \citep{Ilbert2013,Muzzin2013,Davidzon2017}.
They construct the NIR-detected samples (i.e., stellar mass-limited
samples in the first order approximation) in the COSMOS field covering
1.5--2 deg$^2$. Our stellar mass function is much lower than those
from the literature and the discrepancy is larger in HAEs at $z=0.25$
than HAEs at $z=0.40$. This can be also caused again by the EW cut
crucial for the NB emitter selection, as discussed above. 

Our samples cover the stellar mass range from $>10^{11}$ M$_\odot$
down to $\sim10^{8}$ M$_\odot$. The deeper data in both NB and BB are
required to apply the smaller EW cut in the selection of emission-line
and then make the more representative samples of star-forming 
galaxies.  

%%%%%%%%%%%% Figure 20 %%%%%%%%%%%%%%%%%
%%%%%%%%%%%%%%%%%%%%%%%%%%%%%%%%%%%%%%%%

%%%%%%%%%%%%%%%%%%%%%%%%%%%%%%%%%%%%%%%%%%%%%%%%%%%%%%%%%%%%%%%%
%%%%%%%%%%%%%%%%%%%%%%%%%%%%%%%%%%%%%%%%%%%%%%%%%%%%%%%%%%%%%%%%

\section{Conclusions}
\label{sec:conclusions}

The Subaru Strategic Program (SSP) with the Hyper Suprime-Cam (HSC) is
an imaging survey with 300 nights over 5-6 years.
The HSC-SSP provides us with widest and deepest data in five
broadbands, $g, r, i, z,$ and $y$, until the the Large Synoptic Survey
Telescope (LSST) begins to operate, and another characteristic of the
survey is that data from four narrowbands, NB387, NB816, NB921, and
NB101 are available. The first public data release (PDR1) of the
HSC-SSP is now publicly available, and consists of data taken over
61.5 nights.

We present initial results on emission-line galaxies at $z<1.5$
selected with the data from two narrowband filters of NB816 and NB921.
The NB816 (NB921) data are available in the PDR1 over 5.7 (16)
deg$^2$. The narrowband data enable us to select galaxies emitting
H$\alpha$($\lambda6565$\AA) from $z \approx$ 0.25 and 0.40,
[OIII]($\lambda5008$\AA) from $z \approx$ 0.63 and 0.84, and
[OII]($\lambda\lambda3727,3730$\AA) from $z \approx$ 1.19 and 1.47.
Main results that we present in this paper are the following.

\begin{enumerate}

\item This is one of the largest samples of emission-line galaxies at
  $z<1.5$ ever constructed, which includes 
  1,193 (6,861) H$\alpha$ emitters at $z \approx$ 0.24 (0.40) down to the
  luminosities of $\log$(L(H$\alpha$)/[erg s$^{-1}$])$\gtrsim$ 39.4 (40.1),
  2,228 (6,428) [OIII] emitters at $z \approx$ 0.63 (0.84) down to the
  luminosities of $\log$(L([OIII])/[erg s$^{-1}$])$\gtrsim$ 40.4 (40.8), and 
  5,861 (11,016) [OII] emitters at $z \approx$ 1.19 (1.47) down to the
  luminosities of $\log$(L([OII])/[erg s$^{-1}$])$\gtrsim$ 41.1 (41.4).  
  The summary is shown in Table~\ref{tab:Samples}.
  The catalogs are available at the HSC-SSP
  website\footnotemark[\ref{foot:hscsspwebsite}],
  after the paper is published.
  
\item The spatial distribution of the emitters shows large-scale
  structures over $\gtrsim$ 50 Mpc where galaxy clusters discovered
  by a red sequence cluster finding algorithm are embedded
  (Figures~\ref{fig:map_HAEs}--\ref{fig:map_O2Es}).
  This indicates that the samples include star-forming galaxies in
  various environments covering from the core of galaxy cluster to the
  void. The samples of emission-line galaxies are useful to
  investigate the environmental dependence of galaxy evolution at
  $z<1.5$. 

\item Luminosity functions of the emitters are investigated for
  H$\alpha$ emitters at $z \sim$ 0.25 and 0.40, [OIII] emitters at
  $z \sim$ 0.63 and 0.84, and [OII] emitters at $z \sim$ 1.19 and
  1.47 (Figures~\ref{fig:LFobs_HAEs} -- \ref{fig:LFobs_O2Es}).
  The luminosity functions are mostly consistent with the previous 
  studies at the luminosity ranges investigated. However, the number
  densities of our emitters may be slightly smaller than those of the
  previous studies. The slight discrepancy can be caused by the larger
  equivalent width cut in the selection of our emitters than the
  previous studies.
  The luminosity functions in each field surveyed by the HSC-SSP are
  consistent with each other,
  suggesting that more than $5\times10^5$ Mpc$^3$ is required at least
  to overcome the field variance.  

\item Stellar mass functions are investigated for H$\alpha$ emitters
  at $z \sim$ 0.25 and 0.40 (Figure~\ref{fig:MassFunc_HAEs}), and then
  compared with the previous studies for H$\alpha$ emitters at
  $z=0.40$ and star-forming galaxies at $z=$ 0.2--0.5 from the stellar
  mass limited samples. The larger discrepancy is seen at higher
  stellar masses in H$\alpha$ emitters at both $z=0.25$ and $z=0.40$,
  which suggests that there are star-forming galaxies with emission
  lines with small equivalent width that our NB imaging survey
  misses. The deeper data are essential to construct more
  representative samples of star-forming galaxies at the redshifts
  probed.  

\end{enumerate}

The HSC-SSP survey is ongoing. As the survey proceeds further, the
coverage where NB data are available will be by a factor of 1.6 wider
in the D fields and the depth of NB816 and NB921 is 0.5--0.6 (0.2--0.3)
mag deeper in the UD (D) fields than those in the PDR1 data
\citep{HSCSSPDR1}. The wider and deeper data allow us to survey less
massive star-forming galaxies with a lower SFR at higher selection
completeness and derive fainter end of luminosity function more
reliably. We will update the catalogs of emission-line galaxies based
on the upcoming data releases. 

%%%%%%%%%%%% Figure 21 %%%%%%%%%%%%%%%%%
\begin{figure*}
  \begin{center}
    \includegraphics[width=1.0\textwidth]{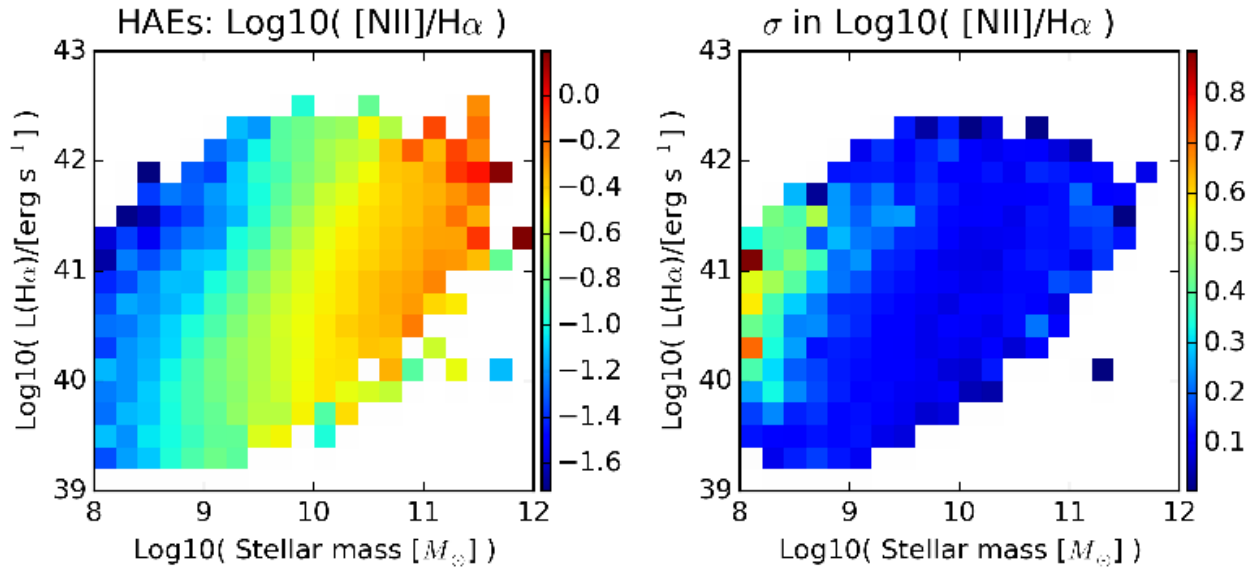}
  \end{center}
 \caption{Emission-line ratio of [NII]/H$\alpha$ for the SDSS analogs
   of HAEs. The galaxies are selected from the SDSS spectroscopic
   catalog by applying the same EW cut as for the NB-selected HAEs.
   The left panel shows the [NII]/H$\alpha$ ratio in a log scale in
   each bin, while the right panel shows the standard deviation of the
   line ratio in a log scale in each bin. 
 }\label{fig:lr_sdss_HAE_N2}  
\end{figure*}
%%%%%%%%%%%%%%%%%%%%%%%%%%%%%%%%%%%%%%%%

%%%%%%%%%%%%%%%%%%%%%%%%%%%%%%%%%%%%%%%%%%%%%%%%%%%%%%%%%%%%%%%%
%%%%%%%%%%%%%%%%%%%%%%%%%%%%%%%%%%%%%%%%%%%%%%%%%%%%%%%%%%%%%%%%

%%%                 %%%  
%%% Acknowledgement %%%
%%%                 %%%  

\begin{ack}
We thank the anonymous referee for providing constructive comments.
MH acknowledges the financial support by JSPS Grant-in-Aid for Young
Scientists (A) Grant Number JP26707006.
This work is supported by World Premier International Research Center
Initiative (WPI Initiative), MEXT, Japan, and KAKENHI (15H02064)
Grant-in-Aid for Scientific Research (A) through Japan Society for the
Promotion of Science (JSPS). 

% acknowledgement to Subaru
This work is based on data collected at Subaru Telescope, which is
operated by the National Astronomical Observatory of Japan.
% acknowledgement for HSC NB filter
The NB816 filter was supported by Ehime University and the NB921
filter was supported by KAKENHI (23244025) Grant-in-Aid for Scientific
Research (A) through the JSPS.

% acknowledgement for HSC SSP
The Hyper Suprime-Cam (HSC) collaboration includes the astronomical
communities of Japan and Taiwan, and Princeton University. The HSC
instrumentation and software were developed by the National
Astronomical Observatory of Japan (NAOJ), the Kavli Institute for the
Physics and Mathematics of the Universe (Kavli IPMU), the University
of Tokyo, the High Energy Accelerator Research Organization (KEK), the
Academia Sinica Institute for Astronomy and Astrophysics in Taiwan
(ASIAA), and Princeton University. Funding was contributed by the
FIRST program from Japanese Cabinet Office, the Ministry of Education,
Culture, Sports, Science and Technology (MEXT), the Japan Society for
the Promotion of Science (JSPS), Japan Science and Technology Agency
(JST), the Toray Science Foundation, NAOJ, Kavli IPMU, KEK, ASIAA, and
Princeton University.
% acknowledgement to LSST
This paper makes use of software developed for the Large Synoptic
Survey Telescope. We thank the LSST Project for making their code
available as free software at \footnote{http://dm.lsst.org}.
% acknowledgement to Pan-STARRS1 (PS1)                   
The Pan-STARRS1 Surveys (PS1) have been made possible through
contributions of the Institute for Astronomy, the University of
Hawaii, the Pan-STARRS Project Office, the Max-Planck Society and its
participating institutes, the Max Planck Institute for Astronomy,
Heidelberg and the Max Planck Institute for Extraterrestrial Physics,
Garching, The Johns Hopkins University, Durham University, the
University of Edinburgh, Queen’s University Belfast, the
Harvard-Smithsonian Center for Astrophysics, the Las Cumbres
Observatory Global Telescope Network Incorporated, the National
Central University of Taiwan, the Space Telescope Science Institute,
the National Aeronautics and Space Administration under Grant
No. NNX08AR22G issued through the Planetary Science Division of the
NASA Science Mission Directorate, the National Science Foundation
under Grant No. AST-1238877, the University of Maryland, and Eotvos
Lorand University (ELTE) and the Los Alamos National Laboratory.
\end{ack}

%%%%%%%%%%%%%%%%%%%%%%%%%%%%%%%%%%%%%%%%%%%%%%%%%%%%%%%%%%%%%%%%
%%%%%%%%%%%%%%%%%%%%%%%%%%%%%%%%%%%%%%%%%%%%%%%%%%%%%%%%%%%%%%%%

%%%%%%%%%%%% Figure 22 %%%%%%%%%%%%%%%%%
\begin{figure*}
  \begin{center}
    \includegraphics[width=0.9\textwidth]{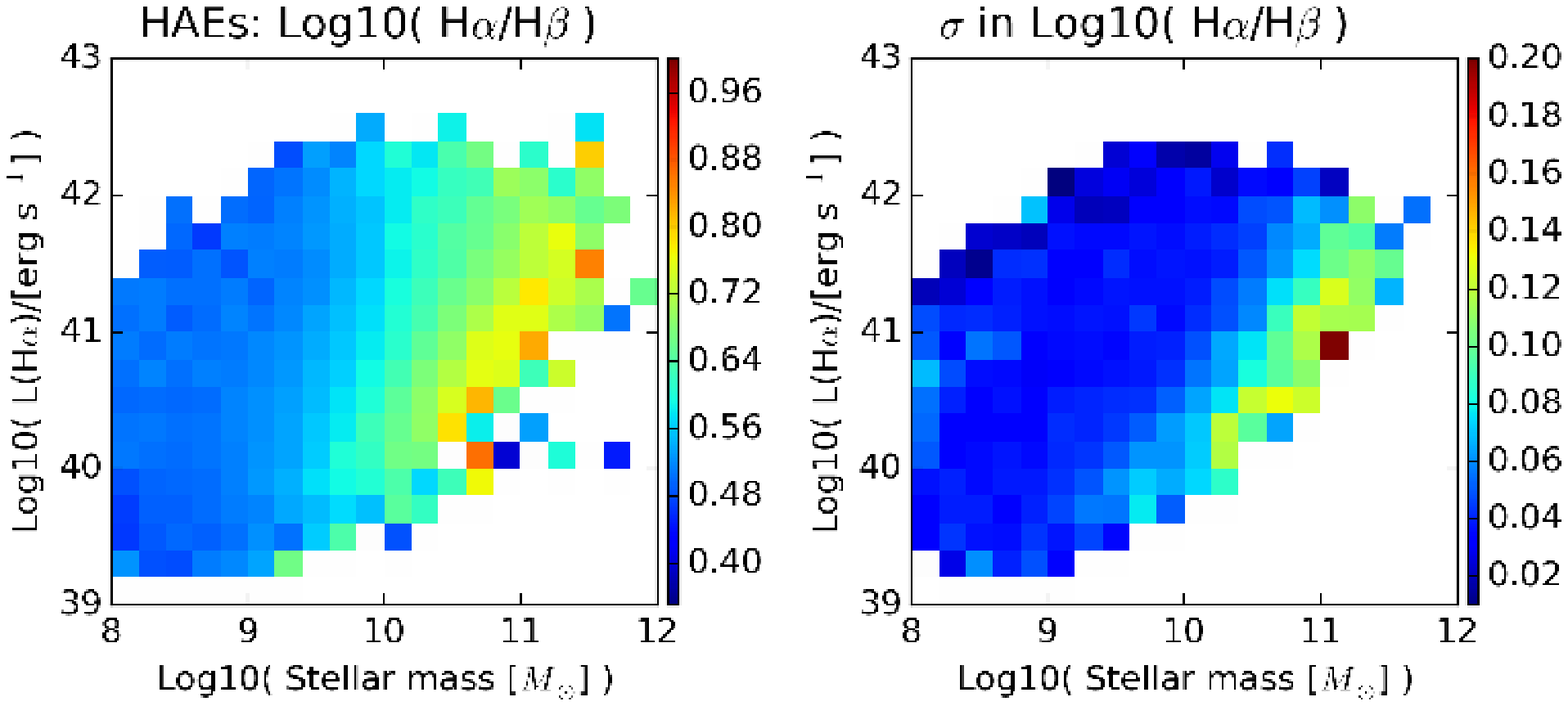}\\[-5mm]
    \includegraphics[width=0.9\textwidth]{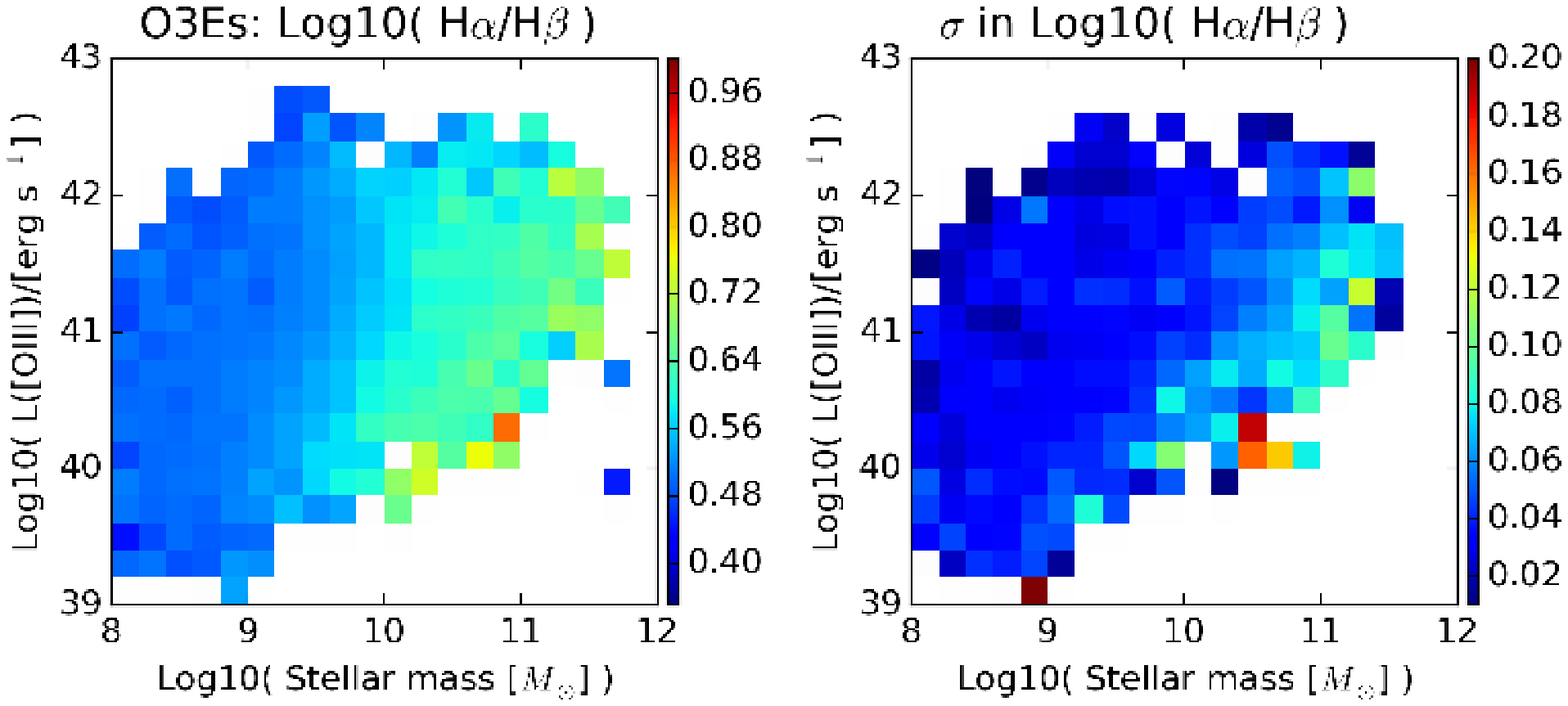}\\[-5mm]
    \includegraphics[width=0.9\textwidth]{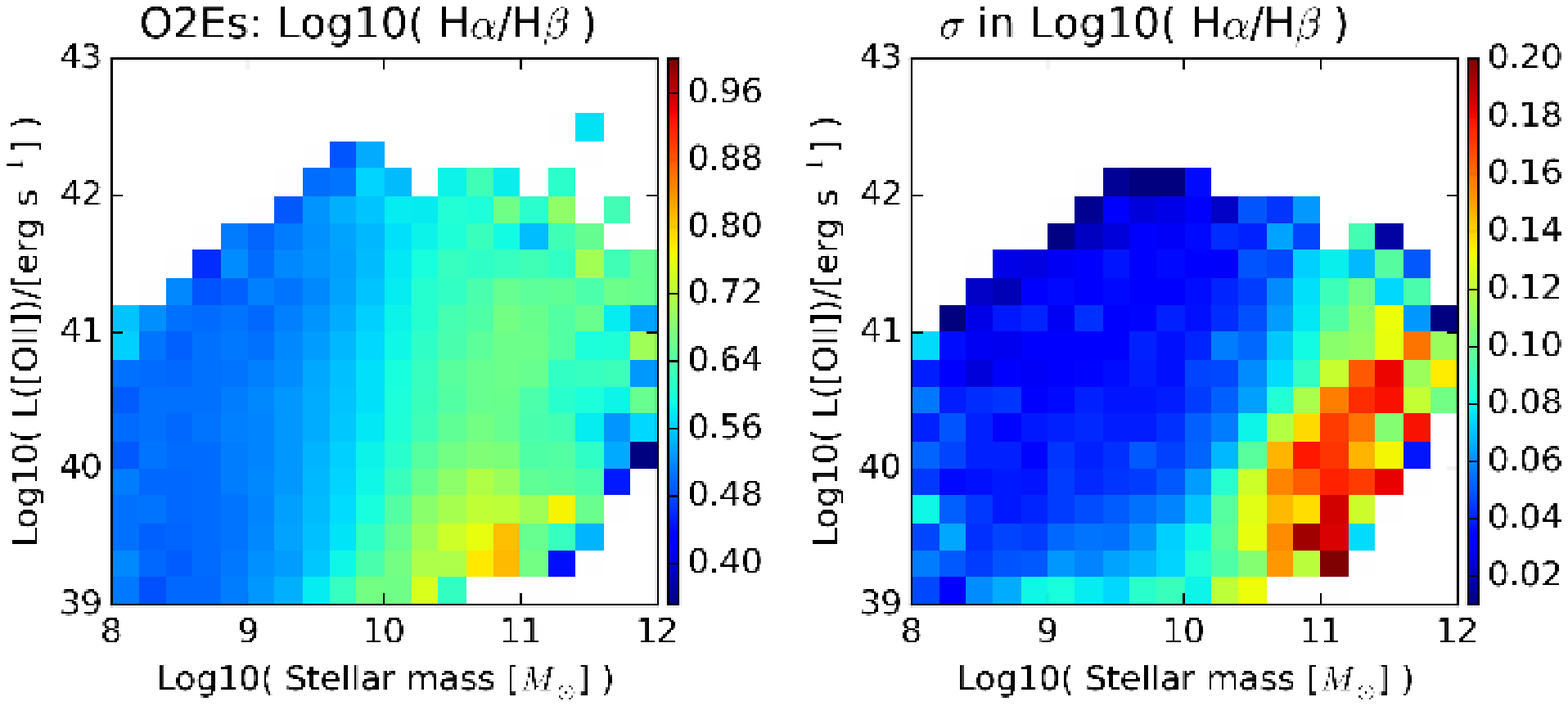}\\[-5mm]
  \end{center}
 \caption{The same as Figure~\ref{fig:lr_sdss_HAE_N2}, but for the
   line ratio of H$\alpha$/H$\beta$. The line ratios for the SDSS
   analogs of HAEs, O3Es, and O2Es are shown in top, middle, and
   bottom panels, respectively.
 }\label{fig:lr_sdss_HAHB}  
\end{figure*}
%%%%%%%%%%%%%%%%%%%%%%%%%%%%%%%%%%%%%%%%

%%%          %%%  
%%% Appendix %%%
%%%          %%%  

\appendix 

\section{Flags used for selection of NB-detected objects}
\label{app:catalog}

hscPipe outputs a lot of flags showing reliability and measurement
in source detection and photometry in the process of the data  
\citep{hscPipe}.
We apply the following flag cut to select NB-detected objects (refer
to \cite{hscPipe} for the meaning of the individual
flags):

\begin{description}  

\item[In NB816 or NB921]\mbox{}\\
{\tt detect\_is\_primary=True,\\
merge\_peak=True,\\
flags\_pixel\_cr\_center=False,\\
flags\_pixel\_saturated\_center=False,\\
flags\_pixel\_bad=False,\\
flags\_pixel\_suspect\_center=False,\\
flags\_pixel\_offimage=False,\\
flags\_pixel\_bright\_object\_center=False,\\
centroid\_sdss\_flags=False,\\
cmodel\_flux\_flags=False}, and\\
{\tt detected\_notjunk=True.}\\

\item[In $i$ ($z$)-band for NB816 (NB921)-detected objects]\mbox{}\\
{\tt flags\_pixel\_cr\_center=False,\\
flags\_pixel\_saturated\_center=False,\\
flags\_pixel\_bad=False,\\
flags\_pixel\_suspect\_center=False}, and\\
{\tt flags\_pixel\_offimage=False}. 

\item[In all BBs]\mbox{}\\
  {\tt flags\_pixel\_bright\_object\_center=False}.

\end{description}

\section{Emission-line ratios for the SDSS analogs of NB emitters}
\label{app:lineratio}

Since the [NII] doublet ($\lambda\lambda$6549.84,6585.23\AA) is close
to H$\alpha$ ($\lambda$6564.61\AA) line, all of the emission lines can
enter the NB filter simultaneously. Also, emission-line fluxes can suffer
dust extinction. To derive intrinsic luminosity of a specific  
emission line, a correction for the neighbor lines and the dust
extinction is required. The SDSS data allow us to estimate such impact
on the emission-line fluxes measured from the NB and BB photometry by
selecting analogs of NB emission-line galaxies and then investigating
the line ratios of emission lines. 

The spectroscopic catalogs are extracted from the MPA-JHU release of
spectrum measurements for the SDSS Data Release 7 (DR7)
\footnote{http://www.mpa-garching.mpg.de/SDSS/DR7/}
\citep{Kauffmann2003,Salim2007,Abazajian2009}. To select the analogs
of NB-selected emission-line galaxies, we apply the same EW cut as for
the NB selection. We then investigate the various line flux ratio as a
function of the observed emission-line luminosity and stellar mass.    

Figure~\ref{fig:lr_sdss_HAE_N2} shows the [NII]/H$\alpha$ ratios for the
analogs of HAEs as a function of the H$\alpha$ luminosity and stellar
mass. The ratios are used to correct the fluxes measured from the
HSC-SSP data for the contribution of
[NII]. Figure~\ref{fig:lr_sdss_HAHB} is the same as
Figure~\ref{fig:lr_sdss_HAE_N2}, but for the H$\alpha$/H$\beta$
ratios, i.e., Balmer decrement for the analogs of HAEs, O3Es and
O2Es. This is used for the correction of dust extinction.
Figures~\ref{fig:lr_sdss_HAE_N2} -- \ref{fig:lr_sdss_HAHB} also show
the standard deviation of the line ratio in each bin to infer an accuracy
of the determination of the line ratios based on the stellar mass and
the luminosity of the targeted emission line.  

%%%%%%%%%%%%%%%%%%%%%%%%%%%%%%%%%%%%%%%%%%%%%%%%%%%%%%%%%%%%%%%%
%%%%%%%%%%%%%%%%%%%%%%%%%%%%%%%%%%%%%%%%%%%%%%%%%%%%%%%%%%%%%%%%

%%%            %%%  
%%% References %%%
%%%            %%%  

\bibliographystyle{apj}
\bibliography{./reference}

\end{document}